\newcommand{\ignore}[1]{}
\documentclass[12pt]{article}
\usepackage[left]{lineno}
\usepackage{amsmath}
\usepackage{amssymb}
 \usepackage{graphicx}
\usepackage{braket}
\usepackage{authblk}
\usepackage{caption,subcaption}
\numberwithin{equation}{section}
\numberwithin{figure}{section}
\graphicspath{{fig/}}
\renewcommand{\baselinestretch}{1.1}
\newlength{\abstractwidth}
\usepackage{color}

 \newcommand{\be}{\begin{linenomath*}\begin{equation}}
 \newcommand{\bea}{\begin{eqnarray}}
 \newcommand{\eea}{\end{eqnarray}}
 \newcommand{\ee}{\end{equation}\end{linenomath*}}
 \newcommand{\eeq}{\end{equation}}
 
 \newcommand{\<}{\langle\,}

 \newcommand{\half}{{\frac{1}{2}}}
 \renewcommand{\>}{\rangle}
\newcommand\dd{\partial}
\newcommand\nn{\nonumber \\}
\newcommand\mS{{\mathbb S}}
\newcommand\mR{{\mathbb R}}
\newcommand\mT{{\mathbb T}}
\newcommand\cO{{\cal O}}

\renewcommand\ell{l}

\title{Lattice Dirac Fermions on a Simplicial Riemannian Manifold }

\author[*]{ Richard C. Brower}
\author[\#]{George T. Fleming}
\author[\#]{Andrew D. Gasbarro}
\author[+]{\\Timothy G. Raben}
\author[+]{Chung-I Tan}
\author[*]{ Evan S. Weinberg}
\affil[*]{Boston University, Boston, MA 02215}
\affil[\#]{Yale University, Sloane Laboratory, New Haven, CT 06520} 
\affil[+]{Brown University, Providence, RI 02912}

\date{\today}
\begin{document}


\maketitle


\begin{abstract}
  The lattice Dirac equation is formulated on a  simplicial
complex which  approximates a  smooth Riemann manifold by introducing a
  lattice vierbein on each site and a lattice spin connection on each link.  Care
  is taken so the construction applies to any smooth D-dimensional
  Riemannian manifold that permits a spin connection.  It is 
  tested numerically in 2D  for the projective sphere
  $\mathbb S^2$ in the limit of an increasingly refined sequence of
  triangles. The  eigenspectrum and eigenvectors are shown to converge rapidly to
the exact result in  the continuum limit. In addition comparison is
made with the continuum  Ising conformal field theory  on
$\mS^2$.  Convergence is tested for the two point, $\<
\epsilon(x_1) \epsilon(x_2) \>$,  and the four point,  $\<
\sigma(x_1)  \epsilon(x_2) \epsilon(x_3 )\sigma(x_4) \>$, correlators
for the energy, $\epsilon(x) = i \bar \psi(x)\psi(x)$,   and twist
operators, $\sigma(x)$, respectively.
\end{abstract}

\setlength{\parskip}{0in}
\thispagestyle{empty}
\setcounter{page}{-1}
\pagebreak
\tableofcontents
\thispagestyle{empty}
\setcounter{page}{0}
\pagebreak
\setcounter{page}{1}
\setlength{\parskip}{.2in}
\newpage

\section{\label{sec:intro}Introduction}

Lattice gauge theory on hypercubic
lattices~\cite{Appelquist:2013sia}  provides a
powerful {\em ab initio} approach to strongly coupled field
theories in flat Euclidean space, $\mathbb R^D$.
However there is  important  non-perturbative physics that would benefit from the extension
of lattice field theory methods to  more general curved  Riemann
manifolds.  One example is a recent proposal to
implement radial quantization  for 
conformal field theories~\cite{Brower:2012vg, Brower:2014gsa,
  Brower:2015zea}. This requires replacing the flat Euclidean manifold,
$\mathbb R^D$, by the cylinder,  $\mathbb R \times
\mathbb  S^{D-1}$, which represents   the  boundary of Anti-de Sitter space
$AdS^{D+1}$ in global coordinates.  Other examples include {\em ab initio}  calculations of the $c$
and $a$ terms, tests of AdS/CFT duality, quantum criticality in
condensed matter  and perhaps  quantum physics near blackholes.

The  conventional lattice regulator  in flat space is  a  sequence of hypercubic  lattices 
on a torus, $\mT^D$, with a uniform lattice spacing $a$, representing an
increasingly larger subgroup of translations as  the cut-off,
$\Lambda_{UV} = \pi/a$,  is removed. Curved manifolds
lack such  uniform sequences of lattices.
For example, on a sphere, the finest uniform
discretization of $\mS^2$ and $\mS^3$   are the  20-cell icosahedron
and the 600-cell tetraplex respectively.  The lack of an infinite sequence of regular
lattices approaching the continuum compounds the problem of
renormalization and symmetry restorations as the cut-off is
removed. This paper is part of research to develop a  general strategy~\cite{Brower:2016moq}, referred to as Quantum Finite Elements ({\bf
  QFE}), to formulate a lattice field theory path integral for  any renormalizable quantum field theory on a smooth
Riemann manifold (${\cal M}, g$) given the target metric tensor, $g_{\mu \nu}(x)$.  Here we
focus on the construction  of  the free lattice Dirac fermion.
The fermion is an especially challenging and interesting example. The spinor
probes the underlying geometry of the manifold through its vierbein
and  spin connection. From the perspective of   Regge Calculus~\cite{Regge:1961px}, the
vierbein and spin connection are sufficient to define a 
simplicial manifold 
in the Einstein-Cartan formulation of lattice
gravity~\cite{Hamber:2009mt}.  Consequently the fermion lattice field
may also provide 
an alternative approach to reconstructing the intrinsic geometry for the Regge
Calculus approximation to the base Riemann manifold.

The organization of the paper is as follows. In Sec.~\ref{sec:Riemann}, to establish our notation and basic formalism, we review the Finite
Element approximation to scalar field theory on a Riemann
manifold. While we borrow heavily from the conventional piecewise
linear form for Regge Calculus (RC) and Finite Element Method (FEM), it is
important to note that these approximations do not by themselves adequately address
our problem. (Readers familiar with finite elements may prefer to
first skip this
introduction and return for notation.)  In Sec.~\ref{sec:Dirac}, we begin the construction for
the Dirac field, emphasizing the new problem of defining the lattice
vierbein and spin connection and removing doublers on the simplicial
complex. In Sec.~\ref{sec:SpinConnection} we formulate an algorithm
for fixing the lattice vierbein and spin connection designed to
converge to any target smooth Riemann manifold $({\cal M}, g)$.  In
Sec.~\ref{sec:numerics} we test the method  for the Dirac fermion and its rate of convergence
for the $\mathbb{S}^2$ sphere compared to the exact continuum theory.  In
Sec.~\ref{sec:projectiveSphere} the simplicial Majorana fermion on
$\mS^2$ is shown to converge to the analytical result for 2-point and
4-point correlation functions for the c = 1/2 minimal model conformal
field theory. In Sec.~\ref{sec:conclusion} we discuss extensions and
future directions in the study of quantum field theories with gauge
and scalar fields. Several technical details are relegated to the
appendices.

\section{\label{sec:Riemann}Review of Scalar Fields  on a Simplicial Lattice}

Lattice field theory on a Riemann manifold $({\cal M} ,g)$ requires a
discrete definition for the metric field, $g_{\mu \nu}(x)$, and the
quantum fields, scalars $\phi(x)$, fermions $\psi(x)$, and gauge
fields $A_\mu(x)$.  Aspects of this problem have been considered
extensively in a number of related fields. One example is Regge
Calculus (RC), which introduces an ensemble of piecewise flat
simplicial lattices as a basis for non-perturbative quantum
gravity~\cite{Regge:1961px}. A second example is FEM, designed to discretize  partial differential equations
and to solve them numerically~\cite{StrangFix200805}. The third
example involves a formal geometrical framework~\cite{2005math8341D}
for a Discrete Exterior Calculus (DEC) on the Delaunay lattice $\cal S$ and
its circumcenter Voronoi dual $\cal S^*$.  We should also emphasize the
classic study of field theory on random lattices by Christ, Friedberg
and Lee (CFL)~\cite{Christ:1982ci,Christ1982,Christ1982a} that in
fact anticipated much of the relevant FEM and DEC formalism for the
simplicial lattice field theory in flat space. 

Each method provides some useful and closely related tools, but they
do not fully address the problems of a rigorous simplicial lattice
representation guaranteed to converge to the continuum for
renormalizable quantum field theories---the ultimate goal of this
research.  Both the RC and the CFL approaches introduce
a random ensemble of  simplicial lattices in order to hopefully restore continuum
symmetries (diffeomorphisms, chiral symmetry, etc.) of the target quantum
field theory. Here we do not advocate this approach.  {\bf
  Instead we impose regularity on a single sequence of increasingly
  refined simplicial lattices designed  to approach the continuum limit
  on a fixed target Riemann manifold. }  Our approach depends on
combining two elements. First the {\bf classical} FEM method provides
the theoretical framework of convergence~\cite{StrangFix200805} in the
IR for all solutions to the equation of motion (EOM or PDEs) smooth
enough to be insensitive to the UV cutoff. Second, counter terms are
added to the FEM Lagrangian to deal with the UV divergences so that
the lattice quantum path integral will converge to the target
renormalizable quantum field theory on the Riemannian manifold. We
refer to the combination of these two steps as the {\bf Quantum Finite
  Element} (QFE) method. While the problem of UV divergences is not
addressed here, the reader is referred to a companion
article~\cite{phi4}, where the one loop QFE counter term is
successfully applied to the 2D $\phi^4$ theory on $\mS^2$ at the
Wilson-Fisher conformal fixed point.

\subsection{Piecewise Linear Finite Elements}

Consider the action for  a free  scalar field in the continuum  on  (${\cal   M}, g$) given by 
\be
S = \half  \int_{\cal M} d^Dx \sqrt{g} [g^{\mu\nu}\dd_\mu \phi(x)\dd_\nu \phi(x) +
  (m^2  + \xi R) \phi^2(x)]   \; ,
\label{eq:scalar}
\ee
with proper distances defined by the metric,
\be
ds^2 =g_{\mu \nu}(x) dx^\mu dx^\nu  \; ,
\ee
and its determinant, $g = \det(g_{\mu \nu})$.  Assume also that the Riemann
manifold is {\bf torsion free} ($\Gamma^\lambda_{\mu \nu} =
\Gamma^\lambda_{ \nu \mu}$) and {\bf metric compatible} ($\nabla_\rho g_{\mu\nu} = 0$) so the Levi-Civita
connection is determined uniquely in terms of  the metric,
\be
\Gamma^\lambda_{ \nu \mu} = \frac{1}{2} g^{\lambda \rho}(\dd_\mu
g_{\nu \rho} + \dd_\nu g_{\mu \rho} - \dd_\rho g_{\mu \nu}) \; .
\ee
The classical action (\ref{eq:scalar}) is diffeomorphism invariant. The  coupling $\xi = 
(D-2)/(4 (D-1))$ to the Ricci scalar curvature  is required for conformal invariance
at zero mass but henceforth we will set the Recci scalar term
(\ref{eq:scalar}) to zero since it is inessential for this FEM review.

The conventional FEM/Regge Calculus approach to a simplicial  approximation can
be broadly broken into three steps.
\begin{itemize}
\item \textbf{Topology:} The D-dimensional target manifold $\mathcal{M}$ is replaced by a simplicial complex $\mathcal{M}_\sigma$ composed of elementary D-simplices, which is homeomorphic to the target manifold.
\item \textbf{Geometry:} The metric on the target manifold
  $(\mathcal{M},g)$ is approximated on the simplicial complex to form a
  ``lattice Riemann manifold''  $(\mathcal{M}_\sigma,g_\sigma)$ by
  assigning lengths $l_{ij}$ on links and extending the metric into the interior of each simplex with piecewise flat volumes.
\item \textbf{Hilbert Space:} The Hilbert space of continuum fields,
  $\phi(x)$, is truncated by expanding  in a finite element basis on
  each simplex, $\phi_\sigma(x) \simeq \sum^D_{i = 0}E^i(x) \phi_i$.
\end{itemize}

 In principle one can construct  a one-to-one map
between points  on the target smooth Riemann manifold (${\cal M},g(x)$) and points on the 
piecewise flat simplicial manifold (${\cal M}_\sigma,g_\sigma(y)$)
introduced in Regge Calculus~\cite{Regge:1961px} that preserves
distance to order $O(a^2)$, where the {\em lattice spacing}, $a$, is
a bound on the simplicial diameters.  There are two approaches to this map, employed  in
 detail in Sec \ref{sec:SpinConnection}: The first approach uses the intrinsic
geometry of the D-dimensional  manifold, and the second a higher
dimensional embedding in flat Euclidean space $\mR^N$ for $N >D$.  

The first approach is more fundamental.  One chooses a collection of points
$x_i$ in $\cal M$ and constructs a simplicial complex for this set. A
discrete metric in the spirit of Regge Calculus is computed by an
approximation $l_{ij}$ to the geodesic distances on each link
$\<i,j\>$. Then each D-simplex is interpolated by piecewise flat
co-ordinates $y$. In general, there are subtleties involved in
achieving a good approximation.  The geodesics are only unique if
neighboring points are sufficiently close. An optimal triangulation
should use the Voronoi construction which requires a reasonable
approximation to the distances. (Note that Regge Calculus avoids this
problem by reversing the logic. The simplicial manifold is assumed to
be given {\em a priori} with  the
target manifold as a  consequence defined in the continuum limit, 
$l_{ij}  = O(a) \rightarrow 0$.)

The second and much easier approach, when it is available, is to start
with an isometric embedding of the D-dimensional Riemann manifold
(${\cal M},g$) into a higher dimensional flat Euclidean space
$ {\mathbb R^N}$. An important example is the $\mathbb S^D$ sphere
discussed in Sec.~\ref{sec:spinstructureS2} . This is easily embedded
as $\vec{r} \in \mR^{D+1}$ such that $\vec{r}\cdot\vec{r} = R_0^2$
with $R_0$ fixed. Then one uses a Voronoi construction of simplices on
a set of discrete sites at $x = r_i$ assigning the Euclidean
distances, $l_{ij} = |r_i - r_j|$, to the edges. This construction
turns out to be invariant under the projective transformation of the
sphere $\mS^D$ to the plane $\mR^D$. In general, if we can find a
smooth isometric embedding, this will guarantee convergence of the
simplicial manifold (${\cal M}_\sigma,g_\sigma$) to the target
manifold (${\cal M},g$) as $a \rightarrow 0$.  

To approximate the Hilbert space, we can expand the field $\phi_\sigma$ in a local FEM
basis~\cite{StrangFix200805}. Properly constructed this convergences to the
continuum field, $|\phi_\sigma(x) - \phi(x)| \rightarrow 0$, as the
diameters $a$ of all simplicial elements vanish. But more importantly, FEM
theorems also impose precise {\em shape regular}
condition~\cite{StrangFix200805} on the simplicial geometry to
guarantee that all solutions of discrete equations of motion (EOM)
converge to the classical solutions of the continuum EOM.  This is a
subtle theoretical problem, which involves the order of the
differential equation, the non-linearities of the PDEs, boundary
conditions, the choice of FEM basis, etc. For free fermions, even in
flat space, there are additional well known difficulties, not
addressed in the FEM literature to our knowledge, due to the notorious
spectrum doubling problem and the need to restore chiral symmetry.

\begin{figure}
\centering
  \includegraphics[width=.7\textwidth]{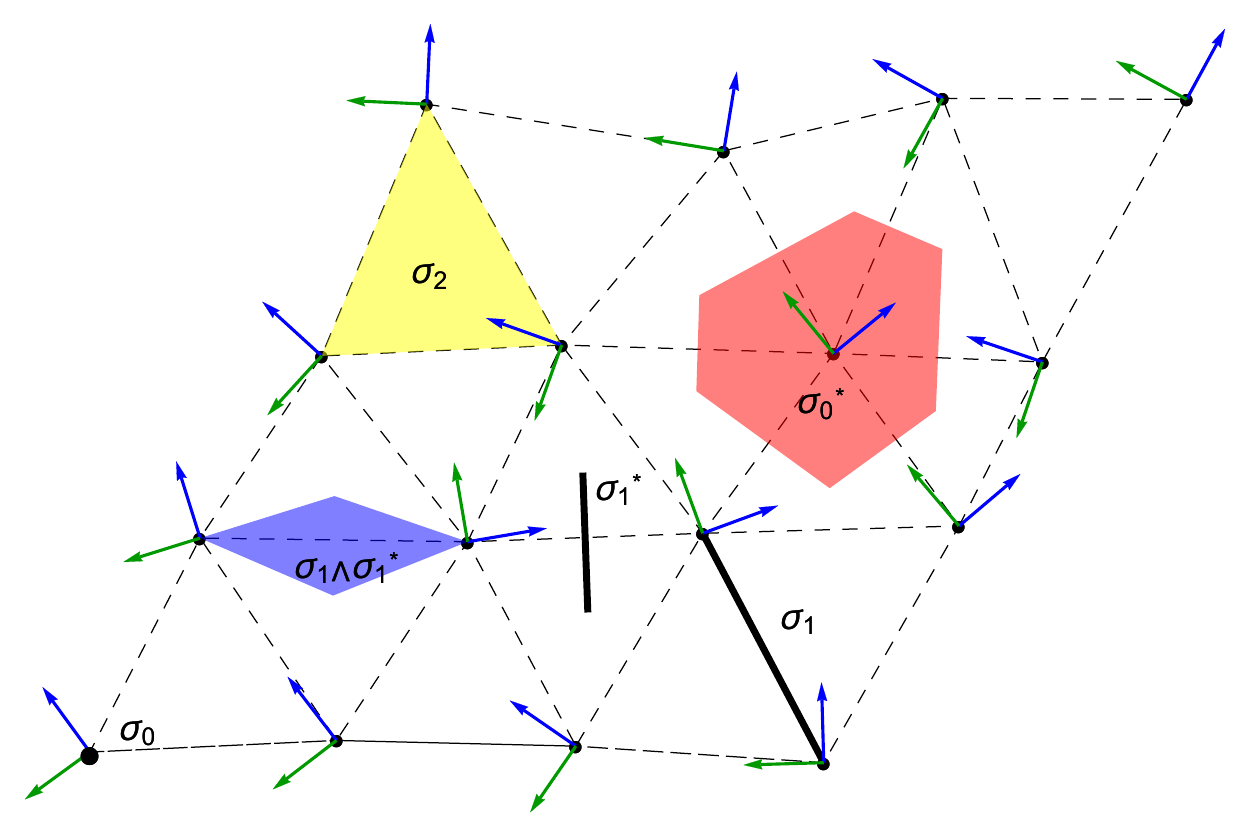}
\caption{\label{fig:lattice} A 2D simplicial complex with  points
  $(\sigma_0)$, edges ($\sigma_1$) and triangles $(\sigma_2)$. At each
  vertex $\sigma_0$ there is   a dual
  polytope in  $\sigma^*_0$ (illustrated in red), and at each link, $\sigma_1$,
there is a dual link $\sigma^*_1$ and its associated hybrid cell $\sigma_1 \wedge \sigma^*_1$ (illustrated in blue). The
arrows at each site represent a random basis for the
local tangent plane. }
\end{figure}

\paragraph{Simplicial Geometry and Notation:} It is helpful to
understand a bit of the formal aspects of each step listed above in
order to establish notation. One builds up the lattice field theory
representation in layers: start with the simplicial complex
$\mathcal S$, then add a metric to get the Regge Calculus, and lastly add
matter fields to construct the simplicial action for the quantum field
theory.  The shared topological and algebraic properties mapped
between each abstract layer is the province of {\em Category
  theory}~\cite{CT}.

A pure simplicial complex $\cal S$ consists of a set of
D-dimensional simplices (designated by $\sigma_D$) ``glued''
together at shared faces (boundaries) consisting of $D-1$ dimensional
simplices ($\sigma_{D-1})$.  The D-dimensional simplex is built
iteratively from lower dimensional simplices,
\be
\sigma_0 \rightarrow \sigma_1 \rightarrow \sigma_2\rightarrow \cdots
\rightarrow \sigma_D
\ee
beginning with $D+1$ sites $\sigma_0(i)$ with $i = 0,1, \cdots, D$
on each simplex, connected together by $(D+1)D/2$ directed links
$\sigma_1(i_1 i_2) \equiv \<i_1,i_2\> $ forming   $D(D+1)(D+2)/3!$
oriented  triangles $\sigma_1(i_1i_2i_3) \equiv \triangle_{i_1 i_2i_3} $, etc. This hierarchy is specified by the boundary
operator,
\be
\partial \sigma_n(i_0 i_1 \cdots  i_n) = \sum^n_{k = 0} (-1)^k
\sigma_{n-1}(i_0 i_1\cdots \widehat i_k\cdots  i_n)\; ,
\label{eq:boundary}
\ee
where $\widehat i_k$ means to exclude this site.  Each simplex
$\sigma_n(i_0 i_1\cdots i_n) $ is an anti-symmetric function of its
arguments. The signs in Eq.~(\ref{eq:boundary}) keep track of the orientation of each simplex.  It is trivial to check that the boundary operator
is closed: $\partial^2\sigma_n = 0$. On a finite simplicial
lattice $\partial$ is  a matrix and its transpose, $\partial^T$, is
the co-boundary operator.   This is a first
 modest step into discrete homology and De Rham cohomology on a simplicial complex.
 
In the next layer,  Regge Calculus introduces a metric by assigning lengths to the
edges $l_{ij} = |\sigma_1(ij)|$, which provides the discrete metric,
$g \rightarrow g_\sigma $,  assuming the interior
of each D-plex is a  flat Euclidean space (e.g.,  piecewise linear coordinates).  This lifts 
the simplex into a metric space. For example, oriented links, $\<i,j\>
= \sigma_1(ij)$, are now associated with  vectors,
$\vec l_{ij}$ and  triangles, $\triangle_{ijk} = \sigma_2(ijk)$, with areas
$A_{ijk}$ and so on. Since the cells are flat, the curvature tensor required
for Einstein gravity in Regge Calculus has singularities
on the boundary, i.e., at vertices in 2D and hinges  for $D > 2$.
Matter fields (or forms) are $n$th-rank tensors, naturally assigned to $\sigma_n$.

Next, it is important to add to our simplicial Delaunay lattice, $\cal
S$,
the circumcenter dual Voronoi lattice, $\mathcal S^*$, composed of polytopes,
$\sigma^*_0 \leftarrow \sigma^*_1 \leftarrow \cdots \leftarrow
\sigma^*_D$
where $\sigma^*_n$ has dimension $D-n$ as illustrated in
Fig.~\ref{fig:lattice}.  A crucial property of this circumcenter
duality is {\bf orthogonality}. Each simplicial element
$\sigma_n \in S$ is orthogonal to its dual polytope
$\sigma^*_n \in S^*$. This orthogonality lies at the heart of defining
the Hodge star $*$ (or alternating symbol
$\epsilon^{i_0 i_1 \cdots  i_D}$).  The circumcenters for the dual
lattice can be found iteratively.  The circumcenter of an edge
$\<i,j\> = \sigma_1(ij)$ is its midpoint, the circumcenter of a
triangle $\triangle_{ijk}= \sigma_2(ijk)$ lies at intersection of
the perpendiculars from the midpoints of the aforementioned boundary edges
$\sigma_1 \in \partial \sigma_2(ijk)$, the circumcenter of a
tetrahedron $\sigma_3$  lies at the intersection of the normals from the circumcenters of its
boundary triangles, etc., as we move into higher dimensions.

Hybrid cells, $\sigma_n \wedge \sigma^*_n$, constructed from
simplices $\sigma_n$ in $\cal S$ and their orthogonal dual $\sigma^*_n$ in
$\cal S^*$ give a proper tiling of the discrete manifold. As a  consequence of
this orthogonality,  the volume $V_{nn*}= |\sigma_n \wedge
\sigma^*_n|$ of the hybrid $\sigma_n \perp \sigma^*_n$ is a simple product,
\be
V_{nn^*} =  \<\sigma_n | \sigma^*_n \> = \int  \sigma_n \wedge
\sigma^*_n   = \frac{n! (D - n)!}{D!} |\sigma_n| |\sigma^*_n|  \; .
\label{eq:SimplicialVolume}
\ee
For future reference, we introduce a simplified notation in lower
dimensions: the
point,  length  of 
links, and area of triangles will be given by
\be
1 = |\sigma_0(i)| \quad, \quad l_{ij} = |\sigma_1(ij)| \quad, \quad
A_{ijk} = |\sigma_2(ijk)|
\ee
respectively and the  D-dimensional hybrid  volumes associated with sites,  links and
triangles will be designated by
\be
V_i  = |\sigma^*_0(i)|\quad, \quad V_{ij} = |\sigma_1(ij) \wedge \sigma^*_1(ij))| \quad, \quad
V_{ijk} = |\sigma_2(ijk) \wedge \sigma^*_2(ijk)|
\ee
respectively. Finally,  when we add matter  fields $\omega$ for scalar ($\phi_i$),
Dirac ($\psi_i$) and gauge fields ($U_{ij}$), we can define a 
discrete exterior derivative ${ d}$ (or finite difference for grad, div
and curl) through a discrete Stokes' theorem   on the simplex,
\be
\int_{\sigma_n} { d}\omega(y)  = \int_{\partial \sigma_n}   \omega(y)
\quad
\mbox{or} \quad \< \sigma_n | {d} \omega\> = \<\partial \sigma_n|\omega\>   \; .
\label{eq:Stokes}
\ee
The Hodge star  takes you to the dual simplex  $\sigma^*_n$ to define the dual operator, $\delta =  *d*$.  The
operators $\delta, d$ automatically  inherit from $\partial, \partial^T$,
respectively the closure property, $d^2 = \delta^2 = 0$. 
While we do not rely heavily on this formalism, it is useful
intuitively to guide our discussion. This formal layered structure, we
believe, is also important for organizing software to implement lattice field theory simulations on 
general simplicial  lattices.
\begin{figure}
\centering
  \includegraphics[width=.7\textwidth]{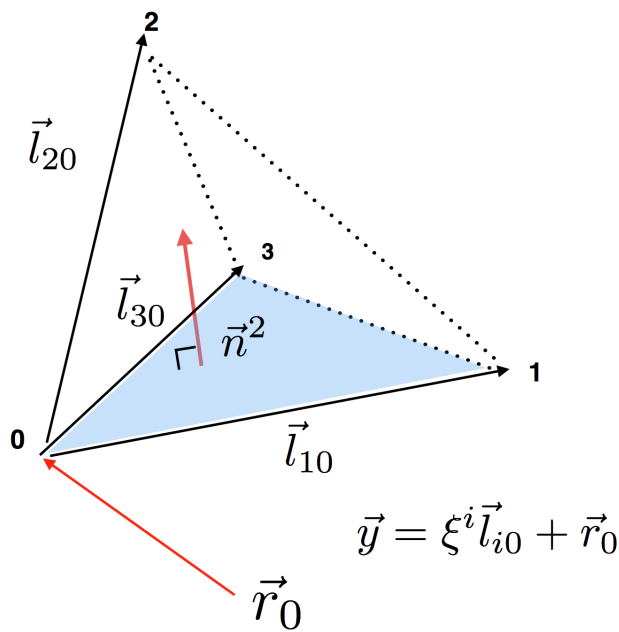}
\caption{\label{fig:3plex } The D-simplex, illustrated for $D = 3$,
  can be defined by D edge vectors 
$ \vec l_{i0}  = \vec r_i - \vec r_0$, picking arbitrarily
the $0$-th vertex. The remaining $D(D-1)/2$ edges are $\vec
l_{ij} = \vec l_{i0} - \vec l_{j0}$. One  dual vector $\vec
    n{\;}^{2}$ normal to  $\sigma_2(013)$  is depicted.}
\end{figure}

\subsection{Simplicial Laplacian for Scalar Fields}
The flat interior of each D-simplex in RC
and FEM is conveniently parameterized as
\be
\vec y  = \xi^0 \vec r_0 + \xi^1 \vec r_1 + \cdots + \xi^{D} \vec
r_{D} = \sum^D_{i=1} \xi^i \vec l_{i0} + \vec r_0 \; ,
\ee
using barycentric coordinates, $0 \le\xi^i \le 1$, with the
constraint $\xi^0+ \xi^1  + \cdots + \xi^{D} =1$.
The vectors on the edges are $ \vec l_{i0} = \vec r_i - \vec r_{0}$. To pick   a unique
coordinate system on ${\cal M}_\sigma$, we can arbitrarily eliminate
$\xi^{0}$, introducing the differentials,
\be
d\vec y = \frac{\partial \vec y}{\dd\xi^i} d\xi^i =  \vec l_{i0}
d\xi^i  \; ,
\ee
where   $ \vec l_{i0}$ are the components of this one form in the
basis   $d\xi^i$  with $i =  1,\cdots, D$ and dual  tangent vectors, 
\be
\vec \nabla =\vec \nabla \xi^i \dd_i = \vec n{\;}^i \dd_i  \; ,
\ee
with components, $\vec n{\; }^i = \nabla \xi^i$ in the basis
$\dd_i$. The flat metric on each simplex  is 
\be
ds^2 = d\vec y \cdot d\vec y = g_{ij} d\xi^i d\xi^j, \quad \quad g_{ij}
=  \vec l_{i0} \cdot \vec l_{j0} = \frac{1}{2}(
l^2_{i0}   + l^2_{j0} -l^2_{ij}) \; .
\label{eq:gSimplex}
\ee
  The standard relations for raising and
lowering indices by the metric tensor ($g_{ij}$) and its inverse 
\be
g^{ij} = \vec n{\; }^i \cdot \vec n{\; }^j \quad \mbox{or} \quad \vec
n{\; }^i \cdot\vec l_{j0} = \delta^i_j
\ee
applies within each simplex.  
Note since the interior of the simplex is flat we 
choose the  notation  $\vec l_{i0}$ and $\vec n{\; }^i$, 
\be
\vec l_{i0} \rightarrow l^a_{i0} = \frac{ \dd y^a}{\dd \xi^i} \quad
\mbox{and}  \quad \vec n{\; }^i \rightarrow n^i_a = \frac{ \dd \xi^i}{\dd
  y^a}   \; ,
\ee
for both upper and lower indices. It is tempting to use the
notation, $\vec l_{0i} \rightarrow \vec e{\;}_i $ and
$\vec n{\;}^i \rightarrow \vec e{\;}^i = g^{ij} \vec e{\;}_j $, but we
reserve this identification with lattice vierbeins for the simplicial
Dirac equation in Secs.~\ref{sec:Dirac} and \ref{sec:SpinConnection}.

The new action on the  simplicial manifold  (${\cal  M}_\sigma,g_\sigma$)  is 
again determined by Eq.~(\ref{eq:scalar}) using the simplicial
metric (\ref{eq:gSimplex}). It is given by a sum over all the D-simplices,
\bea
S_{\sigma} &=& \frac{1}{2} \sum_{\sigma_D} \int_{{\sigma_D}} d^Dy [\vec \nabla \phi_\sigma (y)  \cdot \vec
\nabla \phi_\sigma (y) + m^2 \phi^2_\sigma (y) ] \nn
& =& \frac{1}{2} \sum_{{\sigma_D}} \int_{\sigma_D} d^D\xi  \sqrt{g_\sigma}\; [ g^{ij}_\sigma 
\dd_i \phi_\sigma (\xi)  \dd_j \phi_\sigma(\xi) + m^2 \phi^2_\sigma (\xi) ] \; ,
\label{eq:SimplicialAction}
\eea
where  $\sqrt{g_\sigma}/D! =  |\sigma_D|$ is the volume  in each
D-simplex, or  in 2D the  area $A_{ijk}$ of the triangle
$\triangle_{ijk}$. Finally, we expand  $\phi_\sigma(y)$ in a finite element basis on
{\bf each} simplex, 
\be 
\phi_\sigma(y) \simeq  E^0(y) \phi_0 + E^1(y) \phi_1 +
\cdots +E^{D}(y) \phi_{D}\label{eq:FEM}  \; ,
\ee 
where $E^i(r_j)=\delta^i_j$ so that $\phi_i = \phi(y = r_i)$.  We also
impose the sum rule, $\sum_i E^i(y)=1$, so that the constant field is
preserved.   {\bf For simplicity, our subscript on $\phi_\sigma$,
  $S_\sigma$, etc, implies a restriction to a single simplex,
  $\sigma_D(i_0 i_1\cdots i_D)$.}  The expansion of the field over the entire piecewise flat manifold,
$(\mathcal{M}_\sigma,g_\sigma)$, is given by a sum over all sites, $\phi(x) \simeq \sum_i W^i(y)
\phi_i$, where the $W^i(y)$'s , referred to as {\em tent functions} , are sums over all 
adjacent  elements, $E^i(y)$'s, that have non-zero (unit)  support
at the  site $i$. Once these elements $E^i(y)$ are chosen, explicit integration for the
simplicial action,  Eq.~(\ref{eq:SimplicialAction}), can be carried out,
leading to a quadratic form for the free field action on the values
$\phi_i$. This construction also carries over for  interaction terms, $\phi^n(x)$,   giving higher order polynomials in $\phi_i$ within each simplex.

\begin{figure}[t]
\centering
 \includegraphics[width=.45\textwidth]{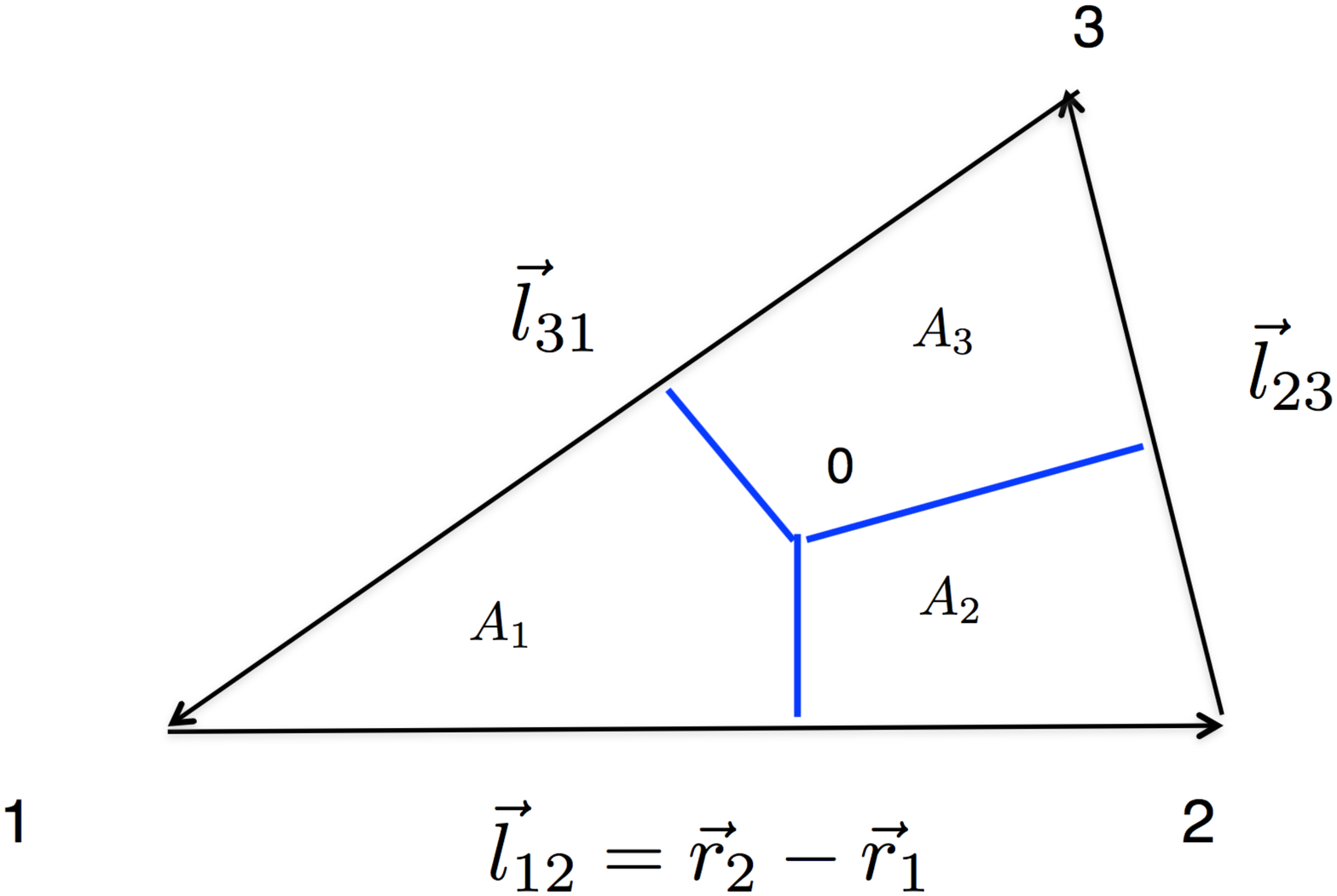}
\includegraphics[width=.45\textwidth]{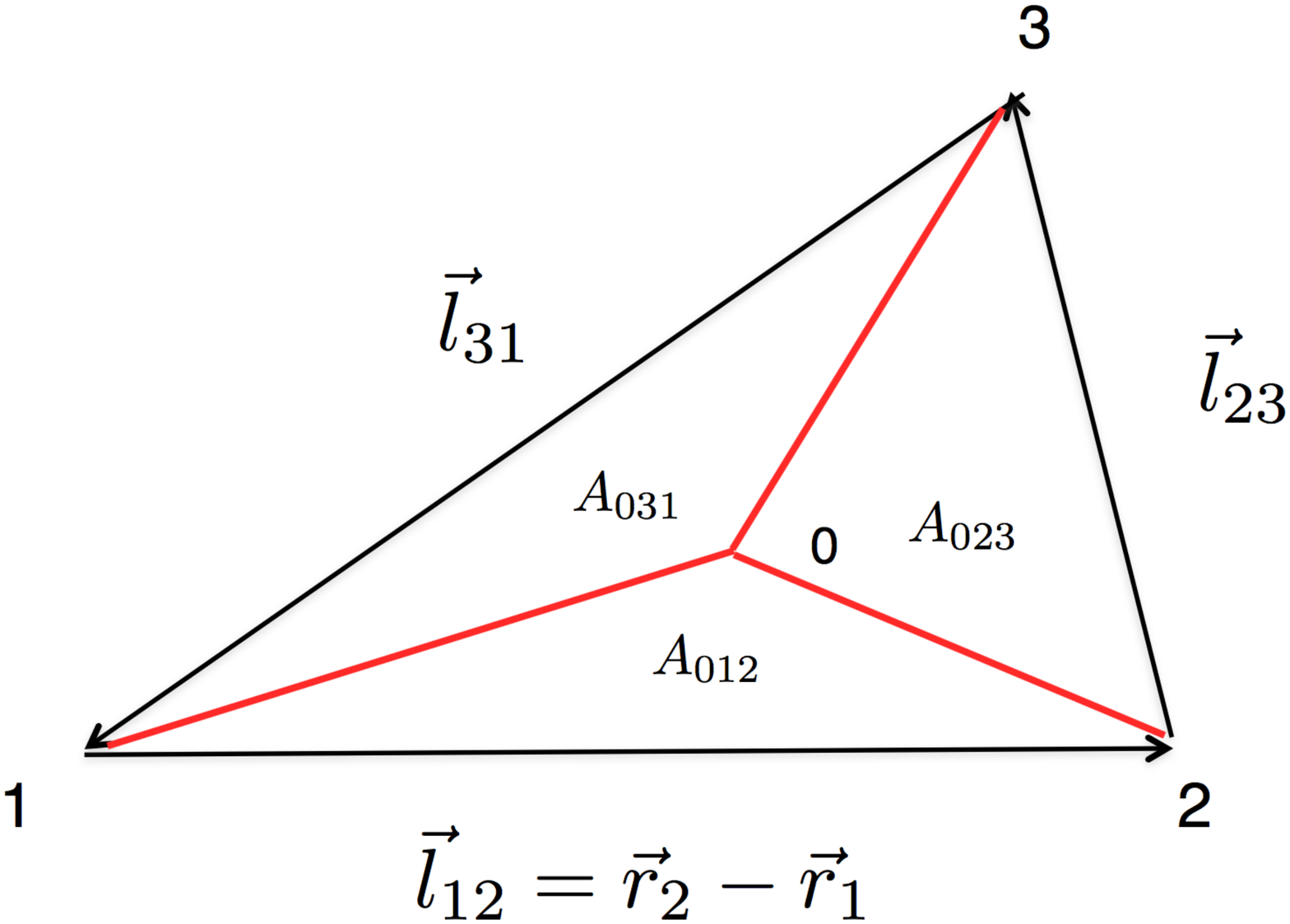}
\caption{\label{fig:lattice2} The geometric contribution of 
linear finite elements to a scalar field in the {\bf Vertex form}
on the left and in the {\bf Link form} on the right.    For
both forms, the triangle $ \triangle_{123}$ is subdivided into
regions meeting  at the circumcenter  $0$, with areas  $A_{i} = |\sigma^*(i)
  \cap \triangle_{123}|$ and   $A_{0ij} = |\sigma_1(ij) \wedge \sigma^*_1(ij)
\cap  \triangle_{123}|$, respectively. }
\label{fig:simplex}
\end{figure}

The simplest  choice is  the {\it linear FEM},
\be
E^i(\xi) = \xi^i, \quad\quad  i=0,\cdots D \; .
  \label{eq:linearFEM}
\ee
Since all the derivatives are constants, the massless action on each simplex,
\be
I_{\sigma} = \frac{1}{2} \int_{\sigma_D}  d^D y \vec \nabla \phi(y)  \cdot \vec
\nabla \phi(y) =  \frac{1}{2} \int_{\sigma_D} d^D\xi  \sqrt{g} g^{ij}
\dd_i \phi(\xi)  \dd_j \phi(\xi)  ,
\label{eq:SimplicialActionB}
\ee
is trivially evaluated, giving
\be
I_{\sigma} 
=  \frac{1}{2D!} \sum^D_{i,j = 1}  \sqrt{g}\; g^{ij} (\phi_i - \phi_0)
(\phi_j - \phi_0) \; .
\label{eq:vertex}
\ee
While this result  (\ref{eq:vertex}) is  correct, one inconvenience is that
our arbitrary choice of eliminating $\xi^0$   appears to  break the symmetry between the
$D+1$ sites.   To fix this we may average over 
the $D+1$ vertices  to yield the correct symmetrized expression, which
will be referred to as the {\bf Vertex Form} (illustrated for $D=2$ in
Fig.~\ref{fig:simplex}) of the simplicial action.

However, a more appealing geometric form can be
found.   A convenient way to derive this is to
relax the constraint $\xi_0 + \cdots + \xi_D =1$ and introduce
an over complete set of $D+1$ dual  vectors, $\vec n{\;}^k
= \vec \nabla \xi^k$, that are perpendicular to the face opposite
the  vertex $k$  and normalized relative to the edge vectors by 
\be
\vec n{\;}^k \cdot \vec l_{ij} = \delta^k_i  - \delta^k_j \; .
\label{eq:dualVectors}
\ee
In this  over-complete basis , the  gradient  is 
$\vec \nabla \phi(y) =  \vec n{\;}^0\phi_0 +  \vec n{\;}^1
\phi_1 + \cdots +  \vec n{\;}^D  \phi _{D} $.
Evaluating the action gives  two equivalent symmetric forms,
\be
I_{\sigma} = \frac{1}{2}  \sum^D_{i,j = 0} |\sigma_D|\; \vec n{\;}^i
\cdot \vec n{\;}^j \phi_i \phi_j = \frac{1}{2}   \sum_{\<i,j\>} |\sigma_D|\;  (-\vec n{\;}^i
\cdot \vec n{\;}^j) (\phi_i - \phi_j)^2  \; ,
\label{eq:ScalarFEM}
\ee
 due to the  constraint,
\be
\vec \nabla (\xi^0 +\xi^1 + \cdots + \xi^D) = \vec n{\;}^0  + \vec
n{\;}^1 + \cdots + \vec n{\;}^D  = 0 \; .
\label{eq:VierbeinZero}
\ee
Recall that $|\sigma_D| = \sqrt{g_\sigma}|/D!$ is the volume of
  D-simplex. 
We refer to this as the {\bf Link Form} (illustrated for $D=2$ in
Fig.~\ref{fig:simplex}).  In two dimensions summing over all the triangles, the
contribution to the lattice action takes an appealing geometric form
\be
S_{\sigma} =  \frac{1}{2}  \sum_{\<i,j\>} A_{ij} \frac{(\phi_i
  -\phi_j)^2}{l^2_{ij}} \; ,
\ee
where in 2D we use the notation
$A_{ij} = |\sigma_1(ij) \wedge \sigma^*_1(ij)|$, instead of $V_{ij}$,
for the dual area Eq.~(\ref{eq:SimplicialVolume}) adjacent to the link
$\<i,j\>$.

\begin{figure}
  \centering 
\includegraphics[width=.5\textwidth]{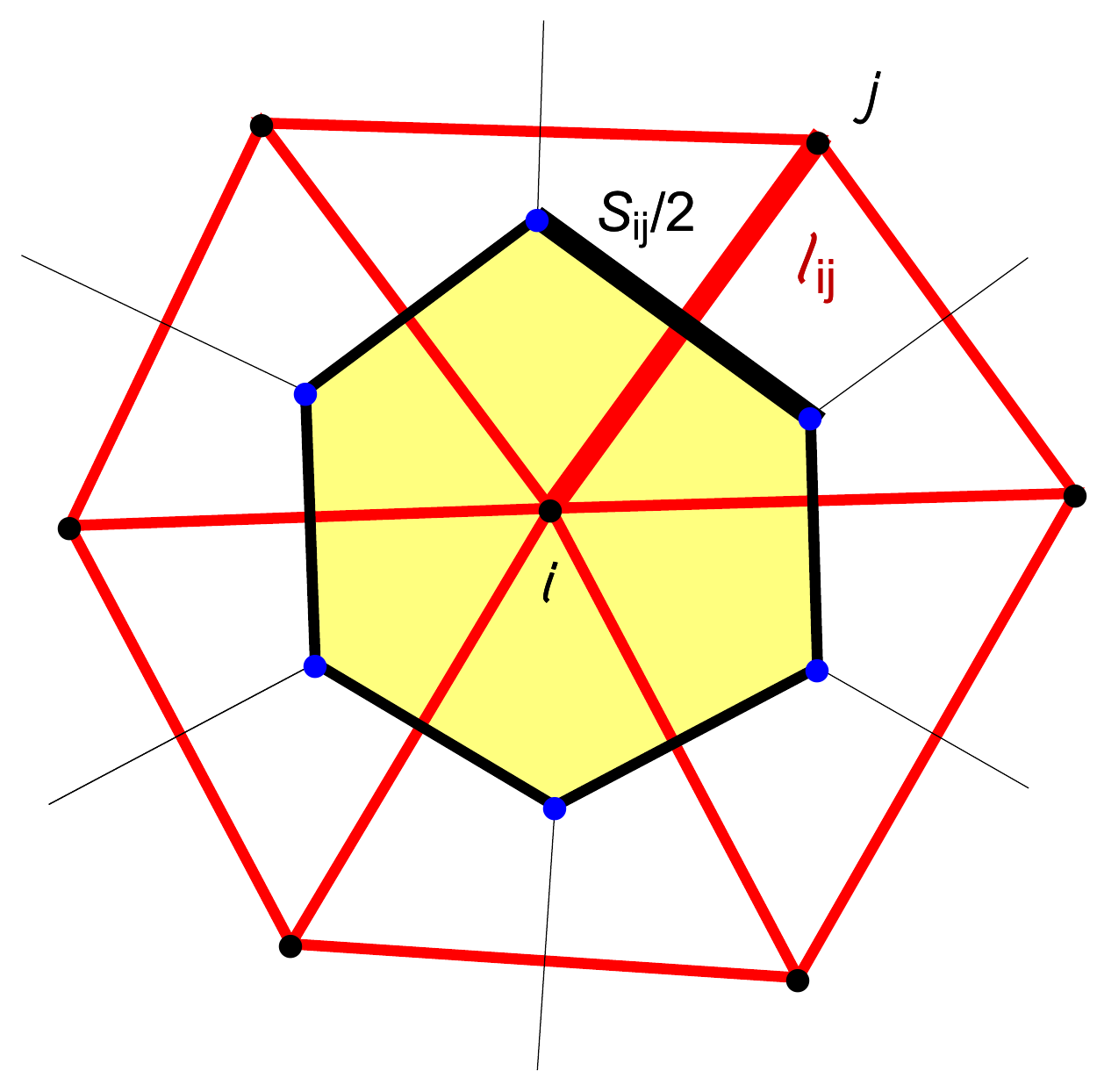}
\caption{\label{fig:LapBelop} The discrete Laplacian at a site $i$ is given by the sum on 
 all links  $\<i,j\>$ (in red)   weighed by gradients $(\phi_i -
  \phi_j)/l_{ij}$ multiplied by the surface  $S_{ij} = 2 V_{ij}/l_{ij}$ (in black) and normalized by the dual volume $|\sigma^*_o(i)| = V_i$ (in yellow).}
\end{figure}

\paragraph{Discrete Exterior Calculus:}
An alternative formalism for constructing the simplicial Laplacian 
relies on an elegant  Discrete Exterior Calculus
(DEC)~\cite{2005math8341D}. For any dimension, the DEC action
for the kinetic term   is  given by
\be
S_{\sigma}[\phi] =\frac{1}{2}  \sum_{\<i,j\>} V_{ij} \frac{(\phi_i 
  -\phi_j)^2}{l^2_{ij}}  + \frac{1}{2} m V_i \phi^2_i \; ,
\label{eq:DECscalar}
\ee
where, as illustrated in Fig.~\ref{fig:LapBelop} in 2D,  
$ V_{ij} = |\sigma_1(ij) \wedge \sigma_1^*(ij)| =l_{ij} S_{ij}/D $  is the product of the
length of the link ($l_{ij}$) times the  volume of the surface, $S_{ij}
=|\sigma_1^*(ij)|$, of the dual polytope normal to the link  $\<i,j\>$.  A local mass term
has been added for future reference even though it does not contribute to
the Laplacian.  {\bf Only in 2D is  the linear FEM  form  (\ref{eq:ScalarFEM})
  equivalent to the DEC form (\ref{eq:DECscalar}).} In 2D the
equivalence follows from the identity,
$ A_{123} \vec n_1 \cdot \vec n_2 = A_{ij} /l^2_{12}$,  often referred to as the co-tangent
 rule.  But for $D > 2$,  it is easy show how this fails by constructing a
a  counter example: Pick a simplex for $D > 2 $ with $\vec n_1 \cdot
 \vec n_2 = 0$ and  $V_{ij} > 0$  that vanish for the FEM construction
 but in non-zero for the linear FEM construction.

The  DEC construction for the discrete Beltrami-Laplace operator, 
\be
\frac{1}{V_i}\frac{\partial S_\sigma[\phi]}{\partial \phi_i} =\frac{1}{V_i} \sum_{j \in \<i,j\>} \frac{ V_{ij}}{l_{ij}} \frac{\phi_i
  -\phi_j}{l_{ij}} \; ,
\label{eq:Laplacian}
\ee
 follows the same basic steps leading to the continuum operator $
- \frac{1}{\sqrt{g}}\partial_\mu \sqrt{g} g^{\mu \nu} \partial_\nu
\phi(x)$. First, we apply the simplicial Stokes' theorem, Eq.~(\ref{eq:Stokes}), to get the discrete gradient (exterior derivative),
\be
d \phi = \frac{1}{|\sigma_1(ij)|} \int_{\sigma_1}   d \phi(x) =
\int_{\partial \sigma_1} \phi(x)/l_{ij} =   (\phi_i - \phi_j)/l_{ij} \;,
\ee
where the scalar (or zero form)  $\phi_i$  and the finite difference (or one form), $d
\phi_i = (\phi_i - \phi_j)/l_{ij}$, are
assigned to sites $\sigma_0(i)$  and  links $\sigma_1(ij)$ respectively.
Next, apply Stokes' theorem  again  on the dual lattice polytope
$\sigma^*_0$ to compute the divergence, $d ( * d \phi_i)$, illustrated in
yellow in Fig.~\ref{fig:LapBelop} for 2D and return to the simplicial
lattice, 
\be
*d*d \phi_i = *\frac{1}{|\sigma^*_0(i)|}\int_{\sigma^*_0} d[ *  (\phi_i -
\phi_j)/l_{ij}] =  \frac{1}{V_i} \sum_{j \in \<i,j\>} \frac{ V_{ij}}{l_{ij}} \frac{\phi_i
  -\phi_j}{l_{ij}} \; ,
\label{eq:BL}
\ee
in agreement with Eq. (\ref{eq:Laplacian}), expressed as 
the sum of  fluxes through   the boundaries $\partial \sigma^*_0(i)$ with
surface area, $S_{ij} /(D-1)!=  V_{ij}/l_{ij} = |\sigma_1(ij)
\wedge\sigma^*_1(ij)|/l_{ij}$. For the local mass term one
would add  $ m \int_{\sigma^*_0}
\phi_i/|\sigma^*_0| = m \phi_i$ to the operator (\ref{eq:BL}). 

\section{\label{sec:Dirac}Dirac Fields on a Riemann Manifold}

The action of the free Dirac fermion on a Riemann manifold,
\be
S =  \int d^Dx \sqrt{g} \bar \psi(x) [ {\bf e}^\mu(x) (\dd_\mu - i
\boldsymbol{\omega}_\mu(x)) + m ]\psi(x) \; ,
\label{eq:dirac}
\ee 
 introduces two new
structures involving spin: (i) The orientation of the spinor in the
tangent plane, ${\boldsymbol e}^\mu(x) = e^\mu_a (x) \gamma^a$,
where $e_a^\nu$ is the inverse (or dual) of the vierbein, $e^a_\mu$,
entering into the metric.  (ii) The spin
connection
$\boldsymbol{\omega}_\mu(x) \equiv {\omega}^{ab}_\mu(x)
\sigma_{ab} /4$, where $\sigma_{ab}/2 = i[\gamma_a,\gamma_b]/4$ are the Lorentz
generators for the Dirac spinor.  The reason for this is because there
are no
finite-dimensional spinor representations of the general covariance
group, so spinor indices are introduced in the tangent space. At each
point $x^\mu$, the flat tangent space can
be spanned by a set of orthonormal coordinates,
$\vec y = (y^1(x), y^2(x),\cdots, y^D(x))$, by expanding the 
cotangent differential,
\be
 d y^a = e^a_\mu(x) dx^\mu = \frac{\partial y^a}{\partial x^\mu} dx^\mu\; .  
\ee
The positive definite metric
\be 
ds^2 =d\vec y \cdot d\vec y= g_{\mu \nu}(x) dx^\mu dx^\nu
= e^a_\mu(x) e^a_\nu(x) dx^\mu dx^\nu \; .  
\ee 
can be  Cholesky factorized in terms of $e^a_\mu(x)$.
Now in addition to invariance under diffeomorphism, there is a local
``gauge'' invariance allowing an arbitrary rotation (or Euclidean
Lorentz transformation), $SO(D)$, in the tangent plane:
$y^a \rightarrow O^a_b y^b$.  This then acts on the spinors as a
gauge invariance in the {\tt Spin}(D) covering group.  

The spin connection and the vierbeins are not independent. For torsion-free and
metric compatible Riemann manifolds, they are related through the
tetrad hypothesis,
\be
\dd_\mu {\boldsymbol e}^\nu(x) + \Gamma^\nu_{\mu\lambda}{\boldsymbol
  e}^\lambda = i[ {\boldsymbol \omega}_\mu,
{\boldsymbol e}^\nu]   \label{eq:TetraHyp1}
\ee
or $[ {\boldsymbol D}_\mu, {\boldsymbol e}^\nu] +
\Gamma^\nu_{\mu\lambda}{\boldsymbol e}^\lambda=   0 $, where
$\boldsymbol{D}_\mu=\dd_\mu - i \boldsymbol{\omega}_\mu$ is   the
``covariant spinor derivative" operator. Expanding in components we
have
\be
\omega^{ab}_\mu  =  e_\nu^a \dd_\mu e_b^\nu +  e^a_\lambda \Gamma^\lambda_{\mu
  \nu} e^\nu_b= \frac{1}{2} e_a^\nu [ \partial_\mu e^b_\nu
- \partial_\nu e^b_\mu +e_b^\rho  e^c_\mu \partial_\nu
e^c_\rho ] -  ( a \leftrightarrow b)\, .
\label{eq:TetraHyp}
\ee
A crucial consequence of the tetrad hypothesis (\ref{eq:TetraHyp1}) is
the anti-Hermitian property of  Dirac operator,
\be
(\sqrt{g}{\bf e}^\mu
{\boldsymbol D}_\mu)^\dag = - {\boldsymbol D}_\mu\sqrt{g}{\bf e}^\mu 
= -  \sqrt{g}( {\bf e}^\mu {\boldsymbol D}_\mu   +
[{\boldsymbol D}_\mu,{\bf e}^\mu] + \frac{1}{\sqrt g} (\dd_\mu
\sqrt{g}) 
{\bf e}^\mu) 
= -  \sqrt{g} {\bf e}^\mu {\boldsymbol D}_\mu \; .
\ee
Consequently the Dirac  spectrum on a general
manifold is  pure
imaginary plus the real mass shift: $ i \lambda+m$ with $-\infty<
\lambda <\infty $.
It is essential when  placing the Dirac equation on a  simplicial
manifold  to provide a lattice realization for this identity.

\subsection{\label{subsec:DiracFEM}The Dirac Finite Element}

The application of classical FEM methods to fermions leads to a series
of difficulties.  First, even in 2D, linear finite elements in flat space do not give
a natural generalization of the scalar FEM expression.   Second, the
well-known problem of species doubling and chiral symmetry
breaking  is not solved by a straight forward application of
FEM. Third, and most troubling, in the Regge
Calculus representation of a  linear simplicial manifold, the curvature  has singularities 
 concentrated at the  vertices  and hinges. It is difficult, if not
 impossible, to place Dirac fields at such singular vertices as there is no well-defined tangent
plane.  We proceed to address the   solution to these
difficulties one by one.

A reasonable ansatz for a simplicial  fermion in flat space is a generalization of
the DEC scalar form in Eq.~(\ref{eq:DECscalar}), 
\be
S_{naive} \simeq  \frac{1}{2} \sum_{\<i,j\>}   \frac{V_{ij} }{ l^2_{ij} }  [\bar \psi_i
\vec l_{ij} \cdot
  \vec \gamma \psi_j  -  \bar \psi_j \vec l_{ij}  \cdot 
  \vec \gamma \psi_i] + \frac{1}{2} m V_i \bar \psi_i \psi_i \;,
\label{eq:canonical}
\ee
also recommend by  Friedberg, T.D. Lee,  and Ren in
Ref.~\cite{Friedberg:1985wr}.  We shall refer to this as the {\bf canonical Dirac form}.  However,
this form is {\bf not} given by the application of linear FEM to the Dirac field.

\begin{figure}
  \centering 
\includegraphics[width=.7\textwidth]{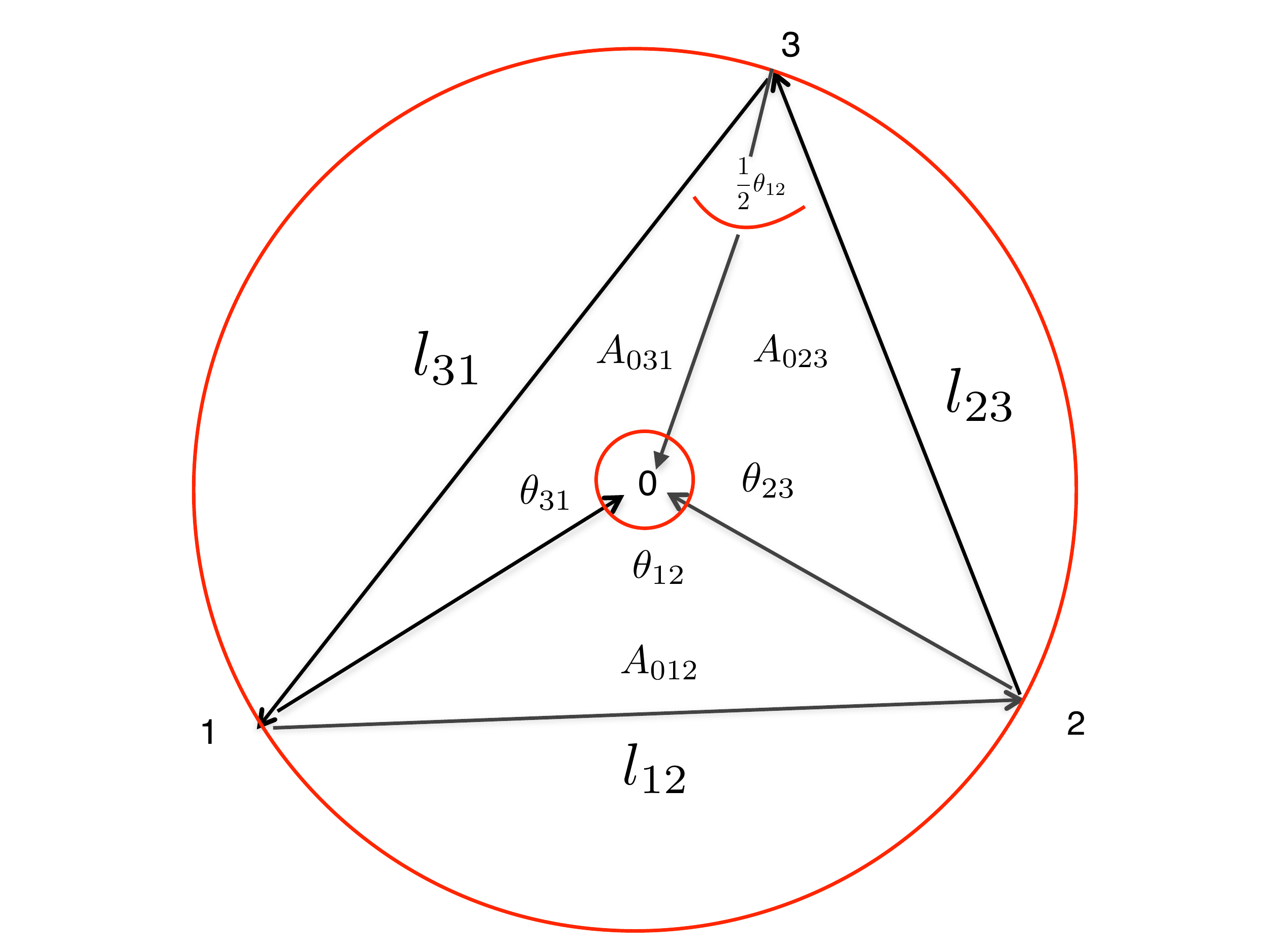}
\caption{\label{fig:DiracFE} A new Dirac finite element  on 
the simplex  splits the each triangle, $\triangle_{123}$, with edge vectors $(\vec l_{12}, \vec
  l_{23} ,\vec l_{31})$ into three
  isosceles sub-triangles that meet at the dual vertex $0$. An interior angle at $0$ opposite a link $\vec l_{ij}$  is designated as $\theta_{ij}$.   }
\end{figure}

Following closely the scalar example  (\ref{eq:ScalarFEM}), 
the linear FEM  evaluation of the Dirac action  on each simplex is 
\be
\int_\sigma  d^Dy [ \psi(y) \vec \gamma \cdot 
\vec \nabla  \psi(y)  =\frac{\sqrt{g}}{2
  (D + 1)!}  \; \sum_{i} \psi_i  \; \sum_{j} \vec n{\;}^j  \cdot \vec
\gamma \psi_j  \; .
\ee
For  anti-Hermiticity to be  enforced, one must explicitly sum over
the oriented and anti-oriented simplex, resulting in 
\be
I_{\sigma} = \frac{1}{2} \int_\sigma  d^D y [ \psi(y) \vec \gamma 
\cdot \vec \nabla  \psi(y) - (\vec \nabla \psi(y))\cdot \vec 
\gamma \psi(y)]  =  \frac{\sqrt{g}}{4 (D+1)!}\sum_{\<i,j\>}\bar \psi_i (\vec
n{\;}^j - \vec n{\;}^i )\cdot \vec \gamma \psi_j \, .
\label{eq:ScalarDirac}
\ee
However, even for  $D = 2$, the linear FEM formula,
\be
S_\sigma = \frac{A_{123}}{6}\sum_{\<i,j\>}\bar \psi_i (\vec
n{\;}^j - \vec n{\;}^i )\cdot \vec \sigma  \psi_j  \; ,
\label{eq:nn}
\ee
{\bf fails} to give the {\bf canonical Dirac form}. Most
peculiarly,  the spin projections $\vec l_{ij}\cdot \vec \sigma$ are
not  aligned with the propagation on the links. Namely the condition $\vec n{\;}^k \cdot (\vec
n{\;}^i - \vec n{\;}^j) = 0$, 
required by alignment,  $\vec l_{ij} \sim \vec n{\;}^i - \vec n{\;}^j $, fails except for an equilateral triangle
where the dual vectors are normal to the opposite sides.  However,  
we have  found a new  {\bf Dirac Finite Element}  prescription that does lead
to the canonical lattice form in 2D by summing over the piecewise
linear elements for each of 3 sub-triangles meeting at the circumcenter of a
general triangle as illustrated in Fig.~\ref{fig:DiracFE}.

The new construction begins by expanding $\triangle_{123}$ in a
new finite element basis,
\bea
\psi(x) &\rightarrow& \psi_\sigma(y) = E^1(y) \psi_1 + E^2(y) \psi_2
+E^{3}(y) \psi_{3}  \; , \nn
\bar \psi(x) &\rightarrow& \bar \psi_\sigma(y) = E^1(y) \bar \psi_1 +
E^2(y) \bar  \psi_2 +E^3(y) \bar  \psi_3 \; , 
\label{eq:DiracExpansion}
\eea
imposing basic properties for field
interpolates, 
\be
E^i(r_j) = \delta^i_j \quad, \quad  E^1(y) + E^2(y) + E^3(y) = 1 \; ,
\ee
so that $\psi(r_i) = \psi_i$ and a constant field is preserved.
We then make the ansatz that the element can be decomposed into
three  elements meeting at the circumcenter.  We introduce ghost
fields, $\psi_0$ and $\bar\psi_0$, at the circumcenter of each
triangle  and  expand the fields $\psi(x), \bar \psi(x)$ as the
sum of 3 piecewise  linear elements, one on each sub-triangle. The
ghost fields are expressed  as a linear
combination of the original lattice values,
\be
\psi_0= c_1\psi_1 + c_2 \psi_2 + c_3 \psi_3  \quad ,\quad \bar
\psi_0= c_1 \bar \psi_1 + c_2 \bar \psi_2 + c_3 \bar \psi_3 \; .
  \label{eq:circumcenter}
\ee
The constraint $\sum_i c_i=1$ is required so that the constant field
is preserved.  This implicitly defines the  new Dirac elements
(\ref{eq:DiracExpansion}) $E^i(y)$ on the full  triangle
$\Delta_{123}$. By a judicious choice of the
coefficients, 
\be
c_k=  \frac{4A_{0ij}}{ l_{ij}^2} \frac{4A_{0ik}}{ l_{ik}^2} =
\cot(\theta_{ik}/2) \cot(\theta_{jk}/2) \; ,
\ee
 this new Dirac FEM construct  leads to  the {\bf canonical Dirac form
   } (\ref{eq:canonical}), with  all couplings along the edges properly aligned.  (See Appendix~\ref{sec:Element} for a detailed proof.) 

One benefit of 
this construction is that this should allow standard FEM convergence theorems
to be applied to our Dirac FEM. However, we
have not yet sought a generalization of  this FEM construction to $D > 2$. Moreover, in spite of the
intuitive appeal of our ansatz, there is no known generalization of the formalism of  exterior
calculus to a single Dirac fermion, analogous to the use of the Hodge star
operator for the Laplace-Beltrami operator. The closest example is the application to  K\"ahler-Dirac
fermion~\cite{Banks:1982iq}. This is an interesting area for future investigation~\cite{Watterson:2005dd}.

\subsection{The Simplicial Spin Connection}

In preparation for curved space, we will first consider the
simplicial complex for a flat manifold after, applying at each site $i$, an arbitrary rotation by a  Lorentz
transformation, $O(D)$,  on the tangent vectors. The
result is to transform each spinor: $\psi_i \rightarrow \Lambda_i
\psi_i $, with $\Lambda_i \in \mbox{\texttt{Spin}}(D)$. The action in
this general gauge
becomes,
\be 
S_{naive}  
 = \frac{1}{2} \sum_{\<i,j\>}   \frac{V_{ij} }{ l_{ij} }  [\bar \psi_i
\vec e{\;}^{(i)j} \cdot \vec \gamma \Omega_{ij}\psi_j  -  \bar \psi_j
\Omega_{ji} \vec e{\;}^{(i)j}\cdot   \vec\gamma \psi_i] + \frac{1}{2} m
V_i \bar \psi_i \psi_i \; ,
\label{eq:naive}
\ee
where  $\Omega_{ij}  = \Lambda^\dag _i \Lambda_j = \Omega^\dag_{ji} $
serves as the  lattice spin connection and $\vec{e}^{(i)j}$ serves as the lattice vierbein. The link variable, 
\be
\Omega_{ij} = e^{\textstyle i l^\mu_{ij} \boldsymbol{\omega}_\mu(x)}\; ,
\ee
 is entirely analogous to the compact Wilson gauge variables,  
 $U_\mu(x) = \exp[i A_\mu(x)]$, for color spinors in 
lattice gauge theories:  $A^{ab}_\mu(x) = \lambda^{ab}_cA^c_\mu(x)$
and $\boldsymbol{\omega}_\mu(x) = \omega^{ab}_\mu(x) \sigma_{ab}/4$
are in the Lie algebra of the color  $SU(N)$ and   $\mbox{\texttt{Spin}}(D)$ groups respectively. The lattice vierbein  is 
\be
{\bf e}^{(i)j}= e^{(i)j}_a  \gamma^a \equiv \vec e{\;}^{(i)j} \cdot \vec \gamma
 = \Lambda^\dag _i \hat l_{ij} \cdot \vec \gamma
 \Lambda_i   \; ,
\ee
where $\hat l_{ij} $  is the out-going unit vector from $i$ to $j$. 
 With $m=0$, the naive  Dirac action  is 
 anti-Hermitian by the virtue of the identity, $\Omega_{ji} {\bf
   e}^{(i)j} = - {\bf e}^{(j)i} \Omega_{ji}$. 
Note that moving the vierbein to the opposite end of the link gives
\be
e^{(j)i}_a  \gamma^a  = - \Omega_{ji} e^{(i)j}_a  \gamma^a \Omega_{ij}\, ,
\label{eq:parallel}
\ee
which is the lattice realization of the tetrad hypothesis.  In Sec.~\ref{sec:SpinConnection}, we demonstrate that
Eq.~(\ref{eq:parallel})  is equivalent to the continuum tetrad hypothesis Eq.~(\ref{eq:TetraHyp1}) as $l_{ij} \rightarrow 0$. Although in flat space, this spin connection is gauge equivalent
to $\Omega_{ij} = 1$, we will show shortly that the parametric form of the action given by Eq.~(\ref{eq:naive}) 
now applies to any manifold with a spin connection by requiring 
the product of the link matrices, $\Omega$, around a closed path to be
a measure of the curvature on the triangle.  Before describing the
algorithm for  determining a non-trivial lattice spin connection in
 Sec.~\ref{sec:SpinConnection}, we will address the problem of species  doubling.

\subsection{The Wilson Term}
\label{sec:doublers}

At this point we have replaced the first derivative continuum
operator,
$\nabla ={\bf e}^\mu \boldsymbol{D}_\mu= {\bf e}^\mu (\dd_\mu - i
\boldsymbol{\omega}_\mu)$,
with the {\em naive} or central difference form on the simplex, gauged
by the compact spin connections in Eq.~(\ref{eq:naive}). This simplicial
discretization preserves the anti-Hermiticity condition of the
continuum,
$\nabla^\dag \sqrt{g} = - \sqrt{g} \nabla $, and therefore it
preserves the 
spectral property,
$ (\nabla + m) \psi_\lambda =( i \lambda + m) \psi_\lambda $, with
$-\infty< \lambda <\infty $ as well. However, this spectrum includes spurious,
or so called doubler, states familiar to the naive fermion on the
hypercubic lattice. The FEM methods do not solve this problem.

To remove these doublers, we  introduce a spinor gauged Wilson term in
close analogy with conventional non-Abelian  flat space lattice  gauge field theory. 
The 4D lattice field theory doublers  are removed by adding an irrelevant
dimension 5  Wilson term  to the fermions action. This discrete
approximation to the  continuum operator is contained in
the square of the covariant Dirac operator,
\be
[\gamma^\mu(\dd_\mu -i  A_\mu)]^2 = (\dd_\mu -i  A_\mu)^2 +
\sigma_{\mu \nu} F^{\mu \nu}\, .
\label{eq:DiracSquared}
\ee
When placed on a regular lattice, the first term is referred to as  the Wilson (or gauge Laplacian) term, while the second is referred to as the clover term.
On a flat manifold, the doublers can be removed by
adding the Wilson term.  The free spectrum in momentum space of the
Wilson term  is proportional to $\sum_\mu (1 -\cos(ap_\mu))/a$ which
is irrelevant at $p \rightarrow 0$ but divergent as $a \rightarrow
0$  for doublers on the edge
of the Brillouin zone. 

 A similar approach can be applied to curved space.  Consider adding
 to the action a second order derivative term,
\be
\int d^Dx \sqrt{g} |\nabla 
\psi|^2 = \int d^Dx \sqrt{g} (\bar \psi \overleftarrow{\nabla} ^\dag)( \nabla 
\psi) = - \int d^Dx  \sqrt{g} \bar\psi \nabla^2\psi \; . 
\label{eq:nablasqr}
\ee
using $\nabla^\dag \sqrt{g} = - \sqrt{g} \nabla $.
 The square of the spinorial Dirac operator,  $\nabla =  {\bf e}^\mu \boldsymbol{D}_\mu= e^\mu_a
  \gamma^a \boldsymbol{D}_\mu$, is give by the Lichnerowicz formula,
\bea
 - \nabla^2   &=&  
  - g^{\mu \nu} (\boldsymbol{D}_\mu \boldsymbol{D}_\nu 
- \Gamma^\sigma_{\mu \nu} \boldsymbol{D}_\sigma)  +  \frac{1}{2} \sigma^{ab} e_a^\mu e_b^\nu \boldsymbol{R}_{\mu \nu}\nonumber\\
&=&- \frac{1}{\sqrt{g}} \boldsymbol{D}_\mu
\sqrt{g} g^{\mu \nu} \boldsymbol{D} _\nu  + \frac{1}{2} \sigma^{ab} e_a^\mu e_b^\nu \boldsymbol{R}_{\mu \nu}\, .
\label{eq:SimDiracSquared}
\eea
The first term on the second line is
nothing but the covariant spinor Laplacian, while the second term is related
to the curvature, 
\be
\boldsymbol{R}_{\mu \nu} = i [ \boldsymbol{D}_\mu, \boldsymbol{D}_\nu]
= i [ \dd_\mu - i\boldsymbol{\omega}_\mu, \dd_\nu -
i\boldsymbol{\omega}_\nu] \; .
\label{eq:CurvatureTensor}
\ee 

We  introduce a lattice version of the covariant spinor Laplacian
as a Wilson term to remove doublers on the simplicial lattice.  This is
just our lattice Laplace-Beltrami operator for the
scalar in Eq.~(\ref{eq:Laplacian}) in a general gauge,
\be 
S_{Wilson Term}= \frac{r}{2} \sum_{\braket{i,j}} \frac{V_{ij}}{l^2_{ij}}
(\bar{\psi}_i-\bar{\psi}_j \Omega_{ji})(\psi_i -  \Omega_{ij}\psi_j)
\; , \label{eq:wilsonterm}
\ee
Again, this canonical form generalizes to simplicial Dirac fermions on a
general Riemann manifold.  Further generalizations to include color gauge
fields and to construct Domain Wall actions are straightforward as
briefly mentioned in the conclusion.

\section{\label{sec:SpinConnection}Lattice Spin Structure}

\begin{figure}[ht]
\begin{center}
\includegraphics[width=0.7\textwidth]{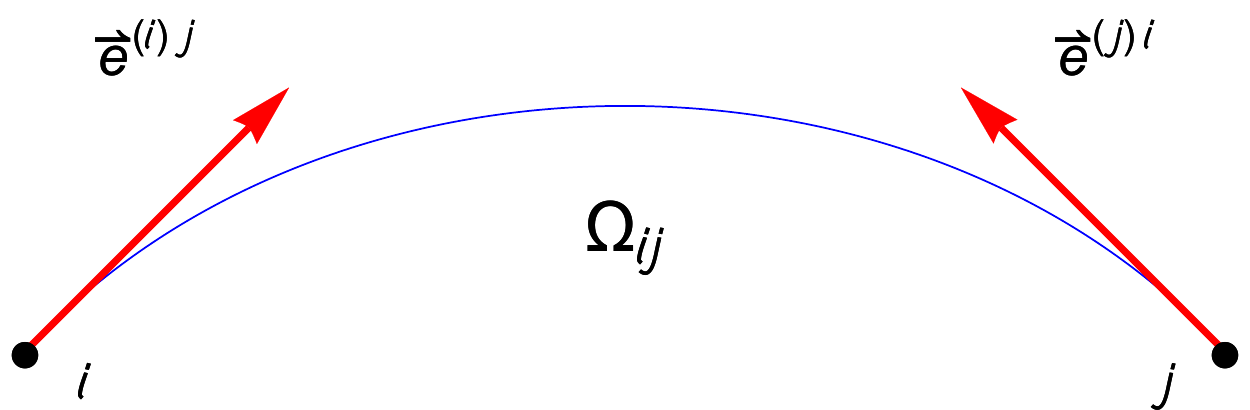}
\caption{\label{fig:LinkGeo} The  tangent vectors $\vec e{\;}^{(i)j} $ and $\vec e{\;}^{(j)i}$ on
opposite sides of the geodesics on the link $\<i,j\>$ are
related by a parallel transport, ${\bf e}{\;}^{(j)i} \equiv  \vec e{\;}^{(j)i} \cdot \vec \gamma =
- \Omega_{ji} \vec e{\;}^{(i)j} \cdot \vec \gamma\Omega_{ij}$.}
\end{center}
\end{figure}

We now present a procedure for fixing the vierbein $\vec
e{\;}^{(i)j}$ and  connection matrix,  
$\Omega_{ij}$, on each link $\< i,j \>$ of the simplicial lattice.
 Once this has been accomplished,  the
  parametric form for  a general Riemann manifold,
\be
S   = \frac{1}{2} \sum_{\<i,j\>}   \frac{V_{ij} }{ l_{ij} }  [\bar \psi_i
{\bf e}^{(i)j} \Omega_{ij}\psi_j  -  \bar \psi_j 
\Omega_{ji}  {\bf e}^{(i)j} \psi_i]  +\frac{m}{2} \bar \psi_i \psi_i \; , 
\label{eq:FermionForm}
\ee
 is unchanged from the flat space formula  (\ref{eq:naive}). The 
 spin connection matrices, $\Omega _{ij} =\Omega^\dag_{ij}$,  are no
 longer equivalent to a pure gauge  transformation.  A successful construction must respect the exact lattice tetrad hypothesis (\ref{eq:parallel}), 
\be
{\bf e}^{(i)j} \Omega_{ij} + \Omega_{ij} {\bf e}^{(j)i} = 0 \; ,
\label{eq:THid}
\ee
in order to ensure that the  naive lattice Dirac
operator, Eq.~(\ref{eq:FermionForm}) is anti-Hermitian  in the massless
limit, or  equivalently the full operator 
including the mass term in Eq.~(\ref{eq:FermionForm}) and the
Wilson term in Eq.~(\ref{eq:wilsonterm}) is  $\gamma_5$-Hermitian. 
This gauge covariant identity in Eq.~(\ref{eq:THid}), arising from parallel
transports of the vierbein along the link, is crucial to the construction.  If we expand
in the lattice spacing, $a$, we can immediately see how it is a discrete version
of the continuum Eq.~(\ref{eq:TetraHyp1}). In
Fig.~\ref{fig:LinkGeo} let $i$ and $j$ be located at  $x^\mu = x^\mu(0)$ and
$x^\mu(1) =  x^\mu(0) + a \hat l^\mu$, respectively,  on the geodesic, $x^\mu(s)$, between them. Introduce a smooth bi-spinor field,
${\bf e}(x) =  t_\nu(x) {\bf e}^\nu(x)$. Expanding ${\bf e}^{(j)i} + \Omega^\dag_{ij} {\bf
  e}^{(i)j} \Omega_{ij}$ term by term, we get
\bea
0 &=&  \hat t_\nu(x + a \hat l){\bf e}^\nu (x + a \hat l) - e^{\textstyle ia   l^\mu \omega_\mu(x)}\;
\hat t_\nu(x) {\bf e}^\nu(x) \; e^{ \textstyle - i a  l^\mu
  \omega_\mu(x) }  \nn
&\simeq& a  \hat t_\nu \hat l^\mu\; \dd_\mu  {\bf e}^\nu (x)+
a  \hat t_\nu \hat l^\mu\Gamma^\nu_{\mu \lambda} (x)  {\bf e}^\lambda (x)  - i a  \hat t_\nu \hat l^\mu [
\omega_\mu(x) , {\bf e}^\nu (x)]   +  O(a^2) \; ,
\label{eq:conteqn}
\eea
which is equivalent to the continuum expression, $(\dd_\mu  +
\Gamma^\nu_{\mu \lambda} (x) ) {\bf e}^\lambda (x)  - i [
\omega_\mu(x) , {\bf e}^\nu (x)] =0$,  to leading order.  In expanding Eq.~(\ref{eq:conteqn}), we have made use of the approximation $\hat
t_\nu(x + a \hat l) - \hat t_\nu(x) \simeq a  \hat t_\nu \hat l^\mu\Gamma^\nu_{\mu \lambda} (x) $  which follows
from the geodesic equation (\ref{eq:geodesics}).

In computing  the spin connection for our target manifold, there are two
crucial  issues we need to address: {\bf i.)} First defining the tangent
plane for the Dirac field at each site. {\bf ii.)} Second resolving the sign ambiguity in
the map from the Lorentz group,  $O(D)$, to the spinor covering group, ${\tt Spin}(D)$.

The first issue is the difficulty of defining tangent plane at the
sites in the conventional piecewise flat Regge Calculus manifold.  The
RC defines the interior of each simplex to be flat so that all
curvature is given by singularities on $D-2$ simplices at the boundary
of the cells referred to as hinges, or vertices in
2D~\cite{Miller:1997wb}.  Since it is impossible to define tangent
planes at the lattice sites of a piecewise linear manifold, previous
attempts to introduce fermions in RC have generally placed the Dirac
fields at the circumcenters of the dual
lattice~\cite{Hamber:2009mt,Hamber:2007fk,Bogacz2001,Bogacz2002,Burda1999,Caselle:1989hd}. However,
this is troublesome for lattice gauge theory. With gauge fields on
links, matter fields (scalar and Dirac) should be on sites to maintain
local gauge covariance as described briefly in  Sec.~\ref{sec:conclusion}.

Our solution is  to re-interpret the RC
manifold as smooth, with well-defined tangent planes at the vertices.  For
example, on the sphere, we can remove the singular
curvature at the sites by replacing  each link 
$\<i,j\>$ by geodesics (great circles in 2D). This allows us to define tangent planes at the vertices.  More
generally, as pointed out by 
Brewin~\cite{Brewin:1996yk}, it is possible to provide a re-interpretation of the
RC geometry. Given the RC data of a simplicial
complex and the set of edges lengths $l_{ij}$, it is possible to construct 
a smooth interpolation of the curvature field, e.g., accurate to
$O(a^2)$ in the continuum limit,  in much the same spirit of higher splines in 1D or higher order FEM for 
matter fields in a general dimension.  This redefinition of the Regge
manifold will be implemented to fix  the lattice vierbein and spin
connection, however, to $O(a^2)$ we can still use the piecewise linear
manifold to compute the pre-factors.

The second issue is determining the spin connection  between the tangent
planes on opposite ends of a link. Under parallel transport,  
one can compute the  rotation $O_{ij}$, an element in the
Euclidean Lorentz group $O(D)$.  However we must also resolve the sign
ambiguity to lift this to the spinor  matrix connection, $\Omega_{ij}$,  in the {\tt Spin}(D) group,
which is the double covering of $O(D)$. The  mapping 
\be 
O_{ij} \implies \pm \Omega_{ij} 
\label{eq:Map}
\ee 
has a sign ambiguity---rotating a Dirac field by
$2\pi$ changes its sign.  The parallel  transport of the 
tangent planes on a link $\<i,j\>$ fixes the $O(D)$
rotation matrix $O_{ij}$ but not the sign in the map as  can be illustrated for
the tetrad hypothesis, Eq.~(\ref{eq:parallel}),
\be
e^{(i)j}_a \gamma^a  = - \Omega_{ij} e^{(j)i}_a \gamma^a \Omega_{ji}
\quad \implies  \quad \vec e{\;}^{(i)j} = - O_{ij}  \vec e{\;}^{(j)i}
\label{eq:discreteTH}
\ee
The sign of the mapping in Eq.~(\ref{eq:Map}) onto ${\tt Spin}(D)$ must be fixed so
that as the simplices are refined the integrated curvature on every
triangle $\triangle_{123}$  vanishes in the continuum limit 
\be 
\Omega_{12}
\Omega_{23} \Omega_{31} \simeq 1 - O(A_{123}) \rightarrow 1 
\label{eq:contplaq}
\ee 
and we approach the continuum Dirac equation on the
Riemann manifold. This  global constraint can be satisfied on a simplicial
complex  only if  the topology of the target manifold admits a
spin structure. We present here two approaches to constructing the lattice
spin connection.

\subsection{Construction by Parallel Transport} 
\label{sec:Parallel}

The first approach assumes that, given the
continuum metric $g_{\mu \nu}(x)$, we have computed the geodesics between
sites connected by links.  The construction follows 3 steps: 
\begin{enumerate}
\item  Choose a random tangent frame at  $i$ and determine the  tangent
  vectors $\vec e{\;}^{(i)j}$  on geodesics to neighbors $j$.
\item  Parallel transport the tangent frame at $i$ to $j$ and
compute the Lorentz $O^{ab}_{ij}$  rotation  in   $O(D)$ to the frame of
 $j$.
\item  Map each Lorentz  rotation  in   $O(D)$ to a pair in ${\tt Spin}(D)$, $O_{ij}
  \rightarrow s_{ij} \Omega^{(+)}_{ij}$,  and choose
  $s_{ij} = \pm 1$,  leading to the minimal curvature on each fundamental
  triangle.
\end{enumerate}

Let us next expand on each of these steps.  Consider a given link
$\<i,j\>$ illustrated in Fig.~\ref{fig:LinkGeo}.  Each site has its
own tangent plane. We choose an orthonormal set of tangent vectors
$\hat t{\;}^a(i)$ in the tangent plane given by
$\vec y = y_a\hat t{\;}^a(i)$.  We assume that the simplicial lattice
is refined to the point that there is a unique geodesic connecting $i$
with $j$.  At each site $i$, determine the outgoing unit tangent
vector $ \vec e{\;}^{(i)j} \equiv e^{(i)j}_a \hat t{\;}^a$ aligned
with the geodesic from $i$ to $j$.  Constructing the geodesic and the
tangent vector to the geodesic requires in general numerical
integration of the geodesic equation,
\be
\frac{d^2x^\lambda}{d s^2} + \Gamma^\lambda_{\mu \nu}(x) \frac{d x^\mu}{ds} \frac{d
  x^\nu}{ds}  = 0. \label{eq:geodesics}
\ee
This gives the geodesic curve $x(s)$ from $x(0) = x_i$ to
$x(s_j) = x_j$ with tangent vectors
$\vec e{\;}^{(i)j} = d\vec x(0)/ds $ and
$\vec e{\;}^{(j)i} = - d\vec x(s_j)/ds $ at each end.

The next step is to perform a parallel transport from the 
frame $i$ to the frame $j$ and determine the rotation between these
two tangent frames: $\hat t{\;}^a(i) = O^{ab}_{ij}\; \hat t{\;}^b(j)$.
The rotation for the gauge link is given by ordered product on the geodesics from $i$ to $j$,
\be
O_{ij}= {\cal P}[e^{\textstyle -  \int^{s_j}_0 ds\dot x^\mu(s)
  \Gamma_\mu(x(s))}] \; ,
\ee
where  $[\Gamma_\mu(x(s))]^\lambda_\nu =\Gamma^\lambda_{\mu \nu}(x) $ is
the matrix in the Lie algebra for $O(D)$.  This guarantees the discrete tetrad
constraint (\ref{eq:discreteTH}).  For simple manifolds, such as
those of particular interest of conformal field theory, the exact
solution to all geodesics can be determined by symmetries, avoiding
numerical integration altogether.  For example, on a sphere
$\mathbb S^n$, all geodesics are defined by great circles.

Finally, for  each link $\langle ij\rangle$, given $O_{ij} = e^{i\theta_{\mu \nu}
  J^{\mu \nu}}$, $-\pi <\theta_{\mu \nu}<\pi$, the last step involves  fixing
the sign ambiguity of the corresponding element $\Omega_{ij}$ in the spinor
group,
\be
\Omega_{ij} = s_{ij} \Omega^{(+)}_{ij},
\ee
where  $s_{ij}=\pm 1$ and $\Omega^{(+)}_{ij} =e^{i\theta_{\mu \nu}
  \sigma^{\mu \nu}/2}  \in \mbox{\texttt{Spin}}(D)$.
\ignore{For example in 2D the map from
a rotation $O(\theta)\in O(2) $  by  $-\pi < \theta<\pi $ is consistent with both $
\Omega^{(+)}(\theta) =  e^{i\theta \sigma_3/2}$ and  $ -
\Omega^{(+)}(\theta) =  e^{i(\theta+2\pi) \sigma_3/2}$.} 
To provide an algorithm to fix the signs on each link, we start by
considering a 2D manifold. We begin by picking a random triangle and fix all $s_{ij}$
to minimize the curvature. Then select an adjacent triangle that
shares a site $\sigma_0(i)$ and one edge $\<i,j\>$ with the first
triangle. There are now two new links whose signs  we again fix to
minimize its curvature. We continue with all the triangles sharing
this site $i$. This completes all triangles whose circumcenters make
up the dual cell $\sigma^*(i)$. Now pick a new site on the boundary of
this cluster and continue.  This algorithm gradually expands the
closed contour around the polytopes of the dual 2D complex $\mS^*$.  As
we will show explicitly for ${\mathbb S}^2$ in
Sec.~\ref{sec:spinstructureS2}, this continues until the last triangle
which has no signs undetermined. 

A failure at the last step means that the manifold does not admit a
spin connection, for example, non-orientable surfaces in 2D without
boundaries.  The existence of a spin-structure only depends on the
topology of the manifold. For example, a sphere has a trivial first
homotopy group, $\pi_1(\mS^2) = 0$, and it admits a unique spin
connection.  The torus has $\pi_1(\mT^2) = {\mathbb Z^2_2}$, with 4
possible spin connections, familiar to string theorists, as
Neveu-Schwarz/Neveu-Schwarz, Neveu-Schwarz/Ramond,
Ramond/Neveu-Schwarz and Ramond/Ramond sectors respectively.  Assume
that one of the allowed multiple spin connections on the manifold is
achieved.  For each non-contractible loop in the dual lattice, one can
introduce appropriate signs on links to exchange periodic and
anti-periodic boundary conditions. This then allows one to introduce
other inequivalent spin connections. More generally, a compact 2D
Riemann surface of genus $g$ admits $2^{2g}$ inequivalent spin
structures.

This  procedure can be generalized to higher dimensions  along
 similar lines. For example, in 3D, we have an expanding closed
 surface. Start with a single tetrahedron and fix the
signs for all edges. Then proceed to pick an edge $\sigma_1(ij)$ and
visit cyclically all the tetrahedrons with circumcenters for
$\sigma^*_1(ij)$  that share this edge. Now there is surface
$\sigma^*_{ij}$ dual to this edge $\sigma_1(ij)$. Again proceed to
select a new edge on a tetrahedron on the boundary and continue
as before.  The 3D classification
concerns the second homotopy group and 4D the third homotopy group,
etc.  Non-trivial homotopy groups give non-contractable surfaces with co-dimensions $D-1$
allowing one to introduce anti-periodic boundaries for multiple spin
connections.   

Determining the $Z_2$ phases, $s_{ij}$,
only depends on the topology.  For an orientable manifold in the
continuum  the topological condition for the existence of
a spin structure is equivalent to the a vanishing of second Stiefel-Whitney
class index~\cite{Isham:1982gk}. On our lattice it  is equivalent to finding the
ground state in a frustrated $Z_2$ gauge theory. The map for $O(D)$
curvature on each triangle to ${\tt Spin}(D)$ results in discrete $Z_2$
gauge theory.  We must  find a solution to 
\be
K_{ijk} \; s_{ij} s_{jk} s_{ki} = 1 \; , 
\ee 
where
$K_{ijk} = {\mbox{Sign}}[{\mbox{Tr}}(\Omega^{(+)}_{ij}\Omega^{(+)}_{jk} \Omega^{(+)}_{ki}) ] = \pm 1 $.
This is equivalent to the existence of an  $E =0$ ground state for 
\be 
E[s]= \sum_{\triangle_{ijk}} ( 1 - K_{ijk} \; s_{ij} s_{jk} s_{ki}) \; , 
\ee
on the simplicial complex. The number of distinct ground states,
mod a $Z_2$ local gauge invariance,  enumerate  inequivalent spinor representations.

\subsection{Construction  by Relaxation}
\label{sec:relax}
Although the algorithm above is straight forward, it is  computationally difficult,
requiring the determination of the geodesic between neighboring lattice
points and performing parallel transports of the frames to compute the
rotations $O_{ij}$. What is needed in general is an alternative
algorithm that converges to $O(a^2)$.  One approach is to compare the 
lattice and continuum spin connections at each site of the simplicial
lattice and minimize a functional to make them match up to $O(a^2)$.

The idea is to  consider the lattice spin connections,
\be
\Omega^{\alpha \beta}_{ij} = \big[e^{ \frac{i}{2}\omega_{ab} \sigma^{ab}}
\big]^{\alpha \beta}  \quad \mbox{with} \quad
\sigma^{ab} = \frac{i}{2} [\gamma^a,\gamma^b] \; ,
\ee
as independent variables, choosing them to
approximate as well as possible the curvature on the target manifold
defined by the metric $g_{\mu\nu}(x)$.  As is well known in lattice
gauge theory, the product of gauge links around a ``plaquette'' 
(a triangle in this case) is an approximation to the integrated curvature
over the surface. On the simplicial  Regge manifold, we
match the discrete curvature and the continuum curvature,
\be
\Omega^{\alpha \beta }_{\triangle_{ijk}}(i) = [\Omega_{ij} \Omega_{jk}
\Omega_{ki}]^{\alpha \beta }  \quad \leftrightarrow \quad S^{\alpha
  \beta }(i) 
= \big[ e^{i {\boldsymbol R}_{\mu \nu}(i)  A^{\mu \nu}_{\triangle_{ijk}} }\big]^{\alpha \beta},
\ee
respectively for  each triangle  with a vertex at a site $i$. The lattice
estimate is just the open Wilson product on $\triangle_{ijk}(i)$ beginning
and ending at a site $i$ and  the continuum estimate is the exponentiation of
the local spinor curvature tensor, $\boldsymbol{R} _{\mu \nu}(i) $ in
Eq.~(\ref{eq:CurvatureTensor}),  projected onto the triangle. To do this, we need  an
estimate for the oriented area of the adjacent triangle which in the case
of an isometric embedding in higher dimensions is given by 
\be
 A^{\mu \nu}_{\triangle_{ijk}} 
= \frac{1}{2} [ (r^\mu_i- r^\mu_j) (r^\nu_k-r^\nu_i) - (r^\nu_i-
r^\nu_j) (r^\mu_k-r^\mu_i)] \; ,
\ee
to $O(a^2)$.  Consequently we can in principle determine the lattice spin connection
by a typical  relaxation algorithm, minimizing a  quadratic form such as
\be 
G(\Omega_{ij}) = \sum_{\triangle,i}    Tr[(S_\triangle(i)  -
\Omega_{\triangle}(i))^\dag (S_\triangle(i)  -
\Omega_{\triangle}(i))] 
\ee 
with respect to the  unitary matrices, $\Omega_{ij}$, in ${\tt Spin}(D)$
on each link $\<i,j\>$.
The sum is over all triangles incident on each vertex $i$.
While this prescription is not unique,  any choice that is  gauge invariant and
converges to $O(a^2)$ in the continuum limit is acceptable.    Again
multiple spin connections can be generated by studying the homotopy of
the simplicial complex.

Lastly, given the gauge matrices, $\Omega_{ij}$, we also need to
construct the tangent vectors 
$\vec  e{\;}^{(i)j} $  from site $i$ to $j$,  consistent with the discrete
tetrad hypothesis constraint, Eq.~(\ref{eq:discreteTH}). It is important to   focus on the fact that  $\vec e{\;}^{(i)j}$ and $\vec e{\;}^{(j)i}$ are now evaluated in two different frames,
\be
e^{(i)j}_a \gamma^a  = - \Omega_{ij} e^{(j)i}_a \gamma^a \Omega_{ji}
\quad \mbox{or} \quad \vec e{\;}^{(i)j} = - O_{ij}  \vec e{\;}^{(j)i}
\; .
\ee
Let us first consider  tangent vectors $\vec
e{\;}^{(i)j} $ at $i$ on the geodesics,  $x(s)$, from $i$ to all neighboring
sites $j$, i.e., $x(0) = x_i$ and  $x_j = x(s_j)$
at ends of the $\<i,j\>$ link with $s_j$ the geodesic
length. The  geodesic equation, Eq.~(\ref{eq:geodesics}), in {\bf  the same local
coordinate system used to compute the curvature $\boldsymbol{R} _{\mu
  \nu}(i) $ at site $i$}, determines the  geodesic to $x_j(s)$ from
$i$ to  each of the neighbors, $j$. The velocities at $i$ are proportional to
the vierbein: $\dot x_j(0) =  d
  x_j/ds|_{s = 0} \sim  e{\;}^{(i) j}$.  

To approximate these velocities, $\dot x_j(0)$, we consider a Taylor 
expansion~\cite{Brewin:2009se} about $s=0$,
\be
 x_j(s) = x_j(0) +  s \dot x_j(0) + \sum^\infty_{n = 2} \frac{s^n}{n!}
\frac{d^n x_j}{ds^n}|_{s = 0} \; ,
\ee
for the geodesic. Then using the geodesic equation
(\ref{eq:geodesics}), the $n$th derivative in the sum  may be
re-expressed as an  $n$th order polynomial in $s \dot x(0)$. After substituting the 
rescaled velocity $v^\lambda= s_j \dot x^\lambda_j(0)$, the series expansion takes the form,
\be
 v^\lambda \simeq \Delta x^\lambda_{ij} +  \sum^\infty_{n = 2} \frac{1}{n!} \widetilde \Gamma^\lambda_{\mu_1,\mu_2,..,\mu_n}[x(0)]
 v^{\mu_1}  v^{\mu_2}\cdots v^{\mu_n}
\label{eq:velocity}
\ee
where have brought the linear term, $v^\lambda$, 
and the difference, $\Delta x_{ij} = x_j(s_j) - x_j(0)$, to the left
and  right hand side of  Eq.~(\ref{eq:velocity}), respectively.  The $n$th tensor
coefficients $\widetilde \Gamma^\lambda_{\mu_1,\mu_2,..,\mu_n}[x(0)]$
are defined~\cite{Brewin:2009se} recursively in terms of 
 derivatives of $\Gamma^\lambda_{\mu \nu}$ and products of lower rank tensors  starting with
$\widetilde \Gamma^\lambda_{\mu \nu}[x(0)] = \Gamma^\lambda_{\mu \nu}[x(0)]$.

This simple maneuver allows us to approximate the tangent vector as a
series in  $\Delta x_{ij} =O(a)$ in the continuum limit.  In leading order, we see that
$ v(0)\simeq \Delta x_{ij} $, corresponding to the fact that, on a smooth manifold, the
straight line is the first approximation.   The next step is to use
this linear approximation in the second order equation to
get a quadratic approximation.   In general  the $n$th approximation takes the form of an $n$th  order polynomial
in $\Delta x_{ij}$  as described in Ref.~\cite{Brewin:2009se}, leading to 
\be
 v^\lambda=s_j \dot x^\lambda(0) \simeq \Delta x^\lambda_{ij} +
 \frac{1}{2} \Gamma^\lambda_{
\mu \nu}[x(0)]
  \Delta x^\mu_{ij}  \Delta x^\nu_{ij} + C^\lambda_{\mu_1\mu_2\mu_3} \Delta x^{\mu_1}_{ij}
  \Delta x^{\mu_2}_{ij} \Delta x^{\mu_3}_{ij}   + \cdots.
  \label{eq:Tangent2}
\ee
The quadratic approximation in $\Delta x_{ij}$ gives $O(a^2)$
errors for the normalized tangent vector, which is sufficient for our
construction. 

After normalizing the velocities, we have an approximation to the
lattice vierbein  $E^{(i)j} \simeq  \dot x(0)/ \dot x(0)| \simeq e^{(i)j}$.  If we repeat this construction at all sites,
adopting   coordinate systems at $i$ and $j$ sites, related  by
$\Omega_{ij}$,   we have an approximate solution to the lattice tetrad hypothesis:
 $\vec E{\;}^{(i)j} + O_{ij}  \vec E{\;}^{(j)i} = O(a^2) $ on each
 link. Remarkably from, this approximation we can construct an exact
 solution to the tetrad hypothesis simply by
averaging the estimate for $\vec
 E{\;}^{(i)j}$ at $i$  with the pullback ($- O_{ij}  \vec E{\;}^{(j)i} $) from
 $j$,
\be
\vec  e{\;}^{(i)j}  = \frac{\vec E{\;}^{(i)j} - O_{ij}\vec E {\;}^{(j)i}}{|\vec E{\;}^{(i)j} - O_{ij}\vec E {\;}^{(j)i}|} \quad , \quad
\vec  e {\;}^{(j)i}  = \frac{\vec E {\;}^{(j)i} - O_{ji}\vec
  E{\;}^{(i)j}}{|\vec E {\;}^{(j)i} - O_{ji}\vec E{\;}^{(i)j}|} \; ,
\label{eq:Eexact}
\ee 
normalized to unit length. The denominators in Eq.~(\ref{eq:Eexact})
  are equal,  so dropping them we can verify the  tetrad
  hypothesis identity on each link $\<i,j\>$ by
\be
\vec  e {\;}^{(i)j} + O_{ij}
\vec  e {\;}^{(j)i} \sim \vec E{\;}^{(i)j} - O_{ij}\vec E {\;}^{(j)i} + O_{ij}
(\vec E {\;}^{(j)i} - O_{ji}\vec E{\;}^{(i)j}) = 0 \; .
\ee
With this construction, we may also replace the area estimate by 
\be
 A^{\mu \nu}_{\triangle_{ijk}} (i)
= \frac{l_{ij} l_{ik}}{2} [   e^{(i)j}_\mu   e^{(i)k}_\nu -
e^{(i)j}_\nu   e^{(i)k}_\mu ] \; ,
\ee
to  order $O(a^2)$. The entire approximation procedure depends
only on a consistent choice of a coordinate system at each site
$i$. However, the accuracy of this approximation can depend on this choice. An attractive
convention which is worth investigating further is to introduce  Riemann normal
coordinates (RNC)~\cite{Brewin:1996yk} at each site $i$, with the metric, 
$ g_{\mu \nu}(x)  = g_{\mu\nu}(x_i)  - (1/3) \Delta x^\lambda \Delta x^\sigma R_{\mu  \lambda\nu
  \sigma} + O(a^3) $
to help in approximating the tangent vectors.

\subsection{\label{sec:spinstructureS2}Spin Structure on  the
  Simplicial $\mathbb S^2$}

In preparation of our numerical tests and as a simple example, we 
present the construction of our 2D simplicial Dirac action on
$\mS^2$. The above procedures can be tested and used on a
sphere, but a far simpler approach is to realize that 
all geodesics are just given by great circles.  Given two points on the
$D-$dimension sphere denoted by unit vectors $\vec r_i$ and
$\vec r_j$, the geodesic is parameterized simply by
$\vec x(t) = (t \vec r_i +(1-t) \vec r_j)/| t \vec r_i +(1-t) \vec
r_j|$  with tangent vectors $ e^{(i)j} = \dot x(0) /|\dot x(0)|$. 
The entire construction is reduced to simple vector algebra in the embedded
space. Other symmetric manifolds have similar embedding methods. 

\begin{figure}[t]
\centering
\includegraphics[width=0.32\textwidth]{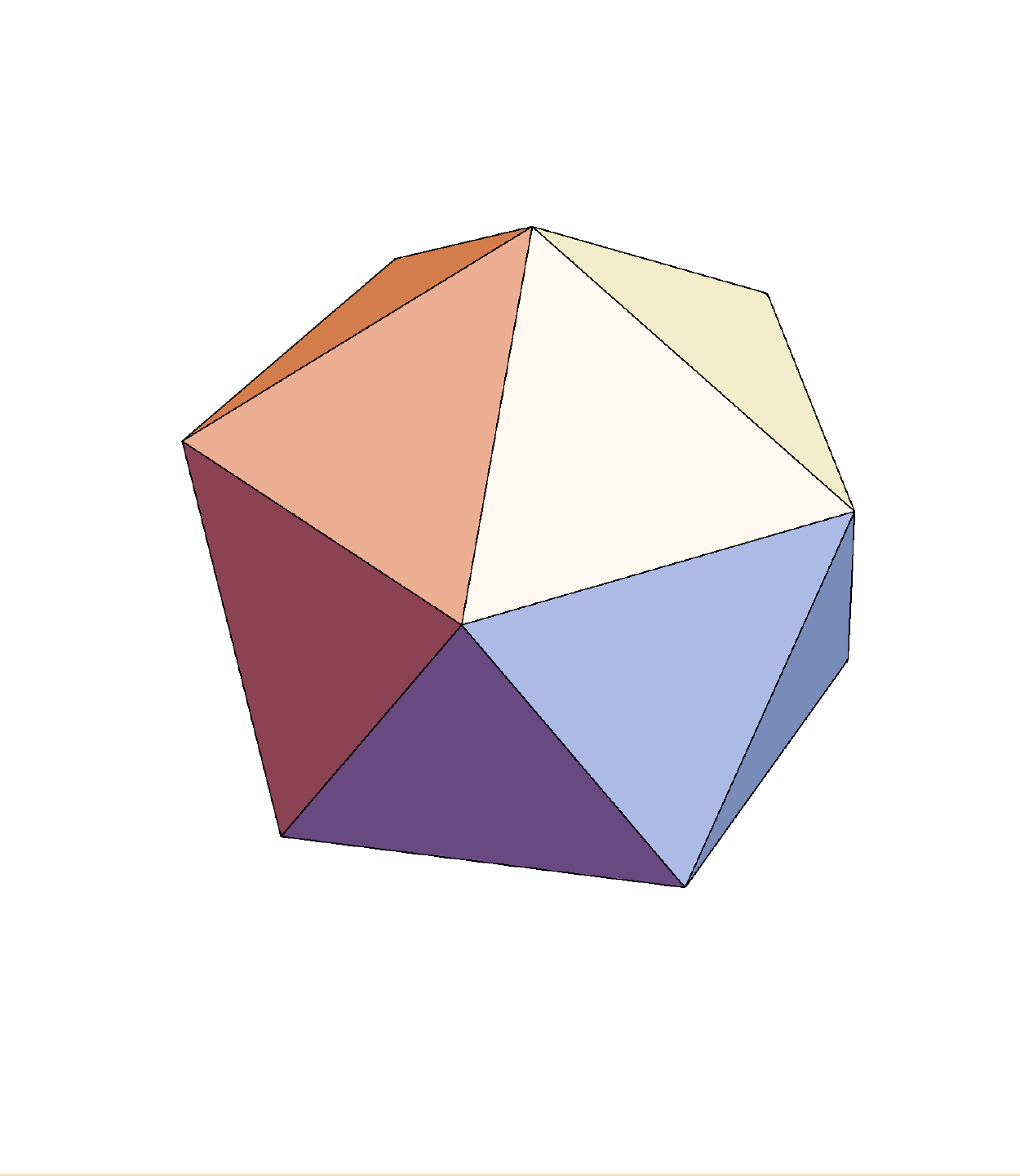}
\includegraphics[width=0.32\textwidth]{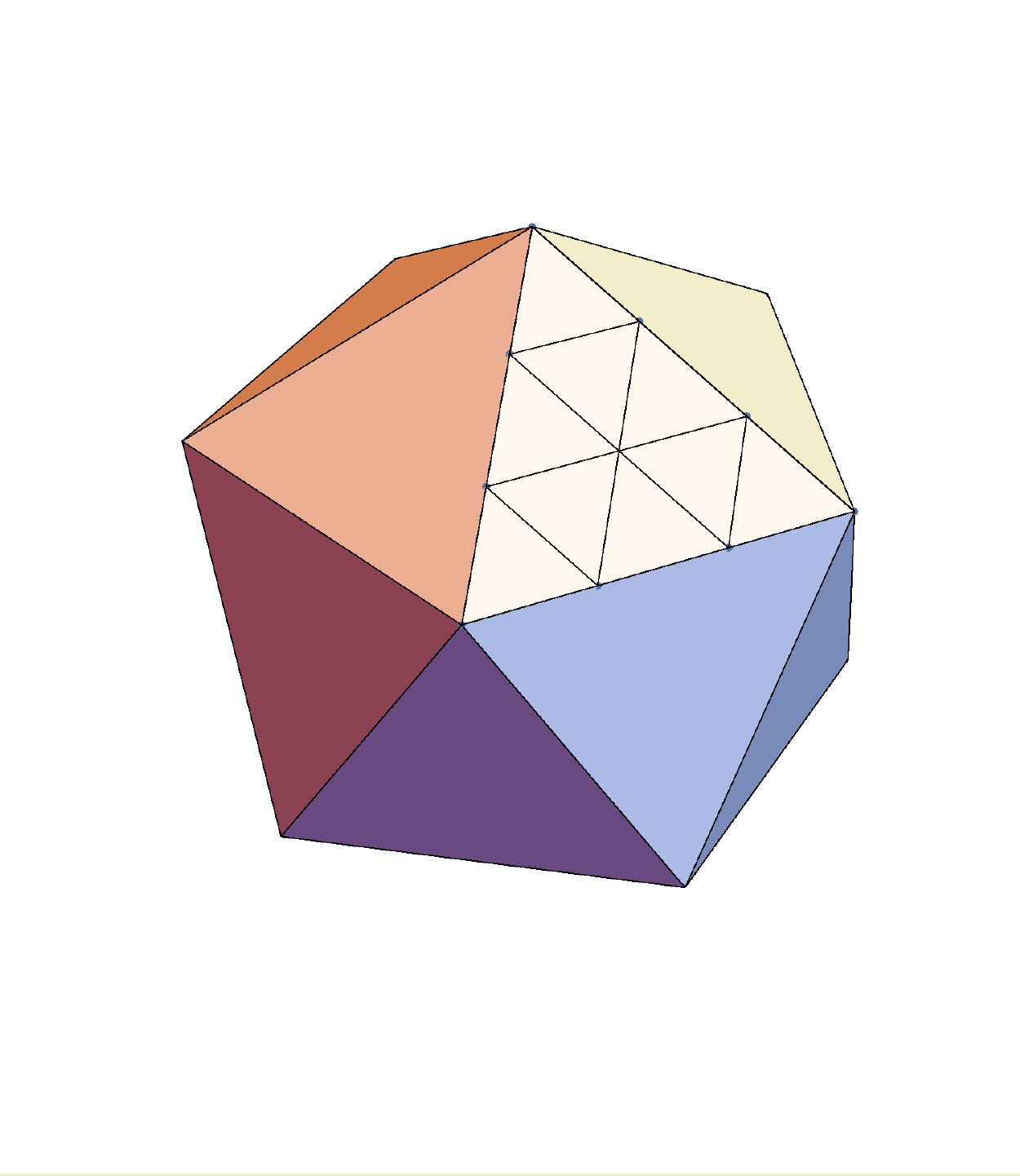}
\includegraphics[width=0.32\textwidth]{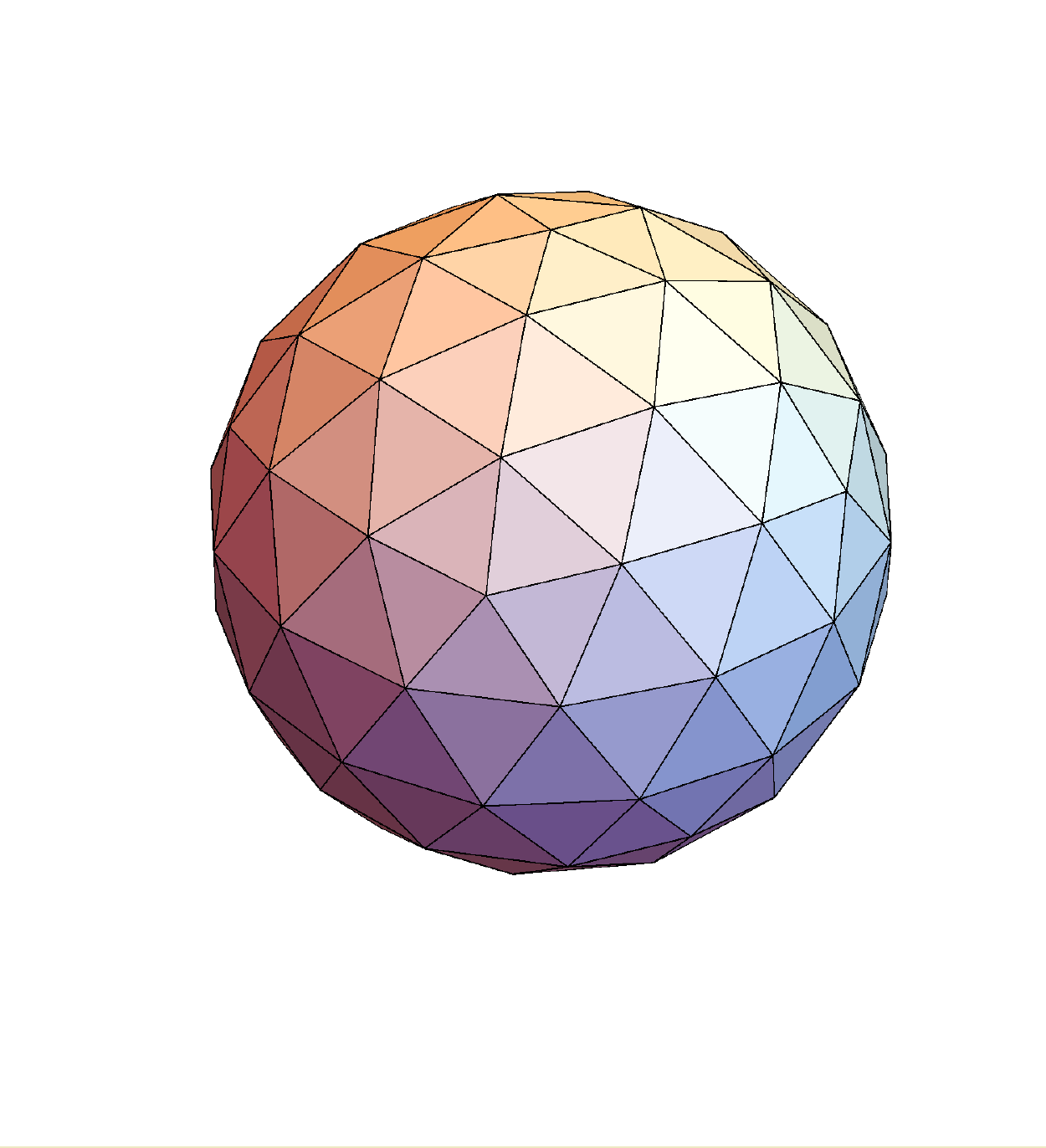}
\vskip -1.0 cm
\caption{\label{fig:icos} The  $L=3$ refinement of
the icosahedron with $V = 2 + 10L^2 = 92$ vertices or sites. The icosahedron on
the left is refined in the middle with $L^2 = 9$ equilateral triangles on 
each face, and  then on the right the new vertices are projected  onto the unit
sphere. The resulting simplicial  complex preserves the icosahedral symmetries. }
\end{figure}

For the $\mS^2$ manifold, our
triangulation~\cite{Brower:2014gsa,Brower:2015zea} starts with an
icosahedron in Fig.~\ref{fig:icos}, which provides the largest
subgroup of the spherical symmetry.  Each one of the 20 faces is then
subdivided into $L^2$ equilateral triangles resulting in a total of
$F = 20 L^2$ triangles. Next, we project each triangle onto the unit
sphere and take as edge lengths the secant distances between vertices
on the sphere, as illustrated in Fig.~\ref{fig:icos} for $L = 3$. This
projection introduces a small deformation of the equilateral
triangles, so to accurately approximate the Lagrangian, we need to
compute the finite element weights.  The topology of the manifold is
determined by the Euler characteristic, $\chi= V-E+F = 2 - 2H = 2$ and
the geometry by the table of lengths $l_{ij}$.

\begin{figure}[ht]
		\begin{center}
		\vspace{10pt}
\includegraphics[width=0.48\textwidth]{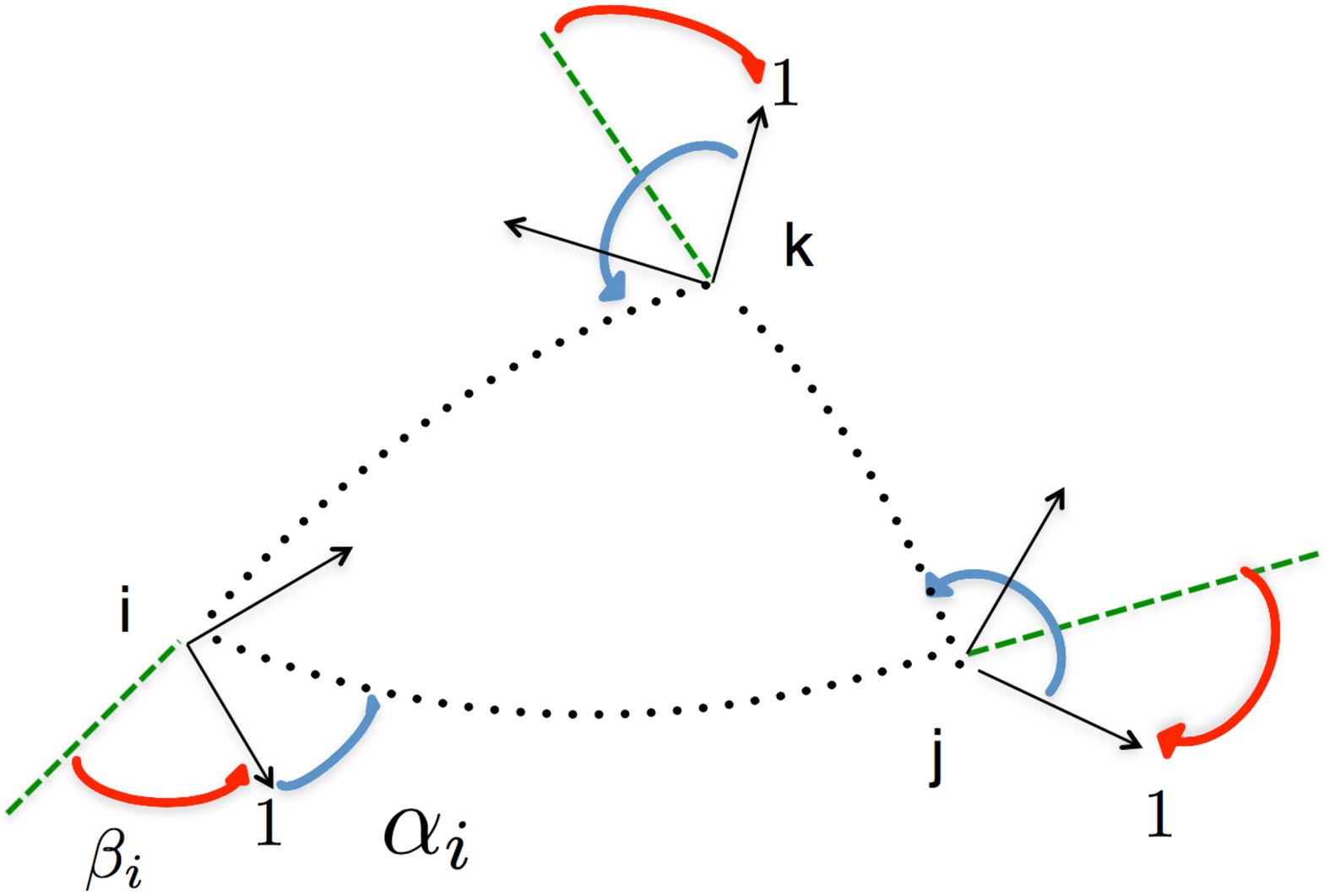}
\includegraphics[width=0.48\textwidth]{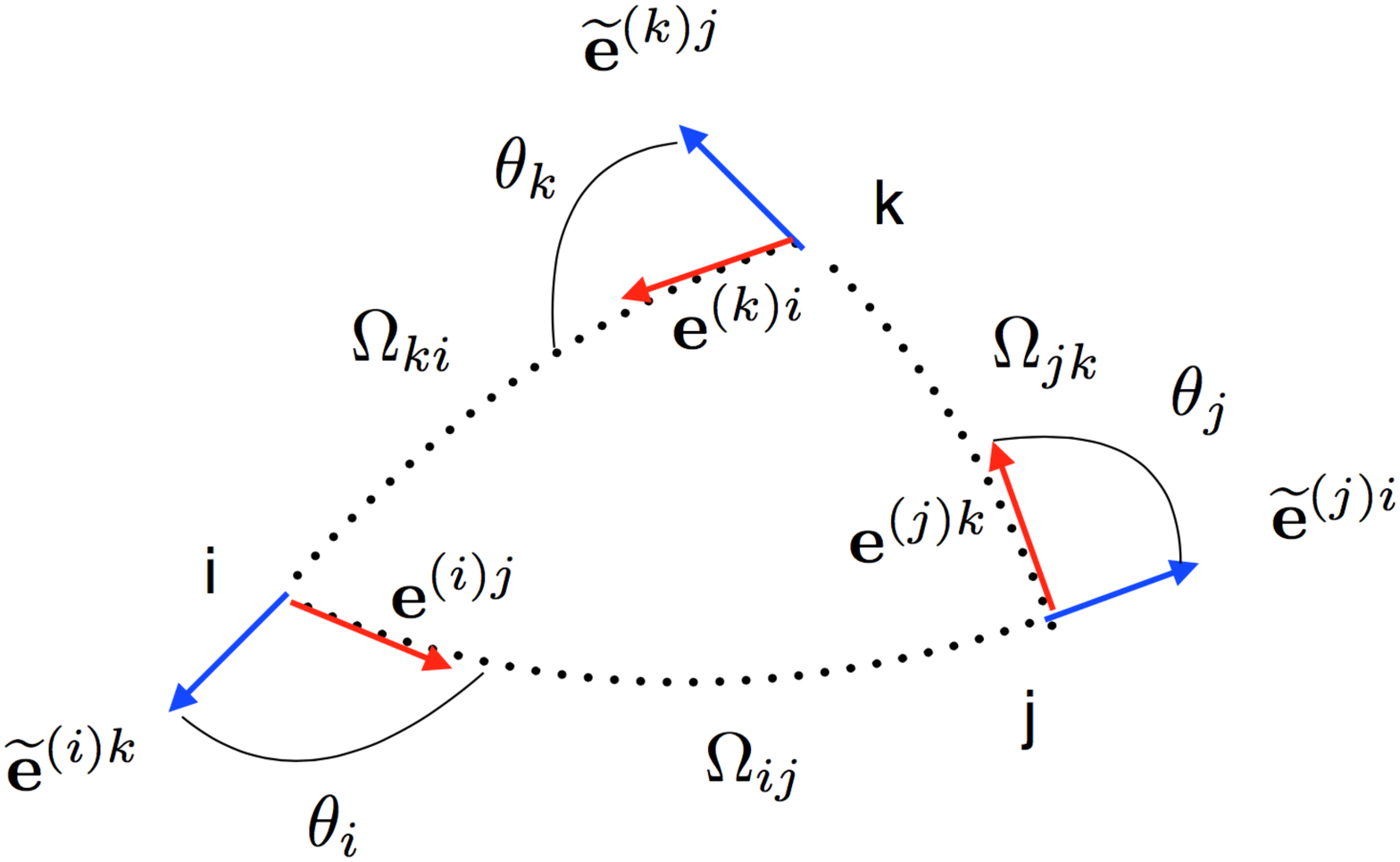}
   	\caption{\label{fig:Triangles} On the   left, vectors in the
          tangent planes, and on the right, the lattice spin connection,
          $\Omega_{12}$ and the outgoing and reflected  vierbeins, ${\bf e}^{(i)j} =
          e^{(i)j}_a  \gamma^a$   and ${\bf \widetilde e}^{(i)j}
          \equiv - {\bf e}^{(i)j}$, respectively.}
   	\end{center}
\end{figure}

The  lattice Dirac action on  $\mS^2$  is 
\be 
S  
 = \frac{1}{2} \sum_{\<i,j\>}   \frac{V_{ij} }{ l_{ij} }  [\bar \psi_i
e{\;}^{(i)j}_a  \sigma^a \Omega_{ij}\psi_j  -  \bar \psi_j
\Omega_{ji} e{\;}^{(i)j}_a   \sigma^a \psi_i] + \frac{1}{2} m
V_i \bar \psi_i \psi_i  + S_{WilsonTerm} \; ,
\ee
where the vierbein
$e^{(i)j}_a \sigma^a = e^{(i)j}_1 \sigma^1 + e^{(i)j}_2 \sigma^2$ are
2-vectors in the tangent plane at site $i$.  For each link $\<i,j\>$,
there is a lattice spin connection,
$\Omega_{ij}(\theta_{ij}) = s_{ij}e^{i \theta_{ij} \sigma_3/2}$,
associated with an Abelian $O(2)$ rotation $O(\theta_{ij})$,
$- \pi <\theta_{ij}< \pi$.  Because we know the exact geodesics on the
sphere are great circles, the geometry for the triangle
$\sigma_2(ijk)$ is fixed by the set of three angles, $\theta_{i}$, as
shown in Fig.~\ref{fig:Triangles}. Once $O(\theta_{ij})$ is specified,
this lattice spin connection, $\Omega_{ij}(\theta_{ij})$, can then be
constructed following the method in Sec.~\ref{sec:Parallel}.

After parameterizing the tangent plane $\vec y = y^a\vec n_a$ at each
site relative to two randomly chosen orthonormal tangent vectors
$\vec n_a$, we can determine  $\theta_{ij}$ by a procedure illustrated
Fig.~\ref{fig:Triangles}. For each triangle we rotate the $1$ axis at
site $i$ into a tangent vector on the arc from $i $ to $j$ by
$\alpha_i$, parallel transport this tangent vector on the geodesic to
site $j$ and rotate it back to the 1 axis at $j$ by $\beta_j$. This
gives $\theta_{ij} = \alpha_i - \beta_j$.  It follows trivially that
$\theta_{12} + \theta_{23} + \theta_{31} =
(\theta_1+\theta_2+\theta_3) \quad {\rm mod}\quad 2\pi$,
where the deficit angle is defined by
\be
\delta_{123} =A_{123}=2\pi - (\theta_1+\theta_2+\theta_3)= 2 \pi -
(\theta_{12} + \theta_{23} + \theta_{31})\quad {\rm mod}\quad 2\pi \; .
\ee
Now the problem is to determine $s_{ij}$ for all links
self-consistently for the entire sphere following the procedure
described in Sec. \ref{sec:Parallel}. As before, choose an arbitrary
triangle and fix the signs, $s_{ij}$, to satisfy constraint to
minimize the integrated curvature (\ref{eq:contplaq}, then move to adjacent
triangles fixing the signs $s_{ij}$ on new edges until you encounter
the last triangle. Now all the edges have fixed signs so there could
be an obstruction.  However, since the deficit angle is additive (or,
for the sphere, the areas are additive), for any closed loop we know
that this last triangle on a unit sphere, when viewed from the
outside, has a deficit angle $\delta \simeq 4\pi- A_\triangle$ in
steradians.  But since $ e^{ 4 \pi i\sigma^3/2} = 1$, the $4\pi$
factor can be dropped and there is no obstruction.

It is a simple algebraic exercise to show this exact consistency
condition on the sphere holds generally for any triangulation of a
surface with the topology of a sphere.  The more general argument is
as follows. Assume the interior angle for the $i$th vertex in triangle
$\triangle_{ijk}$ is given by $\widetilde \theta_i = \pi - \theta_i$
and that all the interior angles on the tangent plane at each vertex
add up exactly to $2 \pi$.  Then the deficit angle is
$\delta(ijk) = \widetilde \theta_i + \widetilde \theta_j + \widetilde
\theta_k - \pi$ and the sum over all angles must give
\be
\sum_F [\widetilde \theta_i + \widetilde \theta_j + \widetilde \theta_k ] = 2\pi V \; .
\ee
Any 2D simplicial triangulation of a closed surface implies
$3 F = 2 E$, so we have the sum rule,
\be
\sum \delta(ijk)   = 2  \pi V - \pi F   = 2 \pi (V - E + F) =4 \pi (1
- H), 
\ee
which for the sphere by Euler's identity gives $4 \pi$.  In fact, this
argument applies to any closed orientable 2D triangulation, or any
surface with an even number of boundaries $B$, such as the cylinder. Even
with an approximate determination of the angles, as for example in our
relaxation algorithm in Sec.~\ref{sec:relax}, the constraint remains  exact.

Finally, we should note a simple interpretation for a 2D complex
 Riemann manifold. In the complex plane, all Riemann  manifolds
can be represented by adding pairs of
square root branch points. For example, a square root branch
point at the origin with a cut out to infinity represents a cylinder
with two open boundaries. When you add an even
number of pairs, these create handles---4 twists for the torus,
etc.  As  we  discuss  in Sec.~\ref{sec:projectiveSphere} for the
simplicial Dirac equation, a pair of branch points  is equivalent to allowing a
pair of $-1$ ``frustrated'' triangles. In the context of Ising CFT, this  corresponds to the
insertion of twist operators.  Just as  square root branches come in
pairs when you flip edges ($s_{ij} \rightarrow - s_{ij}$),  on the
simplicial complex it creates a pair of ``frustrated''
triangles.   This is a nice  illustration of the fact that the existence of
a spin structure  on a  Riemann manifold  is a purely topological  property
that is naturally encoded in the simplicial complex without the need to introduce a metric.

\section{\label{sec:numerics}Numerical Tests for 2D Dirac Fermions }

For simplicity, we restrict our tests to the Dirac fermion (\ref{eq:dirac}) on 
$\mS^2$, which can be easily
solved analytically~\cite{Abrikosov:2002jr}. For future tests, higher dimensional
spherical solutions are also available, for example the 4D sphere in
Ref.~\cite{Abrikosov:2002jr}.
\begin{figure}[t]
		\begin{center}
		\vspace{10pt}
\includegraphics[width=0.43\textwidth,keepaspectratio]{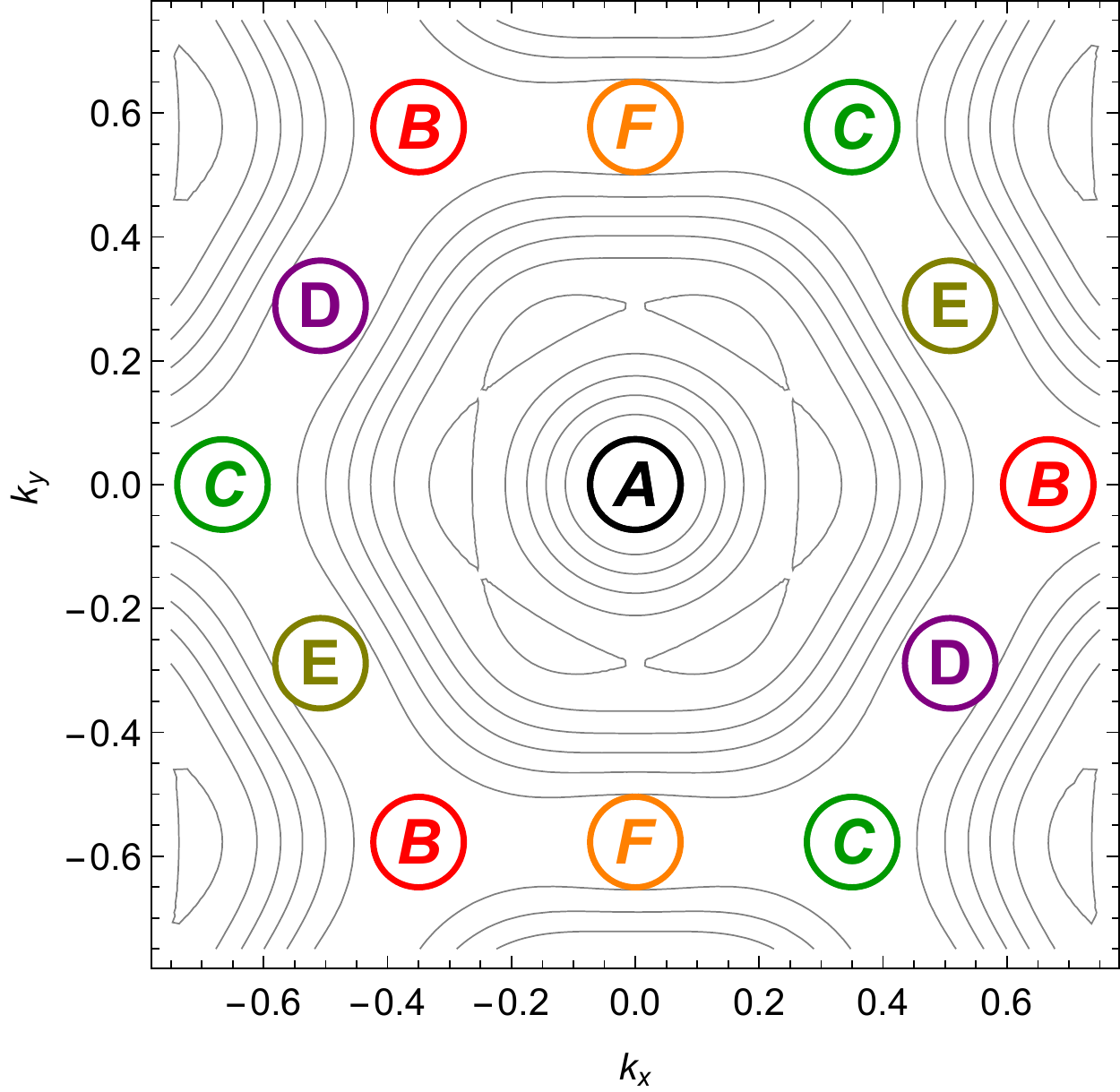}
\hspace{10pt}
\includegraphics[width=0.53\textwidth,keepaspectratio]{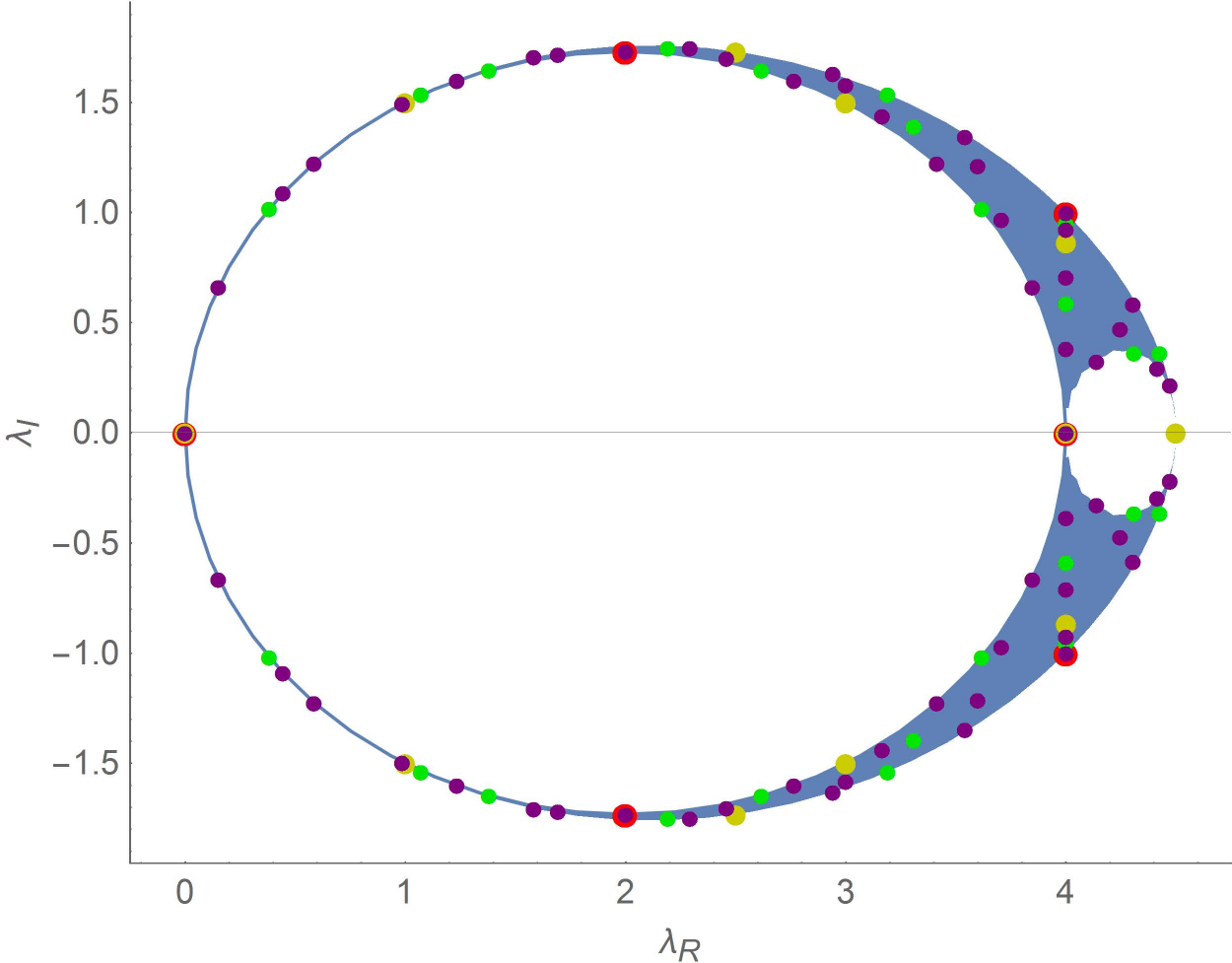}
   	\caption{ On the left, the Brillouin zone for the
          naive Dirac operator on a regular flat triangular lattice. The zero modes are labeled A-F.
          On the right, the infinite
          triangular lattice spectrum with the Wilson term (solid blue) compared to
          small lattices  with  16  (red), 36  (gold), 100
          (green) and 256  (purple) sites.}
          \label{fig:flat}
   	\end{center}
\end{figure}
On $\mS^2$,  the metric is
\be
ds^2_{S^2} =    d\theta^2  + \sin^2 \theta\, d\phi^2  \; .
\label{eq:2d-metric} 
\ee
With   $\vec e_\theta = (1,0)$, $ \vec e_\phi  = (0,\sin \theta)$ and 
$\sqrt{g} = \sin \theta$, the only non-zero components of the spin connection, (\ref{eq:TetraHyp}),
are $\omega^{12}_\phi=-\omega^{21}_\phi = - \cos \theta$. The  action on $\mathbb S^2$ is
\be
S_{sphere} = \int d\phi   d\theta \sin \theta \bar \psi\,   [\sigma^1
( \partial_\theta + \frac{\cot \theta}{2}) + \sigma^2
\frac{\partial_\phi}{ \sin \theta}   + m ] \psi \; . 
 \label{eq:actionsphere}
\ee
The massless Dirac operator, $D =  \sqrt{g} \nabla=\sin \theta[\sigma^1 ( \partial_\theta + \cot \theta/2)
  + \sigma^2(\partial_\phi/\sin \theta)]$, is anti-Hermitian and
  therefore has  pure imaginary eigenvalues $i \lambda$. It also  follows from the $\sigma_3$-Hermiticity property, $\sigma_3\, D\, \sigma_3 =D^\dagger=-D$, that eigenvalues come in
  complex conjugate pairs, 
\be
\lambda = \pm (j+1/2) \; ,
\label{eq:spectrum}
\ee
where $j=1/2, 3/2, \cdots$ are the allowed angular momenta.
Furthermore, for each $j$, the spectrum is $(2j+1)$-fold
degenerate~\cite{Abrikosov:2002jr}, with the degeneracy labeled by $-j\leq m\leq j$. The explicit eigenfunctions in terms of
spherical harmonics are given in Appendix~\ref{app:continuum}. The action is also invariant under $\sigma_1$ conjugation, $\sigma_1 D^* \sigma_1 = D$, or equivalently, together with $\sigma_3$ conjugation, $\sigma_2 D^* \sigma_2 = D^\dagger$. 
  These discrete symmetries are \emph{exactly}
preserved on  our simplicial complex.

For comparison, on the simplicial lattice, our action is
\bea
\label{eq:actionlattice}
S_{Wilson-Dirac} &=& \frac{1}{2} \sum_{\braket{i,j}} \frac{V_{ij}}{l_{ij}} \left(\bar{\psi}_i e^{(i)j}_a \sigma^a \Omega_{ij} \psi_j - \bar{\psi}_j \Omega_{ji} e^{(i)j}_a \sigma^a \psi_i \right) + \sum_i m V_i \bar{\psi}_i \psi_i \nonumber \\
&+& \frac{a}{2}\sum_{\braket{i,j}} \ \frac{V_{ij}}{l_{ij}^2} \left(\bar{\psi}_i - \bar{\psi}_j \Omega_{ji}\right)\left(\psi_i - \Omega_{ij}\psi_j \right)
\eea
with the Wilson term to remove doublers. We have set the coefficient,
$r$, of the Wilson term to the mean lattice spacing on the 
sphere: $r =a$. The Wilson term acts like a mass operator, so now the
eigenvalues have both real and imaginary parts. Defining the lattice
matrix ${\bf D}_{ij}$ by $S_{Wilson-Dirac} =\bar \psi_i {\bf D}_{ij} \psi_j$, $\sigma_3$
Hermiticity is still valid. Therefore, eigenvalues still come in
complex conjugate pairs,
\be
E=\lambda_R \pm i \lambda_I 
\ee
With rotational invariance broken, $\lambda_I$ no longer takes on
exactly   integral values and the $(2j+1)$-fold degeneracy is broken. In the
  limit of zero lattice spacing, $a\rightarrow 0$, one nevertheless
  anticipates the spectrum approaching $\lambda_I\rightarrow (j+1/2)$
  and $\lambda_R\rightarrow 0$, with doublers becoming increasingly
  massive and decoupling from the spectrum.

\begin{figure}
\begin{center}
 \includegraphics[width=.8\textwidth]{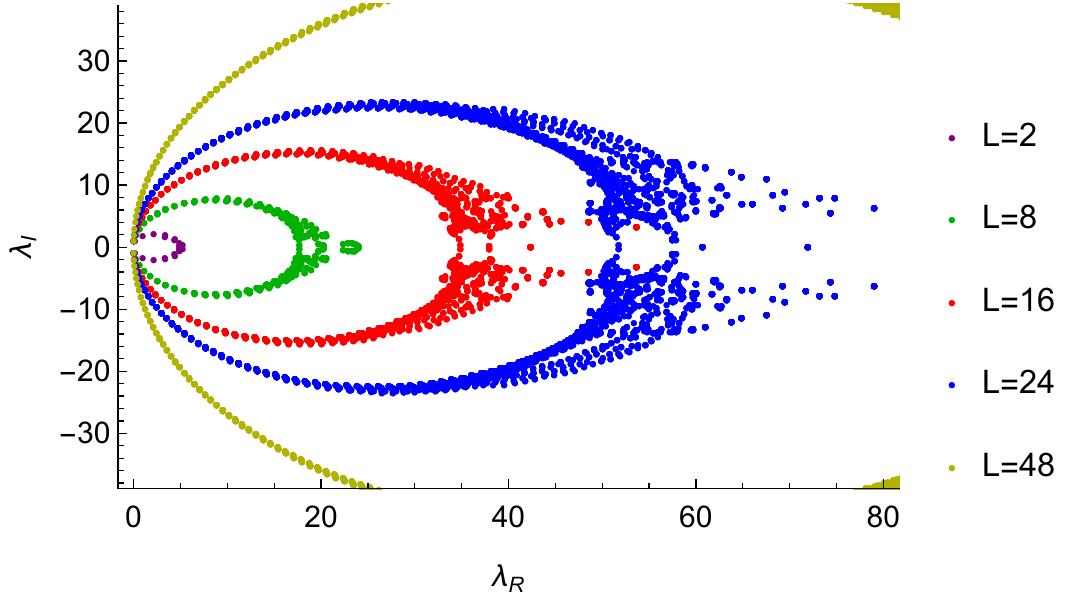}
\caption{ The Wilson-Dirac spectra on the discrete sphere for
  various refinement values of the refinement $L$.}
  \label{fig:spectrum} 
\end{center}
\end{figure}

Before introducing the Wilson term, it is interesting
to see its effect on a flat $L\times L$  regular triangular lattice with
$\Omega_{ij}=1$.  In the absence
of the Wilson term, as depicted by the left figure in Fig.~\ref{fig:flat}, the hexagonal Brillouin zone  actually has 6 copies of the 2-component
spinor zero modes~\cite{Chodos1977}. These   zero modes are labeled as $A, B, \cdots, F$.  The doublers spoil the continuum limit and
even fail to restore Lorentz invariance~\cite{Celmaster:1982ub}. When
the Wilson term is added, the doublers are removed and the spectrum
comes close to the circular complex spectrum of a lattice overlap
operator~\cite{Narayanan:1993sk}, converging rapidly to the
continuum. This is depicted in Fig.~\ref{fig:flat} on the right , with the $L \to \infty$ spectrum in solid blue compared to  small lattices  for $L=4, 6, 10, 16$.

On $\mS^2$, a global view of the Wilson-Dirac spectrum is illustrated
in Fig.~\ref{fig:spectrum}.  Not surprisingly, the qualitative effects
of the Wilson term on $\mathbb S^2$ are very similar to that for the
flat lattice shown in Fig.~\ref{fig:flat}. The apparent difference
between the two figures as a function of the refinement $L$ is due to
our convention.  On the flat plane, we treat the eigenvalues as
discrete dimensionless momenta ( $a p^\mu$), which scale to a
continuum dispersion relation as $a \sim 1/L\rightarrow 0$, whereas on
the sphere we have fixed the radius of $\mathbb S^2$ to one, so the
eigenvalues remain discrete approaching fixed values in the continuum
limit.  Fig.~\ref{fig:spectrum} plots the real vs imaginary parts of
eigen-spectrum for increasing refinement of $L=2,8,16, 24, 48$. In the
limit $L\rightarrow \infty$, the imaginary parts of the low-lying
eigenvalues, $\lambda_I$, approach $\pm (j+1/2)$, while their
corresponding real parts, $\lambda_R$, vanish as $O(1/L)$.

\begin{figure}[t]
\begin{center}
  \includegraphics[width=.9\textwidth]{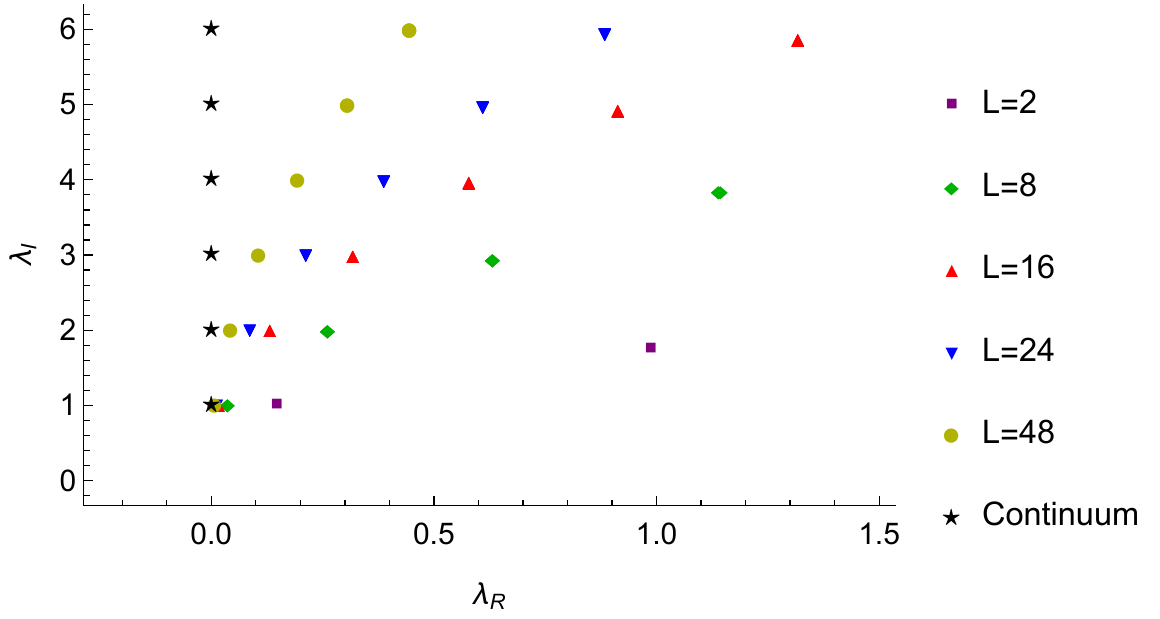}
\caption{As we increase the refinement level $L$, we expect the low-lying eigenvalues to converge to their continuum imaginary integer values. We see that as $L$ increases the real part (due to the Wilson term) approaches zero and the imaginary part approaches an integer. }
  \label{fig:spectrum2}
\end{center}
\end{figure}

\begin{figure}[t]
\begin{center}
\includegraphics[width=0.625\textwidth]{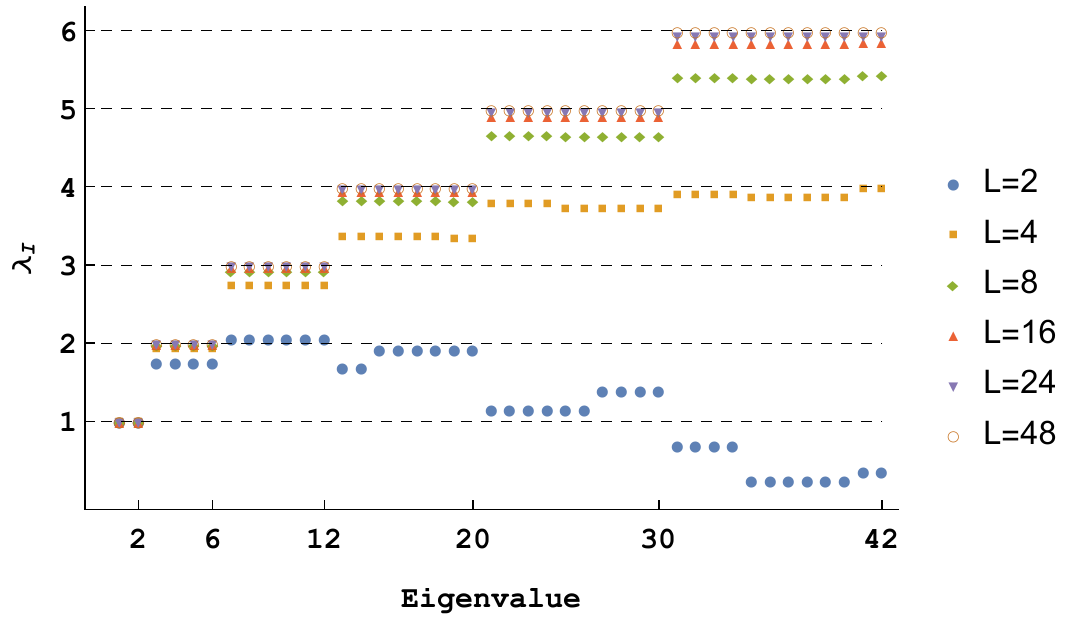}
\hspace{5pt}
	\includegraphics[width=0.325\textwidth]{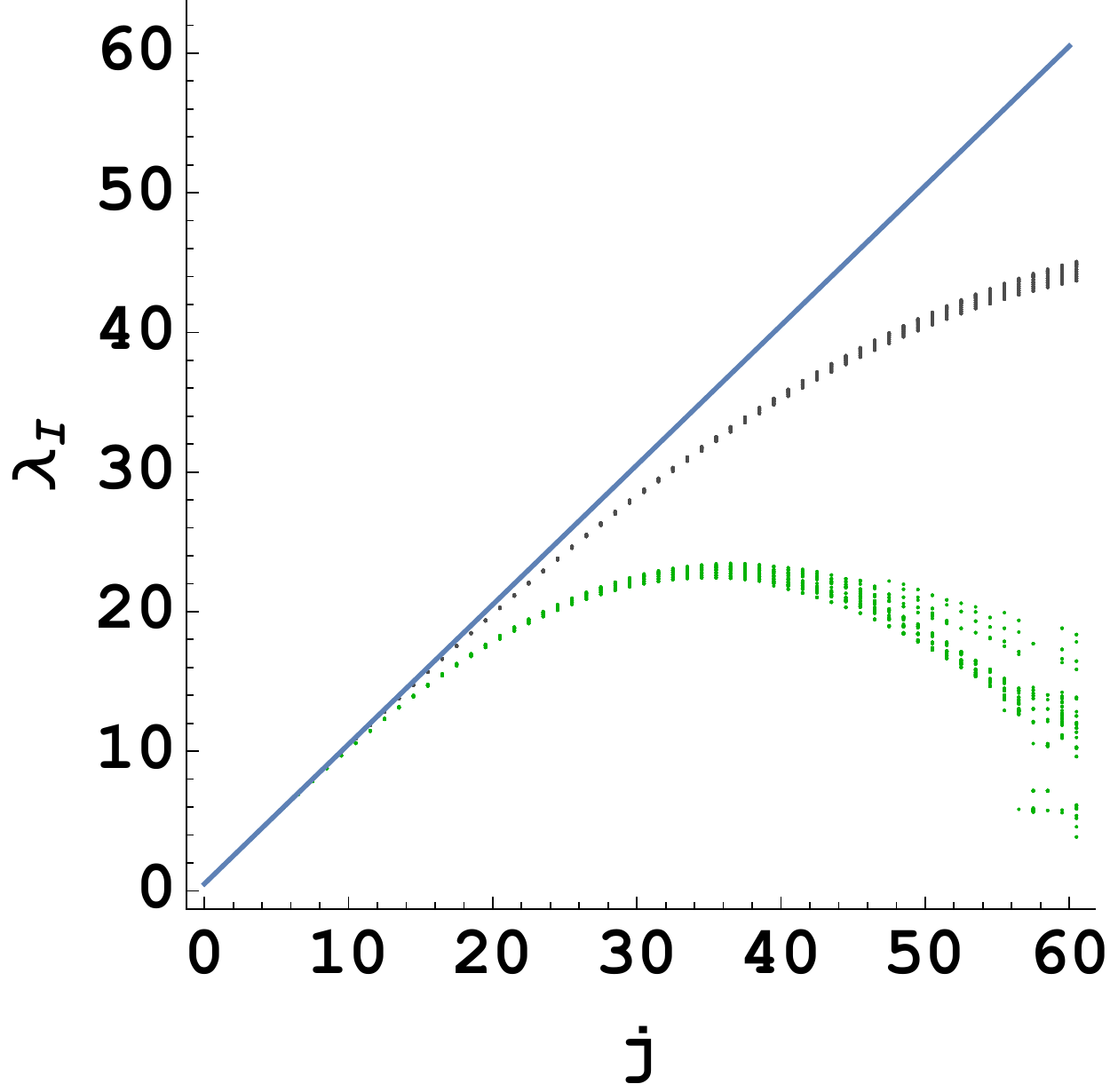}
\caption{On the left, we show the imaginary part of the low-lying  eigenvalues, with degenerate states repeated,   for
various refinements.  On the right, we show how the imaginary part of the eigenvalues 
for $L=24$, in green, and $L=48$, in grey, approach the continuum as a  function of $j$.  
}
\label{fig:globapp} 
\end{center}
\end{figure}

\subsection{\label{sec:Spectrum}Spectrum of the Lattice Dirac Operator}
\label{sec:Spectrum}

There are two approaches to determining the spectrum of the
Wilson-Dirac operator.  The first is to directly evaluate the
eigenvalues of the discrete Wilson-Dirac operator ${\bf D}_{ij}$,
which is limited by the efficiency of eigenvalue routines for sparse
matrices.  The second approach is to assume the eigenvectors are well
approximated by their continuum wave functions evaluated on
the lattice sites, $\psi^{(n)}_i$, and to compute the matrix elements of
the lattice Wilson-Dirac operator,
$\<\bar \psi^{(n)}|{\bf D}|\psi^{(n)}\> \simeq \lambda_{R,n}+
i\lambda_{I,n}$. It  is important to be precise in defining the
spectral problem on the simplicial manifold.  In the continuum 
the spectral problem  is the stationary value of the quadratic form,
\be
I = \int d^Dx \sqrt{g(x)} \bar \psi(x)( \nabla + m  - E )\psi(x)  \l ,
\ee
leading  either to the conventional eigenvalue problem,
 $(\nabla + m) \psi(x) =  E \psi(x)$,  where 
$\nabla ={\bf e}^\mu \boldsymbol{D}_\mu$,  or to the generalized eigenvalue problem
$D\psi(x) = E \sqrt{g(x)} \psi(x)$ where $D = \sqrt{g} ( {\bf e}^\mu
\boldsymbol{D}_\mu+  m)$.  On the simplicial lattice,
based
on the  discrete simplicial quadratic form, $I = \bar \psi_i (D_{ij}
- E V_i \delta_{ij})\psi_j$, is more conveniently given as the
generalized eigenvalue problem,
\be
D_{ij} \psi^{(n)}_j = E_n V_i \psi^{(n)}_i  \quad , \quad  \bar
\sum_i V_i \bar \psi^{(n')}_i \psi^{(n)}_i = \delta_{n',n} \;
\ee
Here the continuum measure, $\sqrt{g(x)}$, is replaced by the Vorioni
dual volume $V_i \equiv |\sigma^*_1(i)|$.  Alternatively one may
rescale by the square root of the measure, redefining the matrix as
$\widetilde D_{ij} = V^{-1/2}_i D_{ij} V^{-1/2}_j$ and eigenvectors as
$\widetilde \psi^{(n)}_i = V^{1/2}_i\psi^{(n)}_i $ to convert it to a
conventional eigenvalue problem.  Either way, properly treating the
measure $V_i$ is critical to a faithful correspondence with the
continuum.
 
\begin{figure}[ht]
\begin{center}
	\includegraphics[width=0.475\textwidth]{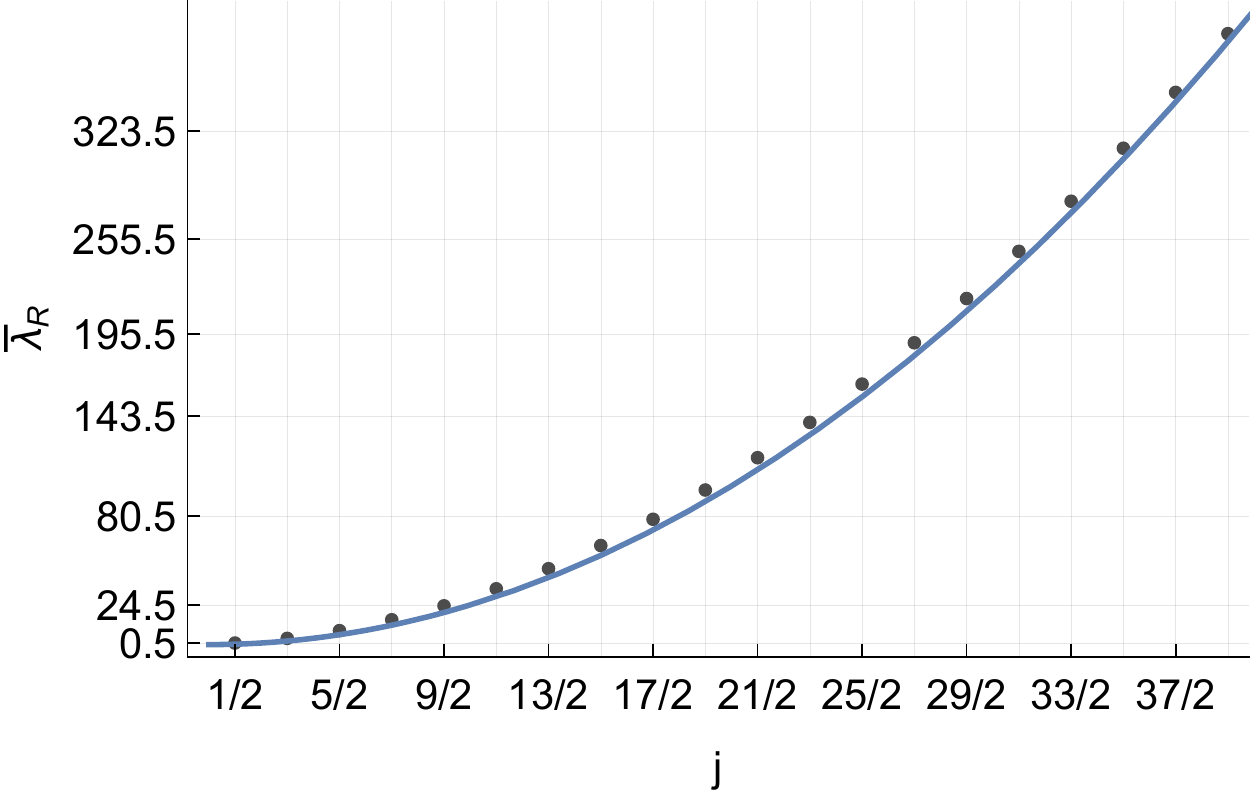}
\hspace{10pt}
	\includegraphics[width=0.475\textwidth]{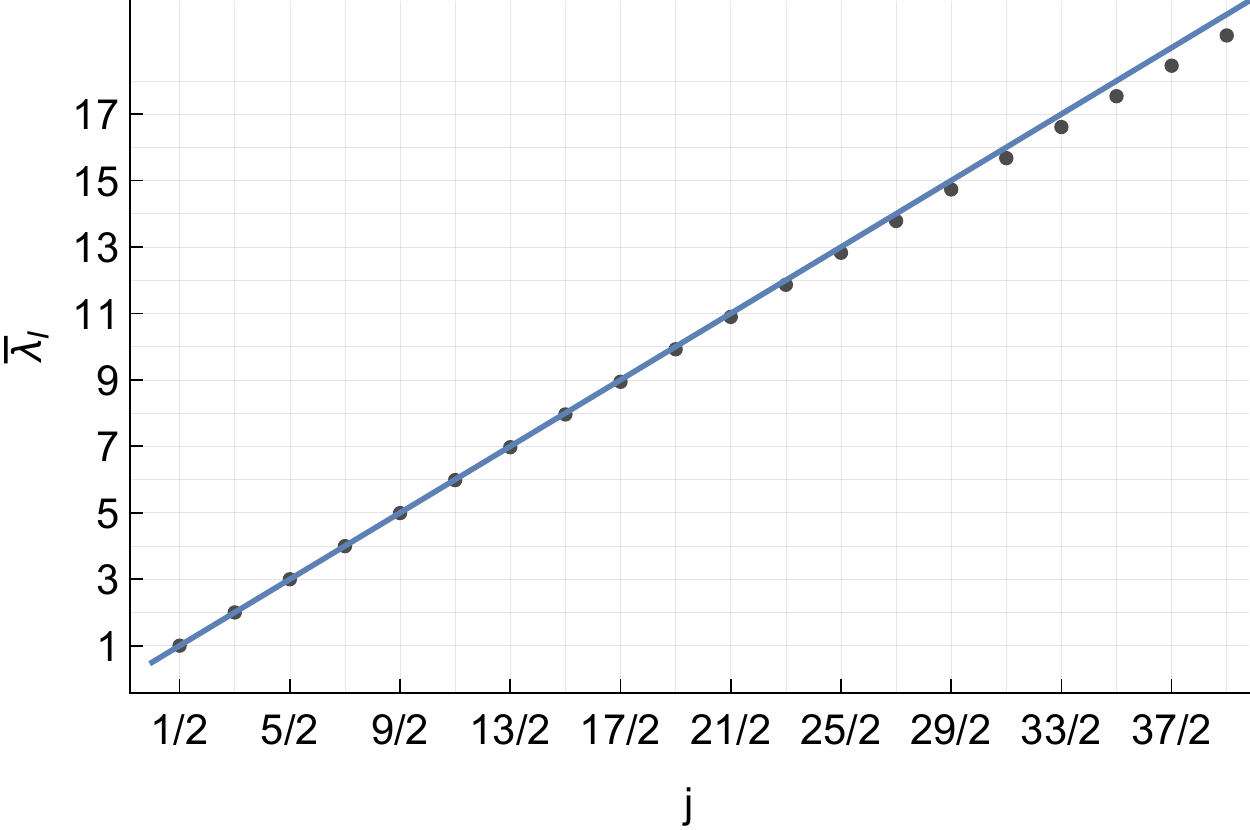}
\caption{ On the left, we show the real part of the eigenvalues for $j \leq 20$, averaged over $m$, at $L=48$. On the right, we show a similar plot for the imaginary part of the eigenvalues. The overlaid curves reflect the asymptotic continuum behavior given in Eq.~(\ref{eq:continuum-lambda}). }
  \label{fig:dispersion}
\end{center}
\end{figure}

\begin{figure}[ht]
\begin{center}
\includegraphics[width=.8\textwidth]{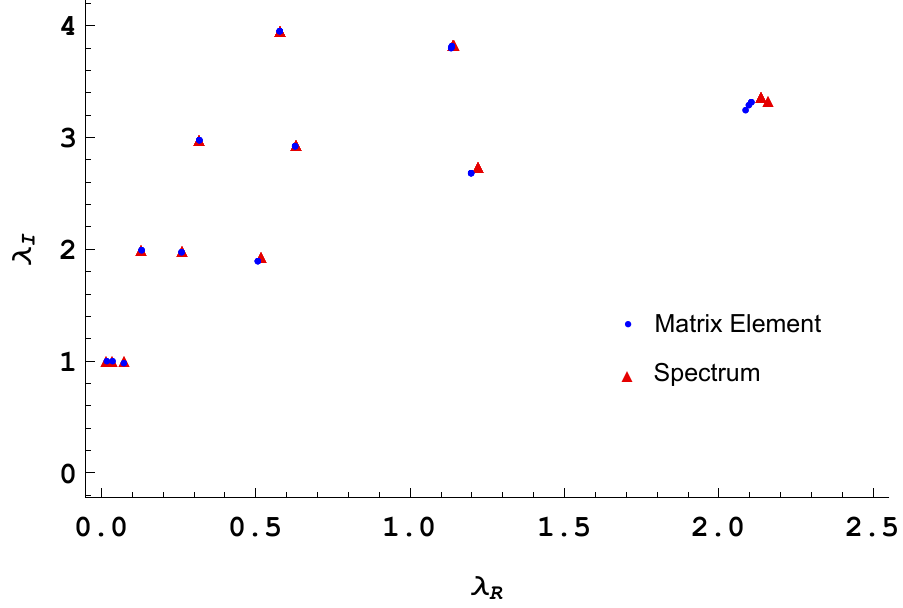}
\caption{ A comparison of the spectrum computed via a numerical eigensolver with the spectrum computed via matrix elements for $L = 4, 8,$ and $16$.}
  \label{fig:waveminCompare}
\end{center}
\end{figure}

\begin{figure}[ht]
\begin{center}
  \includegraphics[width=.85\linewidth]{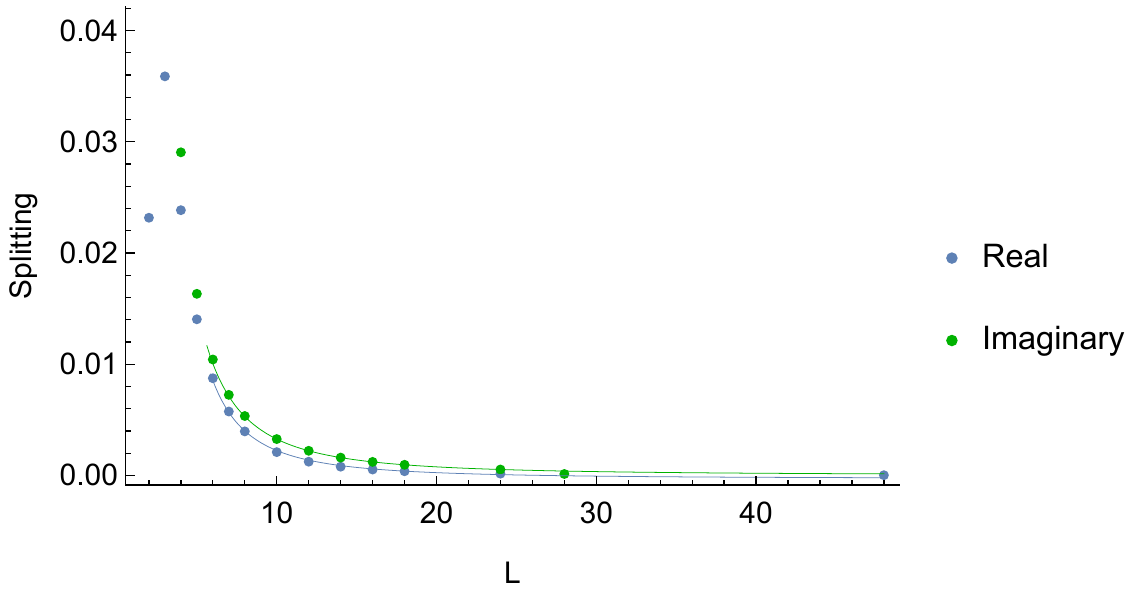}
\caption{The splitting of
  the eigenvalues for $\lambda=4$ as a function of $L$. We note that with increased refinement, the eigenvalues become more degenerate.}
  \label{fig:lambdaspread}
\end{center}
\end{figure}

\paragraph{Lattice Eigenvalues:}  
The low-lying eigenvalues are plotted in  Fig.~\ref{fig:spectrum2}
for a range of refinements $L$. The black
stars on the left side of the plot correspond to the continuum results
for the low-lying eigenvalues of the continuum Dirac operator, which
have integer spacing along the imaginary axis.  The $n$th level has
2$n$ degenerate eigenvalues corresponding to the $2j+1$ values for the
magnetic quantum number, $m$.  On the right side of the plot, we
show the numerically computed spectrum for a range of refinements,
$L = 2,8,16,24,48$, with $\lambda_R<10$. The degeneracy in $m$ is
(partially) broken, but too small to be seen.

In Fig.~\ref{fig:globapp}, we provide a more detailed picture of the breaking of degeneracy in $m$.  The left figure shows the imaginary part of all low-lying eigenvalues, with $1\leq \lambda_I\leq 6$, and $\lambda_R<10$, as $L$ increases.  As 
the lattice is refined, these levels quickly fall into clusters which can be associated with our continuum pattern, labeled by $j$-values, with an approximate degeneracy of  $2j+1$.   The \emph{first three} levels are
\emph{exactly} degenerate due to the symmetry of  the icosahedron under  the subgroup of
rotations.    In the figure on the right, we see that the imaginary part of the spectrum is linear for small $j$ and degenerate. However, for larger
$j$, the degeneracy in $m$ breaks down, as indicated by a spread in
the eigenvalues for fixed $j$, and various levels overlap. 

The dispersion relation including contributions
from both the Dirac and the Wilson term, which should converge
to 
\be
\lambda_I\rightarrow j+1/2\, , \quad\quad  \lambda_R\rightarrow
((j+1/2)^2-1/2)/L, 
\label{eq:continuum-lambda}
\ee
as we approach the continuum. Here the eigenvalues are  averaged over
the $2j+1$ values for the azimuthal angular momentum, $m$.  In Fig.~\ref{fig:dispersion}, for $L=48$, we plot the real and
imaginary parts of the eigen-spectrum as a function of $j$ for
$0< j<20$. For $j\leq15/2$, we performed
unweighted least-squares regression to the imaginary and real parts of
the eigenvalues. For the imaginary and real parts, we find
$\lambda_I(j)= 1.011 ( j + 0.480) -0.00197j^2$ and $\lambda_R(j)= (0.9836j (j +1)- 0.27 +
0.00097j^3)/L$ respectively.  Both are  consistent with the
theoretical expectation given in
Eq.~(\ref{eq:continuum-lambda}) derived in Appendix \ref{app:continuum}.

\begin{figure}[ht]
\begin{center}
  \includegraphics[width=.475\textwidth]{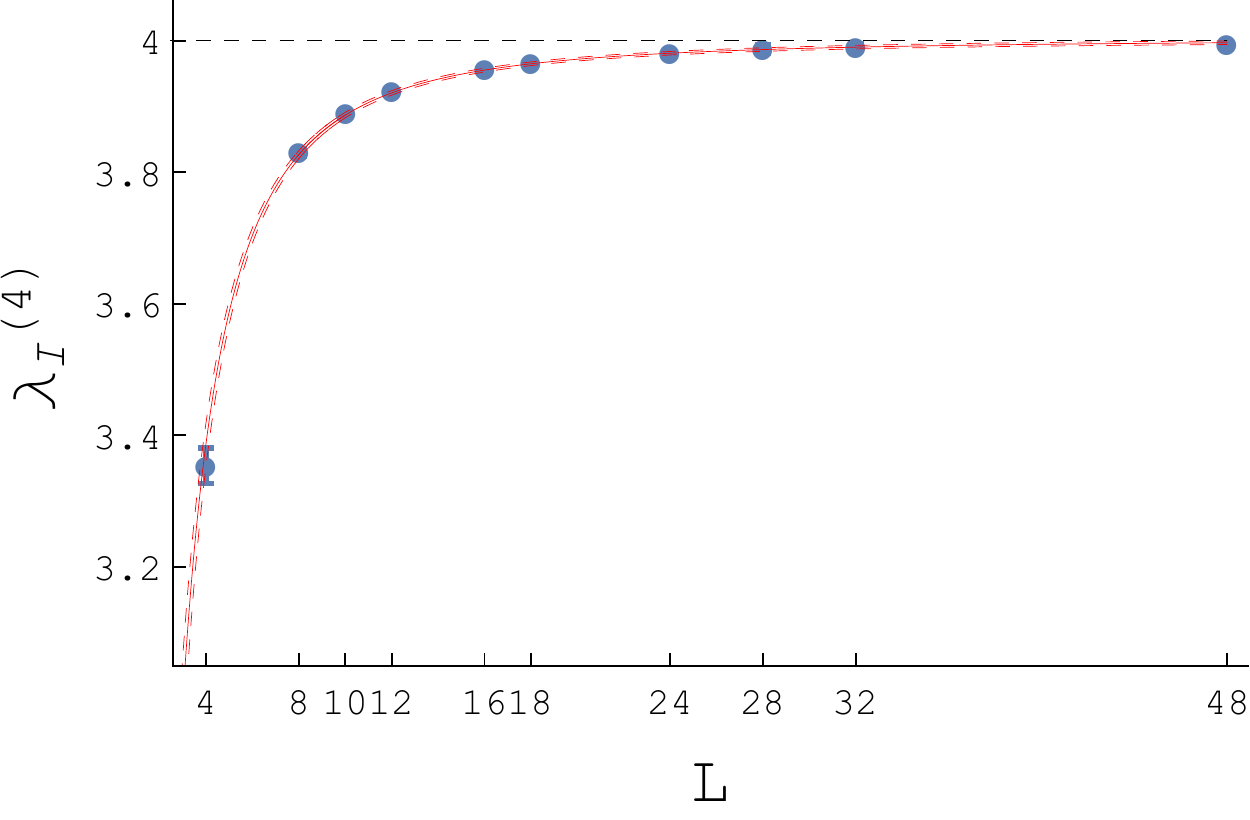}
  \includegraphics[width=.475\textwidth]{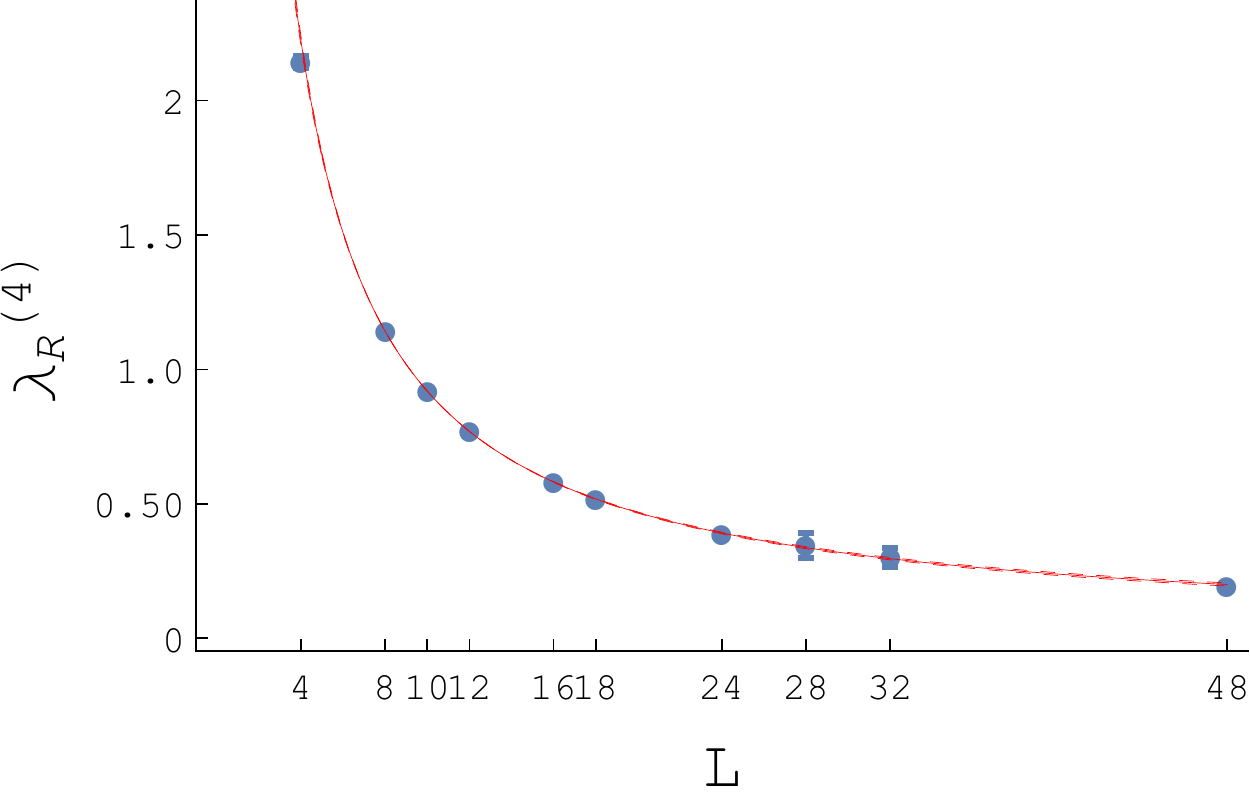}
\caption{On the left, we show the approach of the imaginary part of the $\lambda = 4$ eigenvalues, averaged over $m$, to the continuum as a function of the refinement $L$. On the right, we show the analogous plot for the real part of the eigenvalues.}
  \label{fig:lambdaapproach} 
\end{center}
\end{figure}

\paragraph{Lattice Eigenvectors:}
Given the continuum eigenfunctions restricted to the lattice
$\psi^{(n)}$ lattice, we can in
principle approximate the eigenvalue from matrix elements
$\braket{\psi^{(n)}|{\bf D}|\psi^{(n)}} \simeq \lambda_{R,n}+ i
\lambda_{I,n}$.  This also checks the accuracy of matching lattice
eigenvectors to the continuum (\ref{eq:wfn}). However, before proceeding, one must transform them from the continuum coordinate gauge into the gauge defined by our lattice action.

To fix the gauge, we can take advantage of the exact degeneracy in the magnetic quantum number for the first three levels. For simplicity, we choose the 
two lowest  continuum wave functions, that is, $m = \pm 1/2$ for $j = 1/2$, discretized on to the lattice sites,
$\psi_i = ( \psi^{(1)}(r_i),\psi^{(2)}(r_i))^T$ compared to the
 corresponding lattice eigenvectors of $\Psi_i = (
\Psi^{(1)}_i,\Psi^{(2)}_i)^T$ for ${\bf D}_{ij}$. The desired gauge
transformation at each site can be specified by a  local spinor rotation,
$e^{i\theta_i \sigma_3/2}$,    and  a global 2 by 2 unitary matrix, $\mathbb
U  $,  which mixes the degenerate pair.  These are determined by minimizing the functional 
\be
G(\theta_i,\mathbb U) = \sum_{i} \left|\psi_i - e^{i\frac{\theta_i}{2}
    \sigma_3} \mathbb U\Psi_i\right|^2  = - \sum_{i} \overline \psi_i e^{i\frac{\theta_i}{2}
    \sigma_3} \mathbb U \Psi_i- \sum_{i}  \overline \Psi^{(n)}_i  \mathbb U^\dag  e^{-i\frac{\theta_i}{2}
    \sigma_3}  \psi_i
    \label{eq:minimizing}
\ee
with respect to $\mathbb U $ and $e^{i\theta_i \sigma_3/2}$ on each
site $i$. This enables us to take the matrix element
$\<\bar \psi^{(n)}|{\bf D}|\psi^{(n)}\> $ using the discretize continuum
eigenvector rotated to our lattice frame to estimate the
eigenvalues. In Fig.~\ref{fig:waveminCompare} we compare the lattice
operator eigenvalues to the matrix elements. The two results are in
remarkable agreement, suggesting that the discrete Wilson term has
eigenvectors consistent with the Dirac term.  We found that the
minimum of the function $G$ given in Eq.~(\ref{eq:minimizing})
approaches zero as $1/L^2$. This suggests that the lattice
eigenvectors become an increasingly good approximation of the
continuum eigenvectors as $L$ increases. This is also consistent with
our previous observation that the Wilson term, while crucial to
removing spurious doublers, has a negligible effect on the physical
states as $L \rightarrow \infty$.

\subsection{Rate of Convergence to the Continuum}

There are two ways to test the convergence of the spectrum to the
continuum limit. {\bf i.)}~The restoration of degeneracy in the
magnetic quantum number, $m$, as $L$ increases. {\bf ii.)}~The
behavior of the spectrum, averaged over $m$, as $L$ increases. The
\emph{exact} $\sigma_1$ symmetry results in a pairing of degenerate
eigenvalues for each $j$. Since our discretization exactly preserves 
icosahedral symmetry, the first level which exhibits breaking of the
degeneracy in $m$ is the fourth level. For $\lambda = 4$, there are two
irreducible representations of the icosahedral group, resulting in a
splitting into two groups with two and six members as illustrated on
the left  in Fig. \ref{fig:globapp}.  At higher levels, the
eigenvalues can split into a larger set of irreducible
representations.
 
Restricting our attention to $\lambda = 4$, we define the
splitting in the eigenvalues, independently for the real and imaginary parts, as the difference between the maximum and minimum eigenvalues.  In Fig.~\ref{fig:lambdaspread},
we consider this splitting as $L$ increases. We
perform an unweighted linear regression to the splitting as a function
of $L$. For the imaginary  and real parts, we find the splitting behaves as
$-6\times10^{-5}  + 0.0034 /L + 0.230/L^2$ and $0.0009  - 0.035 /L
+0.44/L^2$, respectively, consistent with restoration of full spherical
symmetry in the continuum. 

Next  we consider how the eigenvalues, averaged over $m$, approach the continuum for $\lambda = 4$. In Fig.~\ref{fig:lambdaapproach}, on the left,
 we fit the eigenvalue to $\lambda_{I,4} =  3.99932+ 0.034/L - 11.67
 /L^2$ consistent with the continuum value,  $\lambda_{I,4} = 4$ .  On
 the right, we see  the real part also approaches the correct
 continuum value, $\lambda_{R,4} = 0$. The convergence of this term is
governed by the Wilson term, which scales with an extra  factor of lattice
spacing compared to the naive Dirac term. We therefore expect it to
converge more slowly, as $O(1/L)$. Our fit gives $\lambda_{R,4} = 0.0025 + 9.19/L$, again consistent with our expectations.

\section{\label{sec:projectiveSphere}The Ising Conformal Field
  Theory on  $\mathbb S^2$}

The exact solution to the 2D Ising model provides a rigorous test of our simplicial construction of the
free fermions on $\mS^2$.  To begin let us review this continuum $c = 1/2$ minimal
model. There are only three Virasoro primaries
${\bf 1}, \sigma, \epsilon$, with an OPE expansion, 
\be \sigma \times
\sigma = {\bf 1} + \epsilon \, ,\quad \epsilon \times \sigma =
\epsilon \, ,\quad \epsilon \times \epsilon = {\bf 1}\; .
 \ee 
It is equivalent to a free Majorana holomorphic, $\psi(z)$, and
anti-holomorphic, $\bar \psi(\bar z)$, field on all 2D Riemann
surfaces~\cite{Christe:1993ij}. In the complex plane, the Riemann surface can be represented by inserting pairs of square root
branch points whose locations corresponds to the $\sigma(z)$ operators. When projected onto our simplicial lattice on 
$\mS^2$, these represent pairs of branch points given by simplicies
with curvature defects  of $-1$. Clearly, these defects must be inserted in pairs by flipping bonds on an invisible
string  between these flipped plaquettes.  Here we compute the 2-point
and 4-point functions, 
\be\label{eq:2and4ptfn} \<
\epsilon(x_1) \epsilon(x_2) \> \, ,\quad{\rm and} \quad  \< \sigma(x_4)
\epsilon(x_3) \epsilon(x_2) \sigma(x_1) \> \, ,
\ee 
where $\epsilon(x) = i \bar \psi(x) \psi(x)$ and $\sigma(x)$ is the twist
operator that introduces the square root branch points. Of course, correlators
with only fermion operators,
 such as \\$\< \epsilon(x_1) \epsilon(x_2) \epsilon(x_3) \epsilon(x_4)
 \> $, are trivially given by the Wick contractions as  products of  2-point functions
$\< \epsilon(x_1) \epsilon(x_2) \>$.   The 
$\< \sigma(x_4) \sigma(x_3) \sigma(x_2) \sigma(x_1) \>$ correlation
function is the partition function on the torus. This is  computed in Ref.~\cite{Brower:2016moq} as a  test of QFE methods for the $\phi^4$
CFT theory on $\mS^2$.

\subsection{Dirac vs Majorana Propagators}  
In the continuum, the 2D Dirac fermion,
\bea
S &=&\int d^2x \bar \Psi[\sigma^\mu \partial_\mu  +m] \Psi =  2 \int dz d\bar
z\bar \Psi [  \sigma_- \partial_{\bar z} + \sigma_+ \partial_z  +m 
] \Psi \; ,
\eea
at  zero mass can be decomposed into two single 
 component Majorana fermions,
\be
S = 2 \int dz d\bar z [  \psi \partial_{\bar z}  \psi + \bar \psi \partial_z
\bar \psi ]= 2 \int dz d\bar z [  \psi \bar \partial  \psi + \bar\psi \partial
\bar \psi] \; ,
\ee
where 
$\Psi = (\Psi_1, \Psi_2)^T  \equiv (\psi, \bar \psi)$  and $\bar \Psi
=\Psi^T \sigma^1 = ( \Psi_2, \Psi_1)^T \equiv (\bar \psi,  \psi) $ 
are  split into a holomorphic and anti-holomorphic parts,
$\psi(z)$ and $\bar \psi(\bar z)$, respectively. The holomorphic
propagator is
\be
\langle \psi(z_1) \psi(z_2)\rangle = \langle z_1,\bar z_1| \bar \partial^{-1}
|z_2,\bar z_2\rangle = \partial \,  \langle z_1,\bar z_1|
(\bar \partial\partial)^{-1} |z_2,\bar z_2\rangle =\frac{1}{2\pi }\frac{1}{z_1-z_2} \, , \label{eq:holo}
\ee
and the anti-holomorphic propagator is $\langle \bar\psi(\bar z_1)\bar \psi(\bar z_2)\rangle =  \langle \psi(z_1) \psi(z_2)\rangle^*$. 
Note that these solutions are regular at $0$ and $\infty$, and
periodic in $\theta \rightarrow \theta  + 2 \pi $ for $z = |z| e^{i\theta}$.  By inserting twist
operators at $0$ and $\infty$, the  propagators,
\be
\langle \sigma(\infty) \psi(z_1) \psi(z_2) \sigma(0) \rangle =\frac{\sqrt{z_1/z_2} + \sqrt{z_2/z_1} }{4\pi }\frac{1}{z_1-z_2} \, ,
\ee
and  $\langle \sigma(\infty) \bar \psi(\bar  z_1) \bar \psi(\bar  z_2) \sigma(0)
\rangle = \langle \sigma(\infty) \psi(z_1) \psi(z_2) \sigma(0)
\rangle^*$ are now  anti-periodic in $\theta$.  To make contact with our simplicial Dirac fermion
requires two steps: first projecting the flat space correlators to the
Riemann $\mS^2$ sphere and second identifying  a single Majorana
component within our 2 component simplicial Dirac fermion.

\subsection{Stereographic Projection for Conformal Fields}

Under a Weyl rescaling of the flat metric,
\be
g_{\mu \nu}(x) = \frac{\dd \xi^\alpha}{\dd x^\mu}
\frac{\dd \xi^\alpha}{\dd x^\nu} =  \Omega^2(x)  \delta_{\mu\nu}    \; ,
\ee  
the conformal correlation functions for
primaries   $\cO_i$ of dimension $\Delta_i$ obey the 
general identity~\cite{Rychkov:2016iqz}, 
\be
\<\cO_1(x_1) \cO_2(x_2) \cdots  \>_{g_{\mu\nu}} =
\large[ \frac{1}{\Omega(x_1)^{\Delta_1}} \frac{1}{\Omega(x_2) ^{\Delta_2}} \cdots  \large] \< \cO_1(\xi_1)  \cO_2(\xi_2) \cdots 
\>_{flat}  \; .
\ee
In particular the map, $\mR^2 \rightarrow \mS^2$, to the projective sphere,
\be
ds^2_{\mS^2} = \frac{2}{(1+z\bar z)^2} dz d\bar z =\cos^2(\theta/2)
ds^2_{\mR^2} \; . 
\ee
introduces the Weyl factor,  $\Omega^2(\theta) =   \cos^2(\theta/2)$, and leads to the identity for the  2 point function
\be
\<\cO_1(x_1) \cO_2(x_2) \>_{\mS^2 }= \frac{1}{[\Omega(x_1)|z_1 -
  z_2|^2 \Omega(x_2)]^{\Delta}} =  \frac{1}{(2 - 2\cos\theta_{12})^\Delta}\, ,
  \label{eq:propagator}
\ee
where $\Omega(\theta_1)|z_1 - z_2|^2 \Omega(\theta_2)  = |\vec r_1 - \vec
r_2|^2 = 2(1 - \cos\theta_{12})  $ with  radial vectors, $r = (r_x+ir_y,r_z)= (\sin\theta
e^{i\phi}, \cos\theta)$  restricted to the  unit sphere embedded in $\mR^3$.  
Just as Poincare invariance on the plane implies that  correlators are
a function of the length (or Euclidean distance on the plane, $|z_1
-z_2|$), rotational invariance on the sphere fixes the correlator  to be  a function of the
geodesic distance, $\theta_{12}$.  In addition scale invariance fixes the
full functional form.

It is often useful to make use of  conformal cross ratios $u$ and $v$, which are also invariant under Weyl transformations,
\be
u =  \frac{x^2_{12} x^2_{34}}{x^2_{13} x^2_{24} } = \frac{r^2_{12}
  r^2_{34}}{r^2_{13} r^2_{24} }  \quad ,\quad 
v = \frac{x^2_{14} x^2_{32}}{x^2_{13} x^2_{24}} = \frac{r^2_{14}
  r^2_{32}}{r^2_{13} r^2_{24}} \; ,
\ee
where $r^2_{ij} = (\vec r_i - \vec r_j)^2 = 2 (1 - \cos \theta_{ij})$.
In moving from $\mR^2$ to $\mS^2$, all conformal factors cancel.
In 2D one also can 
combine the two cross ratios into a single complex number,
\be
\zeta = \frac{(z_1 - z_2) (z_3 - z_4)}{(z_1 - z_3) (z_2 - z_4)} \; ,
\ee
where $z_i = \cot(\theta_i/2) e^{i\phi_i}$ and 
$u = |\zeta|^2, v = |1- \zeta|^2$.

For future reference, we point out that this construction can be generalized  to $\mS^D$ by the replacement
$\hat r = (r_z,\vec r_\perp)= (\cos\theta, \sin\theta\,\hat r_\perp)$, or
if you prefer use rotational symmetry to bring $r_z$ and
$|\vec r_\perp|$ to the $x-y$ plane. One may see this in two
  steps. First, one maps $\mR^D \rightarrow \mR \times \mS^{D-1} $ via radial quantization
   with coordinates
  $(\log r,\vec r_\perp)$, then one maps to the projective sphere
  $\mR \times \mS^{D-1} \rightarrow \mS^{D}$ with coordinates
  $(\cos\theta,\sin\theta \hat r_\perp )$. 

\subsection{Numerical Tests for 2- and 4- Point Correlators}

To numerically compute conformal correlators,  we need to identify the Majorana
components in our simplicial Wilson Dirac fermions.
This is accomplished by including a Majorana mass,  and comparing the
continuum with the lattice form of the Dirac operators,
\be\label{eq:actioncomponents}
M_{z_1,z_2} = 
\begin{bmatrix}
m & \partial \\
\bar \partial  & m
\end{bmatrix}_{z_1,z_2}
\rightarrow 
\begin{bmatrix}
W & \nabla  \\
- \nabla^\dag  & W
\end{bmatrix}_{z_1,z_2}
\ee
On the right, $\nabla$ is the naive central difference operator for a
massless lattice fermion and
$W$ is the Wilson term including the  mass. This identification recognizes that
the Wilson term $W$ plays the role of the mass term in the continuum limit,
in addition to removing the unphysical doublers.  We
compute the inverse for both representations using
the  Schur decomposition.
In the continuum, on $\mR^2$,  we have
the expression,
\be
G(z_1,z_2;m)   = 
\begin{bmatrix}
m^{-1}  +m^{-1} \partial (m^2 - \bar \partial\partial)^{-1} \bar \partial  &  - \partial (m^2 - \bar \partial \partial)^{-1} \\
- (m^2 - \bar \partial \partial)^{-1} \bar \partial  & m (m^2
-\bar  \partial \partial)^{-1} \; ,
\label{eq:SchurContinuum}
\end{bmatrix} 
\ee
for $G(z_1,z_2;m) = M^{-1}_{z_1,z_2}$, which can be compared with the Wilson Dirac lattice propagator,
\be
G(z_1,z_2)  =
\begin{bmatrix}
W^{-1}  +W^{-1} \nabla {\bf \Delta}_s^{-1} \nabla^\dag W^{-1} & - W^{-1} \nabla {\bf \Delta}_s^{-1}\\
{\bf \Delta}_s^{-1} \nabla^\dag W^{-1}  & {\bf \Delta}_s^{-1}
\end{bmatrix}_{zw}  \; ,
\label{eq:SchurLattice}
\ee
where ${\bf \Delta}_s =  W +  \nabla^\dag W^{-1} \nabla$ is the Schur
complement. Taking the zero mass limit of Eq.~(\ref{eq:SchurContinuum}), we
can identify the Majorana propagator  as the
off-diagonal terms in Eq.~(\ref{eq:SchurLattice}), so it follows that on the lattice, we should also
identify these off-diagonal term for the lattice conformal propagators.
Consequently,  in  the zero mass limit, the correspondence,
\be
 G_{12} (z_1,z_2) G_{21} (z_1,z_2) =|G_{12} (z_1,z_2) |^2  \rightarrow 
\langle \psi(z_1)\bar \psi(\bar z_1) \bar \psi(\bar
z_2)\psi(z_2)\rangle 
\label{eq:MajoranaMap}
\; ,
\ee
is established. 

\begin{figure}[h]
		\begin{center}
		\vspace{10pt}
\includegraphics[width=.95\textwidth]{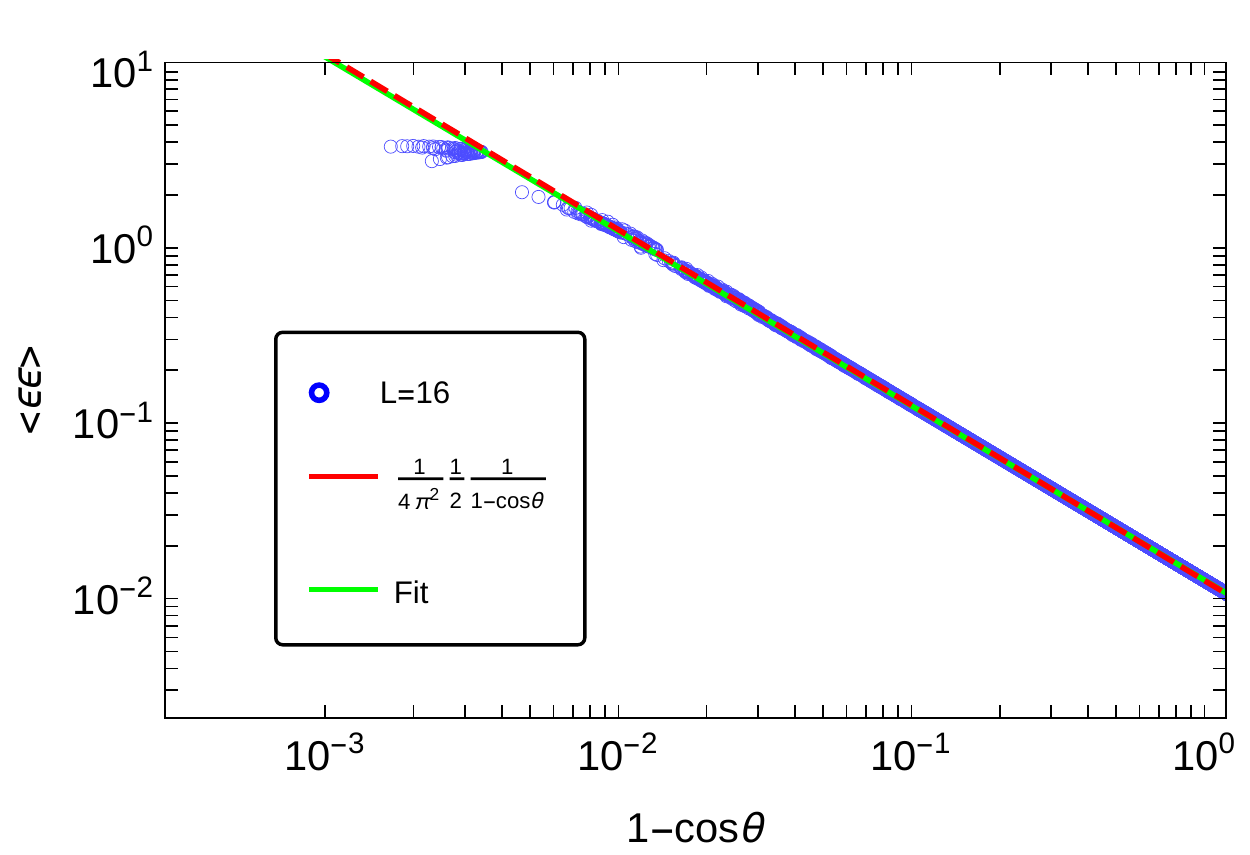}
   	\caption{\label{fig:2pt}Log-Log plot of the two point correlator for $L=16$.}
   	\end{center}
\end{figure}

\paragraph{Lattice $ \epsilon \epsilon$ Correlator:} We will now show
numerically that not only is Eq.~(\ref{eq:MajoranaMap}) correct,  but
the simplicial correlator converges rapidly to the continuum on $\mS^2$,
\be
\langle \epsilon(\vec r_1) \epsilon(\vec r_2)\rangle = \frac{1}{4\pi^2 } 
\frac{1}{2(1 - \cos\theta_{12})} \; .
\label{eq:conformalcorrelator-sphere}
\ee
A comparison of the numerical result versus the analytic result is given in Fig.~\ref{fig:2pt}. 
At very small
distances, cut-off effects give a visible  disagreement with the continuum result,
but otherwise the fit is remarkably good even at
relatively small $L=16$ . It is
important to note that this is a  zero
parameter fit, including the normalization.  Fitting the data to the expected
functional form, we find
$ (a/ 8 \pi^2)\times (1-\text{cos}\theta)^{-\gamma} \simeq  (1.0035/8
\pi^2)\times  (1-\text{cos}\theta)^{-0.996}$. At $L = 16$ the finite lattice errors are less than 1 per cent.

\begin{figure}[h]
\centering
\includegraphics[width=.9\textwidth]{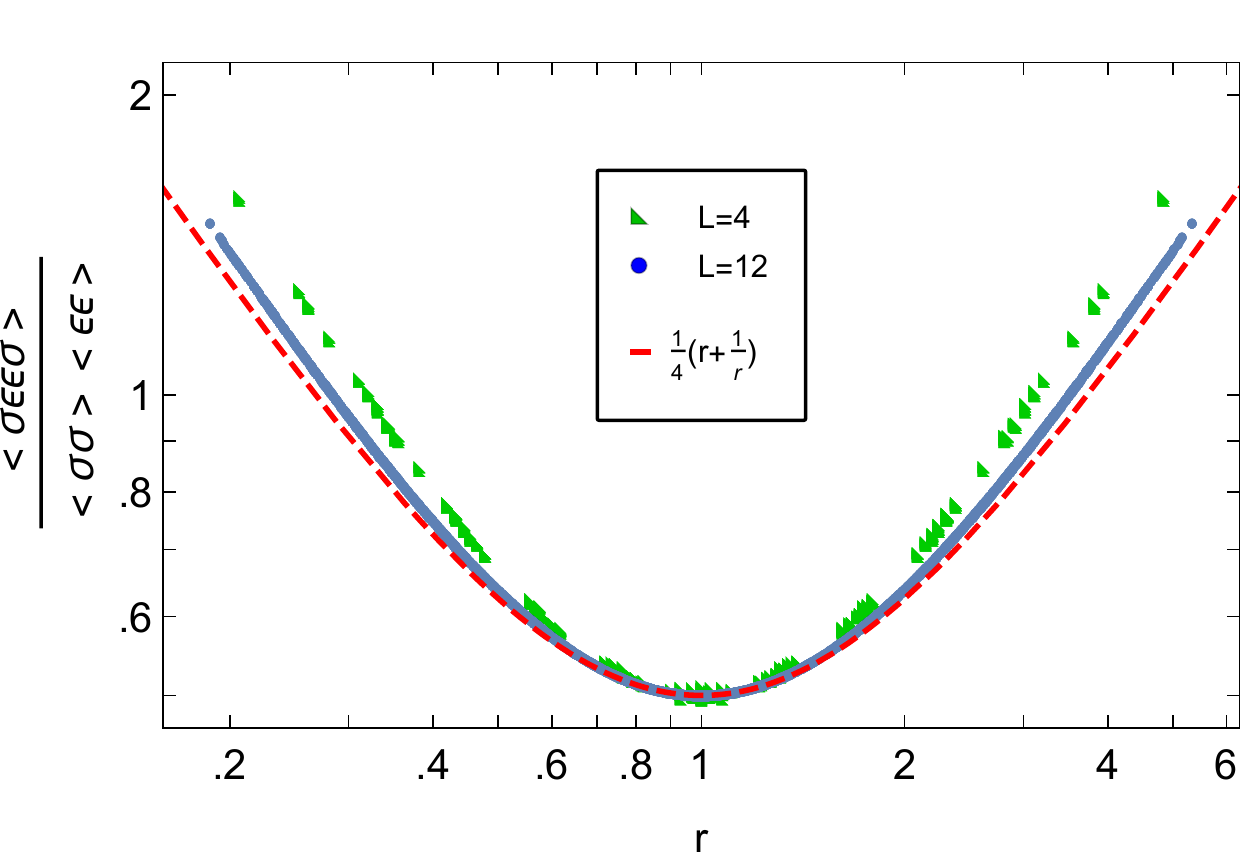}
\caption{\label{fig:4ptRadius} Functional dependence in $r$ of the four
  point correlator for $L = 4, 8, 12$, isolated by subtracting the $\theta$ term.  The dotted red line is the expected continuum behavior.}
\end{figure}

\begin{figure}[h]
\centering
\includegraphics[width=.9\textwidth]{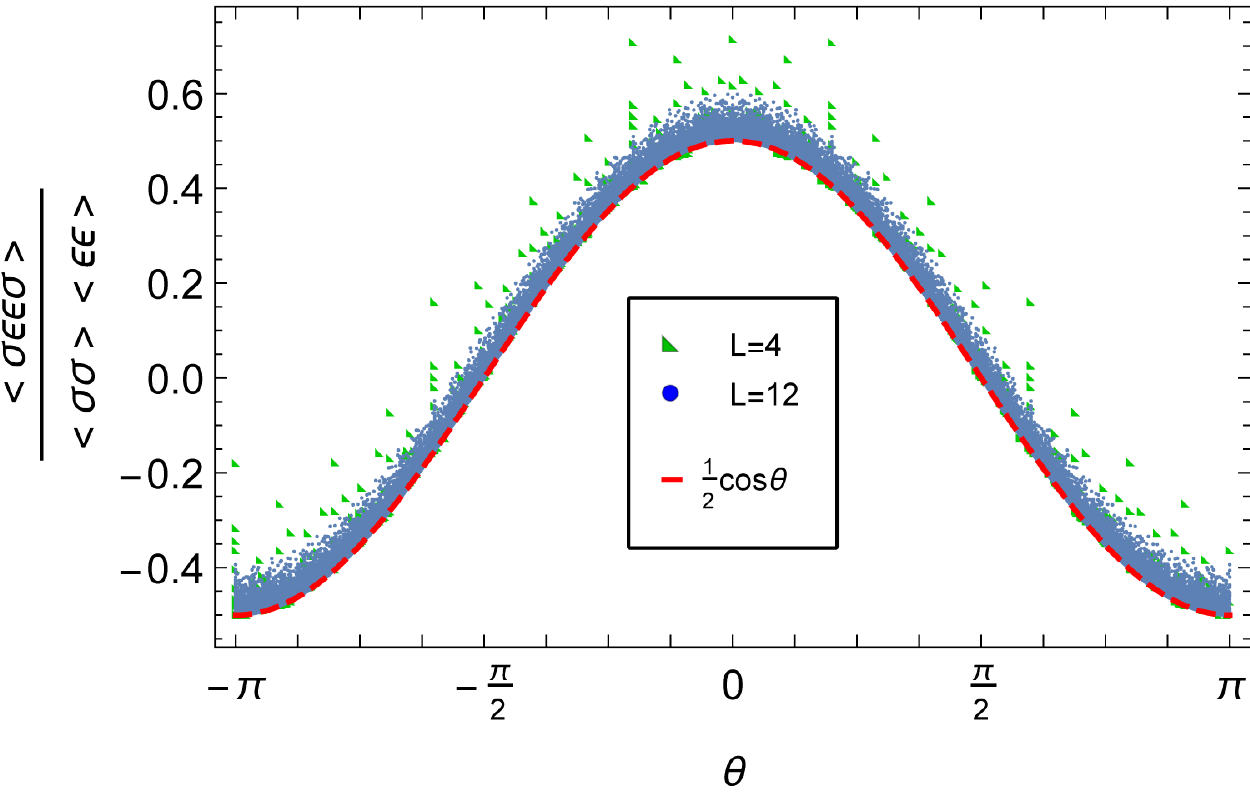}
\caption{\label{fig:4ptTheta} Functional dependence in  $\theta$ of the four
  point correlator for $L= 4, 8,  12 $ isolated by subtracting the $r$ terms.  The dotted red line is the expected continuum behavior.}
\end{figure}

\paragraph{Lattice $\sigma \epsilon \epsilon \sigma$ Correlator:}
To examine the four-point correlator, we need to introduce twist
operators on the lattice.  It is  convenient to introduce the branch points  at the north and south poles of our decorated
icosahedron and to maintain a discrete 5-fold axial symmetry in $\theta$.
To accomplish this, first, the pole points are removed.  This takes our
lattice from the topology of a sphere to the cylinder.  Next, the spin
connection on one link around the poles is flipped in sign.  This
introduces a topological defect at the north and south poles which
corresponds to the insertion of our lattice twist operators.  Finally,
a path is constructed  between the flipped links at the north and south
poles, flipping the sign of the spin connection along the path, so
that the only defects are at the north and south poles.

For a numerical comparison we normalize the lattice four point function by the simplicial lattice two
point function and compare with the analytical form on the sphere,
\be
\frac{\< \sigma(\infty)
\epsilon(z_2)  \epsilon(z_3) \sigma(0) \>}{\< 
\epsilon(z_2)  \epsilon(z_3) \>} 
=\frac{1}{4} (
r + 1/r + 2 \cos\theta_{23}) \; ,
\label{eq:ratiocorr}
\ee
as a function of the conformally invariant co-ordinate: ~$z_2/z_ 3 \equiv r e^{i\theta}$. 
Unlike the 2-point function,
this depends on both the angular separation $\theta_{23}$ and 
the magnitude $ |z_2/z_3|$. When either one of the  $\epsilon$
operators is near the poles, the ratio function in Eq.~(\ref{eq:ratiocorr}) diverges and the lattice
results have strong cut-off effects, which we suppress by restricting the $\epsilon$ fields
to the range between polar angles $[\pi/4,3\pi/4]$. The results can be seen in
Figs.~\ref{fig:4ptRadius} and \ref{fig:4ptTheta}.  First, in
Fig.~\ref{fig:4ptRadius}, we see the $r$ dependence of the ratio
function by subtracting off the $\cos(\theta)/2$ data, and next in
Fig.~\ref{fig:4ptTheta} we see the $\theta$ dependence by subtracting off
the $(r+1/r)/4$ data.  In both cases, the numerical results
converge to the  continuum result shown in red.  The total
data set can be fit to the functional form $a(r+1/r)/4+b
\,\cos(\theta)/2$ with $1/L$ corrections giving $a = 1.0008 + 0.264/L$ and $b = 1.00033 -
0.00566/L$.  In view of the neglecting $O(1/L^2)$ terms in the fit, this is
 consistent with the exact continuum limit ($a =b = 1$).

 \section{\label{sec:conclusion}Discussion and Future Directions} 

We have presented a solution to lattice Dirac fermions on a simplicial
complex approximating a general smooth Riemann manifold.  To achieve
this we borrowed methods from Finite Elements (FEM),  Regge
Calculus (RC) and the language of the Discrete Exterior Calculus (DEC). 
However, our solution required substantial new features to
accommodate the curved manifold going beyond the linear piecewise
implementations prevalent in the literature.  To remove the doublers, we have used the construction of Wilson fermions.
As in  flat space, the operator for this simplicial Wilson fermion
can be used as a kernel for Shamir~\cite{Shamir:1999gv} and M\"obius~\cite{Brower:2012vk} Domain Wall  fermions by introducing a flat extra
dimension of length $L_s$.  Just as in flat space, this should
converge  as $L_s \rightarrow \infty$  to an exact simplicial
lattice chiral  overlap fermion representation~\cite{Narayanan:1993sk}. 

This appears to us to be the first  general solution for simplicial lattice
Dirac fermions on any smooth Euclidean Riemann manifold that is capable of
convergence to the exact continuum limit. To support this conjecture, tests were made for the simplicial lattice 
on a 2D Riemann sphere compared with the exact continuum solutions. While this is obviously far from a proof,
additional tests on higher dimension manifolds will be performed. The proof
of convergence theorems have not yet been attempted. Convergence proofs for
classical FEM and Regge Calculus are far from trivial or
complete~\cite{StrangFix200805}, let alone their extension to the
simplicial fermions
presented here. However, we feel that the geometrical underpinning
of our approach makes our convergence conjecture plausible.

To address the central problem of Quantum Finite Elements (QFE),  interacting quantum field theory on
curved manifolds, we need to introduce  interactions with scalar and gauge
fields. Yukawa terms interacting 
with scalars are not difficult to formulate using  linear FEM truncated to
local terms to represent a minimal set of relevant operators.  The inclusion of
gauge fields interacting with our simplicial lattice fermions is
also straightforward for vector like theories  by replacing
the  spin connection $\Omega_{ij}$ on each link by the product
$\Omega_{ij} U_{ij}$ in the action for the Dirac field,
\be
S_{Wilson} = \frac{a}{2} \sum_{\<i,j\>} \frac{V_{ij} }{ l_{ij}^2 } (\bar \psi_i \hat
e^{j(i)}_a  \gamma^a  \Omega_{ij} U_{ij}\psi_j  -   \bar \psi_j \Omega_{ji} U_{ji}\hat e^{i(j)}_a   \gamma^a
\psi_{i}) 
 +\cdots \; ,
\ee
where $U_{ij}$ is the Wilson compact gauge link
matrix,
\be
U_{ij} = e^{\textstyle \; i  l^\mu_{ij}  A^\mu_{ij}} \; ,
\ee
and $A_{ij} = \lambda^a A^a_{ij} $ is the non-Abelian gauge potential. 
The kinetic term in the action has been considered in
Ref.~\cite{Christ:1982ci}   in flat space, but can be easily introduced  on our simplicial
manifold as well. The  continuum action
\be
S = \half  \int d^Dx \sqrt{g} \; g^{\mu\nu'}  
g^{\mu'\nu} F^a_{\mu \nu}(x) F^a_{\mu'\nu'}(x) \; ,
\label{eq:gauge}
\ee
is replaced by a finite element action as sum over all triangles,
\be
S_\sigma =  \frac{1}{2 g^2 N_c}\sum_{\triangle_{ijk}} 
\frac{V_{ijk}}{A^2_{ijk}} Tr[2 -U_{\triangle_{ijk}}  - U^\dag_{\triangle_{ijk}} ]
\ee
where 
where $A_{ijk} = |\sigma_2(ijk)|$ is the area of the triangle for the
plaquette, $V_{ijk} = | \sigma_2(ijk) \wedge \sigma^*_2(ijk)|$ is
the dual volume element, and the gauge matrix on each plaquette is the 
product $U_{\triangle_{ijk}}  = U_{ij} U_{jk} U_{ki}$. 
 The reader is referred to Ref.~\cite{Christ:1982ci} for the demonstration
that this has the correct continuum limit. 

The quantum field  path integral
on a simplicial lattice requires confronting UV divergences with 
additional counter terms as we will report in  Ref.~\cite{phi4}. Progress has been made for the Wilson-Fisher conformal fixed point in
$\phi^4$ theory by explicitly computing a finite number of UV
divergent diagrams on  the simplicial lattice.  The extension of this
approach  to other super-renormalizable theories appears promising, opening up
a new approach to lattice field theory with a view towards implementing 3D
lattice radial quantization. There are also many other interesting CFTs to explore by
developing code and algorithms similar to those in common use.               
 Our plan is to identify the geometrical properties of counter terms and, if possible,
develop the full QFE path integral in 4D, but we recognize that this is 
a difficult problem. We are optimistic that we will be able to achieve this within
our current QFE methodology via a single sequence of refined simplicial
lattices approaching the continuum Riemann manifold for UV complete
field theories. The guiding principle is to formulate non-perturbative
renormalization schemes similar to methods develop for lattice field theory in flat space
with the geometrical classification of the counter terms required in 
perturbative renormalization on Riemann manifold~\cite{Jack:1983sk,Jack:1984vj}. Other approaches such as a
quenched ensemble of simplicial lattices constrained to the target
manifold as advocated in  the random lattice
program~\cite{Christ:1982zq} for flat space may warrant further
investigation in spite of their increased computational complexity.

 \section*{Acknowledgments}
We would like to thank  Steven  Avery, Peter Boyle, Norman Christ,
Luigi Del Debbio, Martin L\"uscher,
Ami Katz, Zuhair Kandker and Matt Walters for valuable
discussions. R.C.B.  and G.T.F would like to
thank the Aspen Center and Kavali Institutes for their hospitality and
R.C.B thanks the Galileo Gallilei Institute's CFT workshop and the Higgs Centre for
their hospitality during the completion of this manuscript. GTF and
ADG would like to thank the Fermilab Theoretical Physics Summer
Visitors Program for continuing hospitality during the course of this
work.  R.C.B. and E.S.W. acknowledge support by DOE grant DE-SC0015845 and  T.R. and C.-I T. in part by the Department of Energy under contact DE-Sc0010010-Task-A.

\appendix

\section{\label{sec:Element}Dirac Finite Element}

The construction of our new piecewise flat Dirac finite element
described in Sec.~\ref{subsec:DiracFEM}  proceeds in the  following steps. 
We seek a new finite  expansion on each triangle
$\triangle_{123}$  
\bea
\psi  (x) &=&  E^1(x) \psi_1 +  E^2(x)
\psi_2 +  E^3(x)  \psi _3\, ,\nn
\bar \psi  (x)& =& E^1(x) \bar\psi_1 +  E^2(x)
\bar \psi_2 +  E^3(x)  \bar \psi _3\, .
\eea
in terms of  the new Elements, $E^i(x)$, which satisfy 3 conditions:
{\bf (i)}  The faithful interpolation  of the Dirac field requires $E^i(x = r_j) = \delta^i_j$,  at each vertex
$x= r_j$,  {\bf (ii)} the preservation of
constant fields $E^1(x) + E^2(x) + E^3(x) =1$, and {\bf (iii)}  the
lattice Dirac equation propagates on each link $\<i,j\>$ with the spin matrix
$\vec l_{ij} \cdot \vec \sigma$.  Surprisingly, all three constraints
have a simple solution in terms of three sub-triangles with linear elements meeting
at the circumcenter with a  ghost field 
\be
\psi_0= c_1\psi_1 + c_2 \psi_2 + c_3 \psi_3  \quad ,\quad \bar
\psi_0= c_1 \bar \psi_1 + c_2 \bar \psi_2 + c_3 \bar \psi_3 \; .
  \label{eq:circumcenter}
\ee
given as a linear function of the  values at the vertices.
The calculation  requires  computing the action and applying these
constraints to  determine the values of the coefficients $c_i$.

\begin{figure}[ht]
		\begin{center}
		\vspace{10pt}
\includegraphics[width=.8\textwidth]{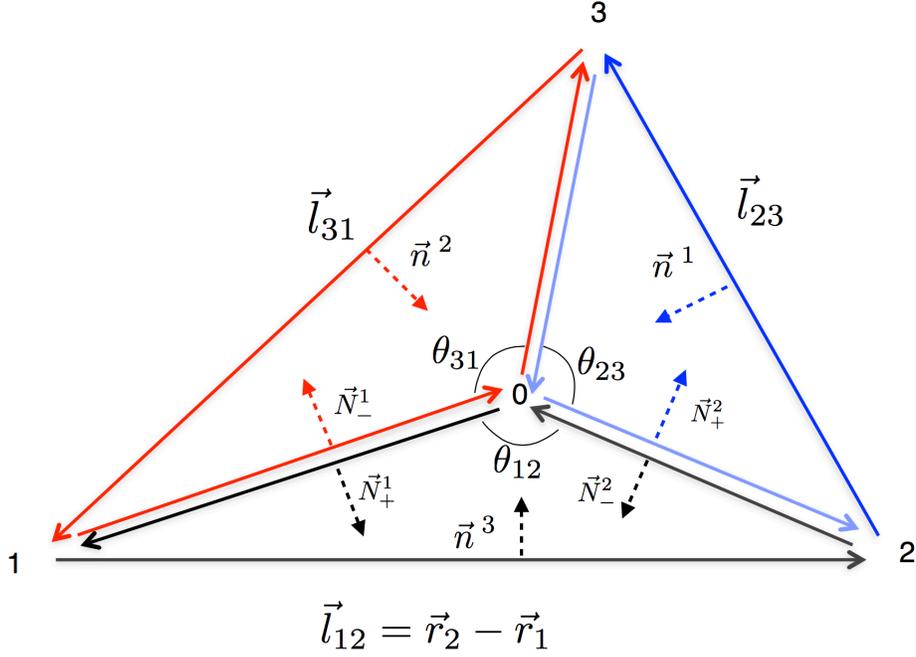}
   	\caption{ \label{fig:DiracFE2} Each triangle on the simplicial
          lattice given by the $\triangle_{123}$ with vertices $\vec
          r_1, \vec r_2, \vec r_3$ is divided into 3 isosceles sub-triangles
          meeting at the circumcenter at $0$. }
   	\end{center}
\end{figure}

The basic algebra relies on the geometry illustrated on the left in Fig. \ref{fig:DiracFE2}
by vectors/dual-vectors, $(\vec l_{ij},\vec n{\;}^k)$.  The simplex
for $\triangle_{123}$  has normal vectors $(\vec n{\;}^1,\vec
n{\;}^2,\vec n{\;}^3)$. In addition, each of the sub-triangle have
the normals $(\vec N_+^{i},
\vec N_0^i, \vec N_-^{i+1})$, where   the normal on the exterior links,  $\vec N_0^i$, 
are just rescaled   from $\vec n{\;}^i $ by $\vec N_0^i = (A_{123}/A_{0
  i,i+1} ) \,\vec n{\;}^i $.  As a consequence we have the sum rules
\be
\vec N_+^{i}+\vec N_0^i +\vec N_-^{i+1} = 0, \quad\quad A_{023} \vec N_0^1 + A_{031}\vec N_0^2 + A_{012}\vec N_0^3 =0.
\ee 
In addition, normals to a shared link for two adjacent sub-triangles
are related by $ A_{0, i-1,i}\vec N^i_- + A_{0, i,i+1}\vec
N^i_+=0$.
All of these relations hold for an arbitrary location for the center
vertex $0$.

Restricting vertex $0$ to the circumcenter leads to three isosceles
sub-triangles and there are now additional geometrical constraints.
Within each sub-triangle, the sum and difference of $\vec N^i_+$ and
$\vec N^{i+1}_-$ are perpendicular and parallel to the opposite link
vector, $\vec l_{i,i+1}$, respectively, i.e.,
\be
\vec  l_{i,i+1} \cdot  (\vec N^i_+ +\vec N^{i+1}_ -)=0 \quad,
\quad  \vec N^i_+ - \vec N^{i+1}_ -  = \frac{2 \vec l_{i,i+1}}{l^2_{i,i+1}} \; ,  
\ee
Applying the linear  FEM  interpolation formula, Eq.~(\ref{eq:nn}), to each sub-triangle 
$\Delta_{0,i,i+1}$  we have,
\bea
S_{0 i,i+1} &=&  \frac{A_{0,i,i+1}}{6} \large[ \bar \psi_i (\vec N^+_i - \vec
N^-_{i+1})\cdot \vec\sigma \psi_{i+1}  + 
\bar \psi_{i+1} (\vec N^0_{i+2}  -\vec N^+_i ) \cdot\vec \sigma \psi_0 
+ \bar \psi_0 (\vec N^-_{i+1}- \vec N^0_{i+2} )\cdot \vec\sigma \psi_i \;
\large]  \nn
&-& c.c.
\eea
and the sum $S_{123} = S_{0,12}  + S_{0,23}  +
S_{0,31}$, with the help of the identity, $(  A_{0,i,i+1} \vec N^+_i
+A_{0,i+1,i+2}   \vec N^-_{i+2} )  =   A_{0,i,i+2}   \vec N^0_{i+1}$,
gives 
\bea
S_{123} &=&\frac{1}{3} \sum_i A_{0,i,i+1}
 [ \bar \psi_i (\vec N^+_i - \vec
N_-^{i+1}) \cdot\vec \sigma \psi_{i+1}  - \bar \psi_{i+1}  (\vec N_+^i - \vec
N_-^{i+1})\cdot\vec \sigma \psi_{i}  ]\nonumber\\
&+&  \frac{1}{3}    \sum_i A_{0,i-1,i+1} [ \bar \psi_{i} \vec \sigma \cdot   
\vec N_0^{i}  \psi_{0}   - \bar \psi_{0} \vec \sigma \cdot   \vec N_0^{i}  \psi_{i}].
\eea

Introducing the expansion for the ghost field $\psi_0$ and $\bar
\psi_0$,  the link $\bar \psi_1 \vec
\sigma \psi_2$ receives contributions from both the $\bar\psi_1 \vec
\sigma\psi_0$ and $\bar \psi_0\vec \sigma \psi_2$ terms. We now require
that each  edge is properly aligned,  
\be
\frac{A_{012}}{3 l^2_{12}}  \vec l_{12}  - \frac{A_{031}}{3} 
\vec N_0^2 c_1 +   \frac{A_{023}}{3}   \vec
N_0^1 c_2 
 \sim \vec l_{12} 
\label{eq:aligned}
\ee
plus permutations for the $\bar \psi_2 \vec
\sigma \psi_3$ and $\bar \psi_3 \vec
\sigma \psi_1$ links.  The first term in each equation is  already in the form we are
seeking.  Now we have what appears to be an over constrained system
for three coefficients $c_i$ satisfying the normalization constraint
$c_1 + c_2 + c_3 =1$.

An efficient approach to solving for these coefficients is to project these equations in
the perpendicular direction by taking the scalar product
with $\vec N_0^3, \vec N_0^1, \vec N_0^2,$ respectively.   After some algebra, using
the identity $\vec l_{ik} \cdot \vec l_{kj} =4 A^2_{123}\vec n{\;}^i
\cdot \vec n {\;}^{j}$, this reduces to a homogeneous matrix equation,
\be
\begin{bmatrix}
\vec l_{31}  \cdot \vec l_{12}  &  - \vec l_{23}  \cdot \vec l_{12} & 0\\
 0& \vec l_{12} \cdot \vec l_{23}& - \vec l_{31} \cdot \vec l_{23}\\
-\vec l_{12}   \cdot  \vec l_{31}& 0  & \vec l_{23}
\cdot  \vec l_{31}
\end{bmatrix}
\begin{bmatrix}
c_1 \\
c_2\\
c_3
\end{bmatrix}
=0
\ee
As the determinant is zero, a non-trivial null vector exists, given by
\be
c_k= c_0 \;\frac{A_{123} }{\vec l_{ik} \cdot \vec l_{kj}}
\label{eq:Null}
\ee
for $ikj = (123)$ and cyclic, where we have expressed the solution
up to a undetermined dimensionless constant $c_0$, which can be
chosen to satisfy the normalization $c_1 + c_2 + c_3 =1$.  After
considerable algebraic manipulation the final solution becomes,
\be 
c_k=  \frac{4A_{0ik}}{ l_{ik}^2} \frac{4A_{0jk}}{ l_{jk}^2} =
\cot(\theta_{ik}/2) \cot(\theta_{jk}/2) \; ,
\label{eq:Normalized}
\ee
where the vertex angle for each isosceles triangle is given by
$\cot(\theta_{ij}/2) = 2h_{ij}/l_{ij} =4 A_{0ij}/l^2_{ij}  $. 

The consistency between Eq.~(\ref{eq:Null}) and
Eq.~(\ref{eq:Normalized}) prior to this normalization requires only
that the ratios $c_i/c_j$ are unchanged which may be verified using
the following set of identities. Let $R$ be the circum-radius such that one
has $l_{ij}= 2 R \sin(\theta_{ij}/2)$. The total area can be
expressed symmetrically as
$A_{123}= l_{12} l_{23} l_{31}/4 R= 2 R^2 \sin(\theta_{12}/2)
\sin(\theta_{23}/2) \sin(\theta_{31}/2) $.
The equation for the scalar product leads to:
\be
 \vec l_{ik} \cdot \vec l_{kj} = 4 R^2 \cos(\theta_{ik}/2
+\theta_{kj}/2) \sin(\theta_{ik}/2) \sin(\theta_{kj}/2))  = 2 A_{123}
\cot(\theta_{ij}/2) .
\ee
 The normalization condition, $c_1 + c_2 + c_3 =1$,  follows from the elegant  identity, 
\be
\tan(\theta_{12}/2) + \tan(\theta_{23}/2) + \tan(\theta_{31}/2)  = \tan (\theta_{12}/2) \tan (\theta_{23}/2) \tan
(\theta_{31}/2) \, ,
\ee
for $\theta_{12} + \theta_{23} + \theta_{31}=2 \pi$. 
Geometrically, this identity reflects the
fact that the area of the triangle equals 
the sum of areas of three sub-triangles, $A_{123}=A_{012}+    A_{023} + A_{031}$. 
Remarkably, with $c_i$ appropriately chosen, the additional two terms
in Eq.~(\ref{eq:aligned}) are
not only aligned with the first one but  the sum of all three provides
precisely the FEM weight for our conjectured Dirac ansatz above. It is appealing that
the use of the dual vertex is necessary to the construction 
analogous to our Discrete Exterior Calculus formulation of the
scalar. Generalizations of this construction 
for $D>2$ using the dual lattice are being sought. 

Let us end with two additional comments.
First, if we choose $c_i=\xi_i^*= A_{0jk}/A_{123}$, we get back
to the naive linear FEM result for the entire triangle $\Delta_{123}$,
which, as stated earlier, does not lead to Eq.~(\ref{eq:canonical}). Second, 
if one  chooses an arbitrary point $0$ inside the triangle, instead of the circumcenter, it is still possible to
adjust the  coefficient $c_i$ so that propagator on the links 
is aligned with $\vec l_{ij}$. However, the magnitude 
does not  agree with our ansatz  in (\ref{eq:canonical}), and it does not
admit a simple geometrical interpretation.

\section{\label{app:continuum}Spectrum of the Dirac Fermion on $\mathbb{S}^2$}

Here we rederive the Dirac operator on $\mS^2$ by starting from the
Dirac fermion
in 4D projected to the 2 sphere. In 4D, consider the change of variables from Cartesian to spherical
coordinates, $x^\mu = (t,\vec r)$, where
$\vec r= (r \sin\theta \cos \phi, r \sin \theta \sin \phi, r \cos \theta)$.
The fermion action, $\bar \psi \gamma^\mu \dd_{\mu} \psi$, can be
re-expressed as 
\be
\gamma^\mu \dd_{\mu}  = \gamma_0 \dd_t + \frac{1}{r} (\gamma^r 
\dd_{\log(r)} + \gamma^\theta \dd_\theta + \frac{1}{\sin\theta}
\gamma^\phi \dd_\phi )  \; ,
\ee
where
$\gamma^r = \hat e_r \cdot \vec \gamma = \sin \theta ( \cos \phi
\gamma^1 + \sin \phi \gamma^2) + \cos \theta \gamma^3$,
$\gamma^\theta= \hat e_\theta \cdot \vec \gamma = \cos \theta ( \cos
\phi \gamma^1 + \sin \phi \gamma^2) - \sin \theta \gamma^3$
and
$\gamma^\phi =\hat e_\phi \cdot \vec \gamma = - \sin \phi \gamma^1 +
\cos\phi \gamma^2 $. The freedom to rotate tangent 
vectors allows one to rotate $\hat e_r$ to
$\hat e^3$. This can be done by first rotating $\hat e_\phi$ to
$\hat e^2$ and then rotating $\hat e_\theta$ to $\hat e^1$.
Equivalently, one rotates the fermion spinors,
$\psi\rightarrow \Lambda \psi$ and
$\bar \psi\rightarrow \bar\psi \Lambda^\dagger $ which 
then rotates $ \gamma^\mu \rightarrow \Lambda^\dag \gamma^\mu \Lambda = O^\mu_\nu \gamma^\nu $,
where $ \Lambda = \Lambda_{12}(\phi) \Lambda_{13}(\theta) = e^{\textstyle  \frac{i}{2}\phi
  \sigma^{12}} \; e^{\textstyle \frac{i}{2} \theta\sigma^{13}} $.  The  gauge transformation picks up an additional term,
$\bar\psi \gamma^\mu \dd_{\mu} \psi \rightarrow \bar\psi O^\mu_\nu
\gamma^\nu \dd_{\mu} \psi + \bar\psi (\Lambda^\dagger\gamma^\mu \Lambda) (\Lambda^\dag
\dd_{\mu} \Lambda) \psi $, or spin connection so the Dirac operator in
this frame is 
\be
 e^\mu_c\gamma^c (\dd_\mu + \frac{1}{4}\omega^\mu_{ab}\gamma^a \gamma^b)
= \gamma_0 \dd_t + \frac{1}{r} [\gamma^3
\dd_{\log(r)} + \gamma^1 \dd_\theta + \frac{1}{\sin\theta}
\gamma^2 \dd_\phi   + \gamma^3 + \frac{ \cot \theta}{2}
\gamma^1] \; .  
\ee
The static approximation removes the $\gamma_0$ reducing it to 3D. The
radial quantization on $\mR \times \mS^2$ rescales the fields ($\psi\rightarrow r^{-1} \psi$,  $\bar
\psi\rightarrow r^{-1} \bar\psi$) , placing the 2D Dirac action on the unit $\mS^2$ given by
\be
S= \int \sin\theta d\theta d\phi~\bar \psi [ \gamma^1
(\dd_\theta+\frac{\cot \theta}{2}) + \frac{1}{\sin\theta} \gamma^2
\dd_\phi  ] \psi \; .
\ee
 This is two
copies of 2 component fermions with action,
\be
S= \int \sin \theta d\theta d\phi \bar \psi (\sigma^\mu\boldsymbol{D}_\mu+m) \psi 
= \int \sin \theta d\theta d\phi\,  \bar \psi [ \sigma^1 ( \partial_\theta + \frac{\cot \theta}{2}) + \frac{1}{\sin\theta}
\sigma^2 \dd_\phi ]\psi,    
\label{eq:sphere}
\ee
in agreement with Eq.~(\ref{eq:actionsphere}),  as promised.  The
term $\frac{1}{2} \cot\theta$ corresponds to a spin connection on $\mS^2$.  
Defining 
\be
 \nabla   = \sigma^1 (\dd_\theta+\cot\theta/2) + \frac{1}{\sin\theta}
 \sigma^2 \dd_\phi \; .
 \ee
we  turn next to the spectrum,  $\nabla \psi   = i \lambda \psi$,  of the massless Dirac
  operator~\cite{Abrikosov:2002jr,Villalba1994,Camporesi1996} on $\mS^2$.

For the positive spectrum, $\lambda_+ >0$, the eigen-functions are designated by $\xi_+(\theta, \phi)$. The 
analysis can be done by the usual procedure, by seperation of
variables and Fourier expansion of the spinor $\xi_+(\theta, \phi)$ in $\phi$,
$\xi_+(\theta,\phi)=\sum_m e^{-im\phi} f_+(\theta)$, leading to a
first order ordinary differential equation in $\theta$ for a
two-component spinor $f_+(\theta)$. It can be shown that spinors in
this gauge are anti-periodic  in $\phi$ and  $m= n+1/2$ takes on
half-integral values.  This leads to a coupled first order ODE
between its upper and lower components, which after one iteration
gives an ordinary second order ODE separately for the upper and
the lower component. 

By imposing a normalizability condition on $f_+(\theta)$, the
discrete spectrum can be found, with eigenvalues,
\be
\lambda_+= j+1/2
\ee
where $j=1/2, 3/2, \cdots$ and $-j\leq m\leq j$.  That is, for each
$\lambda_+$, there is a $(2j+1)$-fold degeneracy due to rotational
invariance.
The corresponding wave functions can be expressed in terms of Jacobi polynomials, $P_n^{(\alpha,\beta)}$, 
\be\label{eq:wfn}
\xi_{+,(j,m)}(\theta,\phi) =  C^+_{jm}\,  e^{i m \phi} \, 
\begin{pmatrix} \sin ^m(\theta/2) \cos ^{m+1}(\theta/2) P_{j-m}^{\left(m-\frac{1}{2},m+\frac{1}{2}\right)}(\cos \theta ) \\
 i\hspace{3pt}\sin ^{m+1}(\theta/2) \cos
 ^{m}(\theta/2)
 P_{j-m}^{\left(m+\frac{1}{2},m-\frac{1}{2}\right)}(\cos \theta )  
\end{pmatrix}.
\ee
The eigenfunctions corresponding to the negative eigenvalues,
$\lambda_-$, can be obtained via
$\xi_{-,(j,m)}(\theta,\phi)=i\sigma_3 \phi_{+,(j,m)}$.  For the
record, we note that these wave functions are normalized so that
\be
\int_{S^2}\text{sin}\theta d\theta d\phi\, \xi^{\dagger}_{\epsilon,(j,m)}\xi_{\epsilon',(j',m')}=\delta_{\epsilon,\epsilon'}\delta_{j,j'}\delta_{m,m'},
\label{eq:normalization}
\ee
 with $C^+_{jm}$ given in Ref.~\cite{Abrikosov:2002jr}. 
 
 By performing a local rotation, it is also possible to express these wave functions in terms of  the usual spherical harmonics, $Y_{l m}$~\cite{Abrikosov:2002jr}. Introducing $\Phi^\pm(j,m)  = V^\dagger \xi_\pm$, where $V^\dagger = e^{i\theta\sigma_2/2} e^{-i\phi\sigma_3/2}$, one finds that 
\bea
\Phi^\pm(j,m) 
 &=& \frac{(1\pm i)  }{2}  \left(\begin{array}{l}
  \sqrt{\frac{(\ell+m)}{4 \ell}}Y_{j^-,m^-}(\theta,\phi) \mp i \sqrt{\frac{(j-m+1)}{4( j+1)}}Y_{j^+,m^-}(\theta,\phi)  \\
\sqrt{\frac{(j-m)}{4 j}}Y_{j^-,m^+}(\theta,\phi)  \pm i  \sqrt{\frac{(j+m+1)}{4( j+1)}}Y_{j^+,m^+}(\theta,\phi) \end{array}\right)
\eea
where $j^\pm=j\pm \half$ and $m^\pm = m \pm \half$.

Finally let's give a direct evaluation of  the  Lichnerowicz 
formula,
\be
- \nabla^2 = - \frac{1}{\sqrt{g}} \boldsymbol{D}_\mu
\sqrt{g} g^{\mu \nu} \boldsymbol{D} _\nu  + \frac{1}{2} \sigma^{ab}
e_a^\mu e_b^\nu \boldsymbol{R}_{\mu \nu} \; .
\label{eq:Lichnerowicz}
\ee
 in Eq.~(\ref{eq:SimDiracSquared}). 
\ignore{
By the tetra hypothesis, this is 
related to the positive definite operator  $(\sqrt{g} {\bf e}^\mu
\boldsymbol{D}_\mu ) (\sqrt{g} {\bf e}^\nu \boldsymbol{D}_\mu )^\dag =
-\frac{1}{\sqrt{g}}\nabla^2$. On the unit radius $\mS^2$, we see that
$\frac{1}{2} \sigma^{ab}   e^\mu_a e^\nu_b R_{\mu\nu}=\frac{1}{2}$. } 
On $\mS^2$ the  operator 
\be
-\nabla^2 = -[\dd^2_\theta   + \cot \theta \dd_\theta +
\frac{1}{\sin^2\theta}  \dd^2_\phi - i \sigma_3
\frac{\cos\theta}{\sin^2\theta}  \dd_\phi  - \frac{1}{4\sin 2\theta} -\frac{1}{4}]. 
\label{eq:LichS1}
\ee
has spectrum $(j + 1/2)^2$, which is naturally the absolute value square of the Dirac
operator spectrum $\pm i(j + 1/2)$.  It follows that  in 2D the
covariant spinor Laplacian alone, which is the first term in the Lichnerowicz formula ~(\ref{eq:Lichnerowicz}),  has
eigenvalues, $(j+1/2)^2-1/2 = j(j+1) -1/4$ in accord with our
numerical evaluation of the Wilson term (\ref{eq:continuum-lambda}) 
in Sec.~\ref{sec:Spectrum}.
 
\newpage

\bibliographystyle{unsrt}
\bibliography{bib/QFE,bib/fermionrefs,bib/RCB}

\begin{thebibliography}{10}

\bibitem{Appelquist:2013sia}
Thomas Appelquist, Richard Brower, Simon Catterall, George Fleming, Joel Giedt,
  Anna Hasenfratz, Julius Kuti, Ethan Neil, and David Schaich.
\newblock {Lattice Gauge Theories at the Energy Frontier}.
\newblock In {\em {Community Summer Study 2013: Snowmass on the Mississippi
  (CSS2013) Minneapolis, MN, USA, July 29-August 6, 2013}}, 2013.

\bibitem{Brower:2012vg}
R.C. Brower, G.T. Fleming, and H.~Neuberger.
\newblock {Lattice Radial Quantization: 3D Ising}.
\newblock {\em Phys.Lett.}, B721:299--305, 2013.

\bibitem{Brower:2014gsa}
Richard~C. Brower, Michael Cheng, and George~T. Fleming.
\newblock {Improved Lattice Radial Quantization}.
\newblock {\em PoS}, LATTICE2013:335, 2014.

\bibitem{Brower:2015zea}
Richard~C. Brower, Michael Cheng, and George~T. Fleming.
\newblock {Quantum Finite Elements: 2D Ising CFT on a Spherical Manifold}.
\newblock {\em PoS}, LATTICE2014:318, 2015.

\bibitem{Brower:2016moq}
Richard~C. Brower, George Fleming, Andrew Gasbarro, Timothy Raben, Chung-I Tan,
  and Evan Weinberg.
\newblock {Quantum Finite Elements for Lattice Field Theory}.
\newblock In {\em {Proceedings, 33rd International Symposium on Lattice Field
  Theory (Lattice 2015)}}, 2016.

\bibitem{Regge:1961px}
T.~Regge.
\newblock {GENERAL RELATIVITY WITHOUT COORDINATES}.
\newblock {\em Nuovo Cim.}, 19:558--571, 1961.

\bibitem{Hamber:2009mt}
Herbert~W. Hamber.
\newblock {Quantum Gravity on the Lattice}.
\newblock {\em Gen. Rel. Grav.}, 41:817--876, 2009.

\bibitem{StrangFix200805}
Gilbert Strang and George Fix.
\newblock {\em An Analysis of the Finite Element Method 2nd Edition}.
\newblock Wellesley-Cambridge, 2nd edition, 5 2008.

\bibitem{2005math8341D}
M.~{Desbrun}, A.~N. {Hirani}, M.~{Leok}, and J.~E. {Marsden}.
\newblock {Discrete Exterior Calculus}.
\newblock {\em ArXiv Mathematics e-prints}, August 2005.

\bibitem{Christ:1982ci}
N.~H. Christ, R.~Friedberg, and T.~D. Lee.
\newblock {Weights of Links and Plaquettes in a Random Lattice}.
\newblock {\em Nucl. Phys.}, B210:337, 1982.

\bibitem{Christ1982}
N.~H. Christ, R.~Friedberg, and T.~D. Lee.
\newblock {Random Lattice Field Theory: General Formulation}.
\newblock {\em Nucl. Phys.}, B202:89, 1982.

\bibitem{Christ1982a}
N.~H. Christ, R.~Friedberg, and T.~D. Lee.
\newblock {Weights of Links and Plaquettes in a Random Lattice}.
\newblock {\em Nucl. Phys.}, B210:337, 1982.

\bibitem{phi4}
Richard~C. Brower, Michael Cheng, Evan Weinberg, George~T. Fleming, Andrew~D.
  Gasbarro, Timothy Raben, and Chung-I Tan.
\newblock {Quantum Finite Elements for Scalar Field Theory on the Riemann
  Sphere}.
\newblock In preparation., 2016.

\bibitem{CT}
Edwin~H. Spanier.
\newblock {Algebraic Topology}.
\newblock {\em McGraw-Hill}.

\bibitem{Friedberg:1985wr}
R.~Friedberg, T.~D. Lee, and Hai-cang Ren.
\newblock {FERMION FIELD ON A RANDOM LATTICE}.
\newblock {\em Prog. Theor. Phys. Suppl.}, 86:322, 1986.

\bibitem{Banks:1982iq}
Tom Banks, Y.~Dothan, and D.~Horn.
\newblock {GEOMETRIC FERMIONS}.
\newblock {\em Phys. Lett.}, B117:413--417, 1982.

\bibitem{Watterson:2005dd}
Steven Watterson and James Sexton.
\newblock {Distributing the chiral and flavor components of Dirac-Kahler
  fermions across multiple lattices}.
\newblock {\em PoS}, LAT2005:277, 2006.

\bibitem{Miller:1997wb}
Warner~A. Miller.
\newblock {The Hilbert action in Regge calculus}.
\newblock {\em Class. Quant. Grav.}, 14:L199--L204, 1997.

\bibitem{Hamber:2007fk}
Herbert~W. Hamber.
\newblock {Discrete and continuum quantum gravity}.
\newblock 2007.

\bibitem{Bogacz2001}
L.~Bogacz, Z.~Burda, J.~Jurkiewicz, A.~Krzywicki, C.~Petersen, and
  B.~Petersson.
\newblock {Dirac operator and Ising model on a compact 2-D random lattice}.
\newblock {\em Acta Phys. Polon.}, B32:4121--4168, 2001.

\bibitem{Bogacz2002}
L.~Bogacz, Z.~Burda, C.~Petersen, and B.~Petersson.
\newblock {Spectrum of the Dirac operator coupled to two-dimensional quantum
  gravity}.
\newblock {\em Nucl. Phys.}, B630:339--358, 2002.

\bibitem{Burda1999}
Z.~Burda, J.~Jurkiewicz, and A.~Krzywicki.
\newblock {Wilson fermions on a randomly triangulated manifold}.
\newblock {\em Phys. Rev.}, D60:105029, 1999.

\bibitem{Caselle:1989hd}
M.~Caselle, A.~D'Adda, and Lorenzo Magnea.
\newblock {Regge Calculus as a Local Theory of the Poincare Group}.
\newblock {\em Phys. Lett.}, B232:457--461, 1989.

\bibitem{Brewin:1996yk}
Leo Brewin.
\newblock {Riemann normal coordinates, smooth lattices and numerical
  relativity}.
\newblock {\em Class. Quant. Grav.}, 15:3085--3120, 1998.

\bibitem{Isham:1982gk}
C.~J. Isham and C.~N. Pope.
\newblock {A Spinor Field Representation of the Stiefel-whitney Class}.
\newblock {\em Phys. Lett.}, B114:137--140, 1982.

\bibitem{Brewin:2009se}
Leo Brewin.
\newblock {Riemann Normal Coordinate expansions using Cadabra}.
\newblock {\em Class. Quant. Grav.}, 26:175017, 2009.

\bibitem{Abrikosov:2002jr}
A.~A. Abrikosov, Jr.
\newblock {Dirac operator on the Riemann sphere}.
\newblock 2002.

\bibitem{Chodos1977}
A.~Chodos and J.~B. Healy.
\newblock {Spectral Degeneracy of the Lattice Dirac Equation as a Function of
  Lattice Shape}.
\newblock {\em Nucl. Phys.}, B127:426, 1977.

\bibitem{Celmaster:1982ub}
William Celmaster and Frank Krausz.
\newblock {LOSS OF CONTINUUM LORENTZ INVARIANCE IN GAUGE THEORIES ON A
  TRIANGULAR LATTICE}.
\newblock {\em Nucl. Phys.}, B220:434--446, 1983.

\bibitem{Narayanan:1993sk}
Rajamani Narayanan and Herbert Neuberger.
\newblock {Chiral determinant as an overlap of two vacua}.
\newblock {\em Nucl. Phys.}, B412:574--606, 1994.

\bibitem{Christe:1993ij}
Philippe Christe and Malte Henkel.
\newblock {Introduction to Conformal Invariance and its Applications to
  Critical Phenomena}.
\newblock {\em Lect. Notes Phys.}, M16:1--260, 1993.

\bibitem{Rychkov:2016iqz}
Slava Rychkov.
\newblock {EPFL Lectures on Conformal Field Theory in $D \ge 3$ Dimensions}.
\newblock 2016.

\bibitem{Shamir:1999gv}
Yigal Shamir.
\newblock {Better domain wall fermions}.
\newblock In {\em {Lattice fermions and structure of the vacuum. Proceedings,
  NATO Advanced Research Workshop, Dubna, Russia, October 5-9, 1999}}, pages
  27--39, 1999.

\bibitem{Brower:2012vk}
Richard~C. Brower, Harmut Neff, and Kostas Orginos.
\newblock {The M\"obius Domain Wall Fermion Algorithm}.
\newblock 2012.

\bibitem{Jack:1983sk}
I.~Jack and H.~Osborn.
\newblock {Background Field Calculations in Curved Space-time. 1. General
  Formalism and Application to Scalar Fields}.
\newblock {\em Nucl. Phys.}, B234:331--364, 1984.

\bibitem{Jack:1984vj}
I.~Jack and H.~Osborn.
\newblock {General Background Field Calculations With Fermion Fields}.
\newblock {\em Nucl. Phys.}, B249:472--506, 1985.

\bibitem{Christ:1982zq}
N.H. Christ, R.~Friedberg, and T.D. Lee.
\newblock {Random Lattice Field Theory: General Formulation}.
\newblock {\em Nucl.Phys.}, B202:89, 1982.

\bibitem{Villalba1994}
Victor~M. Villalba.
\newblock {The Angular momentum operator in the Dirac equation}.
\newblock {\em Eur. J. Phys.}, 15:191, 1994.

\bibitem{Camporesi1996}
Roberto Camporesi and Atsushi Higuchi.
\newblock {On the Eigen functions of the Dirac operator on spheres and real
  hyperbolic spaces}.
\newblock {\em J. Geom. Phys.}, 20:1--18, 1996.

\end{thebibliography}

\end{document}